\documentclass[11pt,a4paper]{article}

\usepackage{jheppub}
\usepackage[usenames,dvipsnames]{xcolor}
\usepackage{amssymb} 
\usepackage{MnSymbol}
\usepackage{pgfplots}
\pgfplotsset{compat=1.17}
\usetikzlibrary{pgfplots.fillbetween}
\usepackage{amsmath}
\usepackage{mathtools}
\usepackage{amsfonts}  
\usepackage{dsfont}
\usepackage{pdfpages}
\usepackage{verbatim}
\usepackage{stmaryrd}
\usepackage{graphicx}
\usepackage{tabularx}        
\usetikzlibrary{positioning}
\usepackage{hhline}
\usepackage{multirow}
\usepackage[most]{tcolorbox}

\usepackage{float}
\usepackage{bbm}
\usepackage{tensor}
\usepackage{mathrsfs}
\usepackage{textgreek} 
\usepackage{euscript}
\usepackage[smalltableaux]{ytableau}
\ytableausetup{centertableaux}
\usepackage{tikz}
\usetikzlibrary{arrows.meta}
\usetikzlibrary{decorations.pathreplacing, decorations.markings,calc,shapes.misc,decorations.pathmorphing,patterns.meta, math}
\usetikzlibrary{decorations.pathmorphing}
\usetikzlibrary{decorations.markings}
\usepackage{slashed}
\usepackage{datetime}
\usepackage{hyperref}
\usepackage{braket}
\hypersetup{
    pdfencoding=unicode,
	colorlinks=true,
	urlcolor=Maroon,
	linkcolor=RoyalBlue,
	citecolor=Maroon,
	pdftitle={},
	pdfauthor={},
	pdfdisplaydoctitle=true,
	pdfstartview=FitH,
	linktocpage=true
}
\usepackage{pgfplots}
\pgfplotsset{compat=1.17}

\usetikzlibrary{shapes}
\usetikzlibrary{arrows}
\usetikzlibrary{decorations.pathmorphing}
\usetikzlibrary{decorations.markings}
\usetikzlibrary{shapes.misc}
\tikzset{snake it/.style={decorate, decoration=snake}}
\tikzset{cross/.style={cross out, draw=black, minimum size=2*(#1-\pgflinewidth), inner sep=0pt, outer sep=0pt},
cross/.default={1pt}}
\usetikzlibrary{shapes.geometric}
\tikzset{
    partial ellipse/.style args={#1:#2:#3}{
        insert path={+ (#1:#3) arc (#1:#2:#3)}
    }
}

\usepackage{subcaption}

\newcommand{\ba}{\begin{align}}

\newcommand{\be}{\begin{equation}}
\newcommand{\ee}{\end{equation}}
\def\bd{\begin{tikzpicture}}
\def\ed{\end{tikzpicture}}

\newcommand\Res{\mathop{\text{Res}}}

\allowdisplaybreaks

\def\XXint#1#2#3{{\setbox0=\hbox{$#1{#2#3}{\int}$}
     \vcenter{\hbox{$#2#3$}}\kern-.5\wd0}}

\definecolor{light-gray}{gray}{0.75}

\newcommand\dd{\mathrm{d}}

\renewcommand\d{\text{d}}

\newcommand{\e}{\mathrm{e}}

\newcommand{\ii}{\mathrm{i}}

\renewcommand{\le}{\leqslant}

\renewcommand{\leq}{\leqslant}
\renewcommand{\geq}{\geqslant}

\newcommand\tpsi{\widetilde{\psi}}
\newcommand\tF{\widetilde{F}}

\newcommand\fb{\mathsf{b}}
\newcommand\fQ{\mathsf{Q}}
\newcommand\fk{\mathsf{k}}

\newcommand{\tepsilon}{\widetilde{\epsilon}}

\newcommand{\btepsilon}{\overline{\widetilde{\epsilon}}}

\newcommand{\tvarphi}{\widetilde{\varphi}}

\newcommand{\bpsi}{\overline{\psi}}
\newcommand{\btpsi}{\overline{\widetilde{\psi}}}

\newcommand{\talpha}{\tilde{\alpha}}

\newcommand{\n}{\mathcal{N}}
\newcommand{\rsr}{\langle r,s\rangle}

\definecolor{bleudefrance}{rgb}{0.19, 0.55, 0.91}

\definecolor{vert}{rgb}{0.1367 0.543 0.1367}

\definecolor{pink}{rgb}{1.0, 0.13, 0.32}

\newcommand{\ie}{{\it i.e.~}}
\def\eg{{\it e.g.~}}

\begin{document}

\vspace*{2.5cm}
\begin{center}
{{\fontsize{15.8pt}{19pt}\selectfont\textbf{\textsc{Toward the Structure Constants of
$\mathcal{N}=2$ Liouville Theory\\
}}}}
\vspace*{.4cm}

\vspace*{1.7cm}

{\bf
\mbox{
Beatrix M\"uhlmann$^{\n=(2,0)}$ and
Themistocles Zikopoulos$^{\n=(0,2)}$}
}

\vspace*{0.6cm}

{\footnotesize
$^{(2,0)}$ School of Natural Sciences, Institute for Advanced Study, Princeton, NJ 08540, USA\\
$^{(0,2)}$ Perimeter Institute for Theoretical Physics, Waterloo, ON N2L 2Y5, Canada\\
}

\vspace*{0.4cm}

{\footnotesize\textsf{
beatrix@ias.edu,
tzikopoulos@perimeterinstitute.ca
}}

\vspace*{0.6cm}
\end{center}

\vspace*{1.5cm}
\noindent
\begin{abstract}\noindent
We discuss the structure constants of spacelike $\mathcal{N}=2$ Liouville theory on the two-sphere.  Due to the absence of a particular $\fb \leftrightarrow \fb^{-1}$ self-dual symmetry, where $\fb$ is the theory's coupling, the standard analytic bootstrap toolkit that was used to solve the bosonic and $\n=1$ theories cannot be implemented in a straightforward way. Our approach, instead, relies on the fact that $\mathcal{N}=2$ Liouville theory is dual by mirror symmetry to the $\mathrm{SL}(2)_\mathsf{k}/\mathrm{U}(1)$ supercoset, whose target space is the $2$d fermionic black hole.~Leveraging this duality, we obtain explicit expressions for the winding number preserving and violating structure constants on the supercoset side, and test them on the Liouville side through a semiclassical analysis finding agreement up to one-loop order.~We also discuss the chiral rings of the theory and evaluate correlators of $\frac 12$-BPS operators.
\end{abstract}

\newpage

\tableofcontents

\section{Introduction}

Liouville theory occupies a central role in modern theoretical physics. It has deep relations with two-dimensional quantum gravity \cite{Knizhnik:1988ak, David:1988hj, Distler:1988jt,Polchinski:1989fn}, string theory \cite{Polyakov:1981rd, Seiberg:2004at, Moore:1991zv} (see also the recent works \cite{Collier:2023cyw, Collier:2024kwt, Collier:2024lys, Collier:2024kmo, Collier:2025lux, Blommaert:2025eps}), higher-spin theory \cite{Fateev:2007ab}, random matrix models \cite{Brezin:1990rb, Douglas:1989ve, Gross:1989vs} and four-dimensional $\n=2$ gauge theories~\cite{Alday:2009aq}, just to mention a few. 

As an irrational and non-compact conformal field theory (CFT), it is one of the few interacting examples that can be solved exactly, with its three-point structure constants given by the celebrated Dorn-Otto-Zamolodchikov-Zamolodchikov (DOZZ) formula \cite{Dorn:1994xn, Zamolodchikov:1995aa, Teschner:1995yf}. In particular, in \cite{Teschner:1995yf} it was shown that the three-point function of Liouville theory can be determined solely by analytic conformal bootstrap arguments. More specifically, exploiting the existence of degenerate representations (which lead to BPZ equations and the associated shift equations \cite{Belavin:1984vu}) together with crossing symmetry, analyticity, and a particular self-dual symmetry under $\fb\leftrightarrow \fb^{-1}$, where $\fb$ can be thought of as the coupling of the theory, uniquely determines the DOZZ formula~\cite{Teschner:1995yf}. The existence of the self-dual symmetry plays an important role in guaranteeing the uniqueness of the DOZZ structure constant. More recently, probabilistic approaches have further provided a rigorous mathematical construction of Liouville theory, leading to a proof of the DOZZ formula and establishing new links between quantum field theory and probability theory~\cite{Guillarmou:2020wbo, Kupiainen:2017eaw, Chatterjee:2024phq}.

Supersymmetric extensions of Liouville theory retain many of these remarkable features, such as exact solvability, while exhibiting additional structure \cite{Rashkov:1996np, Poghossian:1996agj, Belavin:2007gz, Muhlmann:2025ngz, Rangamani:2025wfa}. In particular, a closely related to Teschner's trick mechanism operates in $\mathcal N=1$\footnote{Throughout this work, $\n=c$ denotes $\n=(c,c)$ supersymmetry.} Liouville theory, which also enjoys the same self-dual symmetry. The resulting shift equations again admit a unique solution, completely determining the theory \cite{Rashkov:1996np, Poghossian:1996agj, Belavin:2007gz}.

\begin{figure}
\begin{tikzpicture}
    \begin{scope}
\draw[thick] (0,0) -- (7,0);
        \draw[thick, white] (0,0) --(0,1);
        \draw[thick] (3.5,0) -- (3.5,1);
        \filldraw[fill=violet!15, draw=white, thick] (0,0) rectangle (3.5,.95);
           \filldraw[fill=green!15, draw=white, thick] (3.5,0) rectangle (7,.95);
          \draw[thick] (0,0) -- (0,1.5);
        \draw[thick] (-1,0)--(3.5,0);
        \draw[thick] (-1,1) -- (7,1);
        \draw[thick] (-1,.95) -- (7,.95);
        \draw[thick] (3.5, 1.5) -- (3.5,0);
        \draw[thick] (3.5,0)--(7,0);
        \draw[thick] (8.5,0) -- (14,0);
        \draw[thick] (7,0) -- (7,1.5);
        \filldraw[fill=violet!20, draw=white, thick] (11.,-.09) rectangle (14.,.09);
        \filldraw[fill=green!20, draw=white, thick] (8.5,-.09) rectangle (11.,.09);
        \draw[thick] (11.,-.2) --(11.,.2);
         \draw[thick,->] (10.5,-1)--(10.5,1.8);
         \draw[] (13.5,1.4) -- (13.5,1.7);
         \draw[] (13.5,1.4) -- (13.8,1.4);
         \draw[->] (8.4,0)--(14.1,0);
 \node[scale=.7] at (11,-.35) {$c=3$}; 
 \node[scale=.8] at (13.65,1.55) {$c$}; 
 \node[scale=.9] at (-.55,.5) {$\mathrm{CFT}$}; 
  \node[scale=1] at (1.75,.5) { \cite{Hosomichi:2004ph} $~\&~\mathrm{this~paper}$}; 
  \node[scale=1] at (5.25,.5) {$\mathrm{?}$}; 
 \node[scale=.9] at (-.55,1.25) {$\mathcal{N}=2$}; 
 \node[scale=.9] at (1.75,1.25) {$\mathrm{spacelike~Liouville}$};
  \node[scale=.9] at (5.25,1.25) {$\mathrm{timelike~Liouville}$};
    \end{scope}
\end{tikzpicture}
\caption{We show the main references for $\mathcal{N}=2$ Liouville theory.~We distinguish spacelike and timelike Liouville and restrict to bulk correlation functions. }
\label{fig:figure intro}
\end{figure}
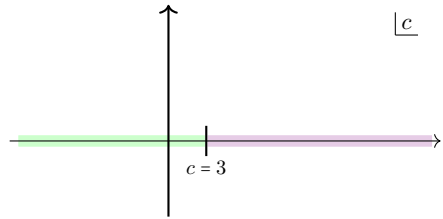

The situation changes qualitatively for $\mathcal N=2$ Liouville theory.~Unlike its $\mathcal N=0,1$ counterparts, the central charge of the theory does not receive quantum corrections due to a non-renormalization theorem, and simply reads
\begin{equation*}
    c=3+6\fQ^2,\quad \fQ= \frac{1}{\fb}~.
\end{equation*}
As a result, the theory does not exhibit the self-dual symmetry, and hence a naive application of Teschner's trick in this case is no longer sufficient to determine the structure constants uniquely, with the crossing symmetry constraints allowing an infinite family of solutions.\footnote{Nonetheless, as we explain in the discussion section, a more sophisticated version of the analytic bootstrap may still be applicable.} Moreover, compared to the $\n=0,1$ instances of the theory, the spectrum of $\n=2$ Liouville contains bound states, with the ones of lowest energy being BPS, as well as operators with spin. Another difference is that the transition from the spacelike to the timelike regime of the theory, \ie from $c\in \mathbb{R}_{>3}$ to $c\in \mathbb{R}_{<3}$,
takes a purely real $\fb$ to a purely imaginary $\hat\fb \equiv \ii \fb$ without crossing a complex $\fb$ regime (see figure \ref{fig:figure intro}), in contrast to what occurs for the Liouville theories with less supersymmetry. These features make $\mathcal N=2$ Liouville considerably richer -- and more subtle --  than the other members of the Liouville family, and raise the question of how its exact correlation functions should be determined.

In the present work,\footnote{See also~\cite{Hosomichi:2004ph} for a discussion of the bulk structure constants from the $\n=2$ Liouville path integral.} we approach this problem via utilizing the fact that $\mathcal N=2$ Liouville theory is mirror symmetric to the supersymmetric $\mathrm{SL}(2,\mathbb{R})_\fk/\mathrm{U}(1)$ Kazama-Suzuki~\cite{Kazama:1988qp} coset model \cite{Mukhi:1993zb,Ooguri:1995wj, Giveon:1999px,Hori:2001ax},
\begin{equation*}
    \mathcal N=2~\mathrm{Liouville~theory}
    \qquad\overset{\text{Mirror}}{\longleftrightarrow}\qquad
    \mathcal N=2~
    \frac{\mathrm{SL}(2,\mathbb{R})_\fk}{\mathrm{U}(1)}
    ~\mathrm{supercoset}~.
\end{equation*}
The target space of the supercoset is the fermionic cigar geometry corresponding to the $2$d Euclidean black hole~\cite{Witten:1991yr,Dijkgraaf:1991ba}, which has been studied extensively, most notably in the context of little string theory~\cite{Giveon:1999px,Giveon:1999tq,Giveon:2003wn,Chang:2014jta}. This correspondence can be viewed as the supersymmetric version of the Fateev-Zamolodchikov-Zamolodchikov (FZZ) \cite{Fzz} duality between bosonic sine-Liouville theory and the bosonic $\mathrm{SL}(2,\mathbb{R})/\mathrm{U}(1)$ coset model (see \eg refs.~\cite{Kazakov:2000pm,Jafferis:2021ywg} for an overview of this bosonic duality). 

The $\mathrm{SL}(2)_\fk/\mathrm U(1)$ supercoset model can be straightforwardly constructed from the bosonic $\mathrm{SL}(2,\mathbb R)_\fk$ WZW model. The latter has been explored in detail in the study of strings in AdS$_3$~\cite{Maldacena:2000hw,Maldacena:2000kv,Maldacena:2001km,Eberhardt:2019qcl, Dei:2021xgh, Dei:2021yom} (see also \cite{Kovensky:2026usc}), with its correlation functions being closely related to those of Teschner's $H^+_3$ model~\cite{Teschner:1997ft,Teschner:1999ug, Teschner:2001gi}, another exactly solvable, irrational, non-compact two-dimensional CFT. As we explain in section~\ref{sec:supercoset}, given the above ingredients one can obtain explicit expressions for the two and three-point functions of the supercoset.~To be more specific, from the perspective of the supercoset's cigar target the angular momentum of the strings is conserved while the winding number is not, since the $S^1$ in the cigar is contractible and strings can unwind at the tip. The $\mathrm{sl}(2)_\fk$ representation theory imposes constraints on winding number violation~\cite{Maldacena:2001km}; for instance, the two-point function must preserve winding number, while the three-point function may violate winding number conservation by at most $\pm 1$ units. The winding number preserving two and three-point functions can be readily deduced from the corresponding correlators of the $H^+_3$ model; the winding number violating three-point functions require a more elaborate treatment involving the evaluation of a particular four-point function with one BPS operator insertion. This was originally proposed in the bosonic case by FZZ in unpublished work, and here we will present explicitly the details of this construction.

Having obtained the relevant three-point functions, in section~\ref{sec:Liouville} we leverage the duality dictionary between the supercoset and $\n=2$ Liouville theory to arrive at explicit expressions for the angular momentum preserving/violating three-point functions of the latter.\footnote{Since the two theories are related by mirror symmetry, their bosonic targets are T-dual and hence winding number gets exchanged with angular momentum, and vice versa.} The target space of $\n=2$ Liouville is topologically a cylinder and can be thought of as containing a string momentum condensate in the strongly coupled region, which is responsible for violating angular momentum conservation. In section~\ref{sec:Liouville_semiclassical}, we compare our findings to the action formulation of $\n=2$ Liouville, by first identifying, within the proposed structure constants, the perturbative poles of correlation functions predicted by a zero mode treatment of the path integral. We then provide a semiclassical computation of various non-BPS correlation functions by expanding the Liouville path integral up to one-loop order, and find exact agreement (up to normalization dependent factors) with the proposed structure constants in both angular momentum preserving/violating sectors. The way the semiclassical computations have to be preformed to match the exact answer is instructive, including contour rotations of negative modes and summation over complex saddle points.~Finally, in section~\ref{sec:BPS_chiral_rings} we turn to $\frac 12$-BPS correlation functions, evaluating the chiral ring coefficient of the chiral-chiral ring which behaves according to standard expectations.

 We include various calculations and additional details in the appendices; in particular, appendix~\ref{ap:special_functions} contains useful properties of the Ypsilon function, appendix~\ref{app:bestiary} the details of the evaluation of several integrals, appendix~\ref{ap:H^+_3_model} a detailed review of various properties of the $H^+_3$ model and appendix~\ref{ap:N=2_SCFTs} a primer on aspects of $2$d SCFTs with $(2,2)$ supersymmetry.\\

Overall, we find remarkable similarities with the bosonic theory. In particular, the same special functions --  the Ypsilon functions -- also appear in the structure constants of the $\mathcal{N}=2$ theory. As a matter of fact, using the results from the bosonic Liouville-$H^+_3$ correspondence~\cite{Ribault:2005wp,Ribault:2005ms}, correlation functions of $\n=2$ Liouville theory can in principle be obtained from correlation functions of the bosonic theory, via a rather complicated integral transform.\\

Curiously, $\mathcal N=2$ Liouville theory with boundary has, by contrast, received considerable attention. The bulk one-point functions on the disk were derived from the modular properties of characters and matched against the conformal bootstrap in \cite{Ahn:2003tt}. Using the representation theory of the $\mathcal N=2$ superconformal algebra, the consistent Cardy boundary states were classified into the FZZT and ZZ-brane analogues of bosonic Liouville boundary states, together with an additional family of branes arising from the extra $\mathrm{U}(1)$ R-current \cite{Hosomichi:2004ph,Eguchi:2003ik,Eguchi:2004yi, Ahn:2004qb}. Notably, unlike the bulk three-point functions discussed in the present work, the boundary data appears to be fixed already by bootstrap and modular invariance arguments.\\

Our analysis of $\n=2$ Liouville is partly motivated also by the fact that, in the timelike regime, the theory constitutes a tractable $2$d model of de Sitter quantum gravity \cite{Anninos:2021ene,Collier:2025lux}. Interestingly, in two-dimensions the usual no-go arguments relating de Sitter space and supersymmetry are circumvented, making supersymmetric extensions of Liouville theory a promising setting for studying quantum aspects of cosmological backgrounds \cite{Anninos:2023exn}. In particular, as proposed in~\cite{Anninos:2023exn}, the sphere path integral of timelike $\n=2$ Liouville theory might be amenable to supersymmetric localization, offering a promising alternative perspective on the Gibbons-Hawking entropy~\cite{Gibbons:1977mu} of the de Sitter horizon. Here we take a preliminary step in that direction by attempting to understand the spacelike theory, both from a CFT as well as a path integral viewpoint. As the title of the paper suggests, this work is intended as a first step toward determining the bulk structure constants of $\mathcal{N}=2$ Liouville, rather than a complete treatment of the theory, and we hope that it will serve as an invitation to further investigations of $\n=2$ Liouville theory.\\

\section{$\mathrm{SL}(2)_\mathsf{k}/\mathrm{U}(1)$ supercoset model}\label{sec:supercoset}

In this section we review the $\mathrm{SL}(2)_\mathsf{k}/\mathrm{U}(1)$\footnote{Throughout this work $\mathrm{SL}(2),\,\mathrm{sl}(2)$ stand for $\mathrm{SL}(2,\mathbb{R}),\,\mathrm{sl}(2,\mathbb{R})$.} supercoset model, as well as its structure constants and $2$d fermionic cigar target.

\subsection{Basics of the supercoset construction}\label{sub:supercoset_construction}

Let us review the construction of the Kazama-Suzuki $\mathrm{SL}(2)_\mathsf{k}/\mathrm{U}(1)$ supercoset model~\cite{Kazama:1988qp,Giveon:1999px,Giveon:1999tq,Giveon:2003wn}, see also~\cite{Chang:2014jta}. The first step is to construct the $\n=1$ $\mathrm{sl}(2)_\fk$ WZW model at level $\fk$.\footnote{This construction is general; a $\mathfrak{g}_\fk$ WZW model with $\n=1$ supersymmetry can be constructed via combining a bosonic $\mathfrak{g}_{\fk\mp h^\vee}$ WZW model, where $\mp$ corresponds to the group $G$ being compact or non-compact and $h^\vee$ is the dual Coxeter number, with $\dim{\mathfrak{g}}$ adjoint Majorana fermions. The supercurrent of the theory is as in eq.~\eqref{eq:N=1_WZW_supercurrent}.} This can be achieved by starting with the bosonic $\mathrm{sl}(2)_{\fk+2}$ WZW at level $\fk+2$, with currents $j^a$ obeying the standard OPE,
\begin{equation}
    j^a(z)j^b(w)\sim \frac{(\fk+2)\eta^{ab}}{2(z-w)^2}+\ii\frac{\epsilon^{abc}j_c(w)}{z-w} \ ,
\end{equation}
where $a=1,2,3$ and $\eta^{ab}=\text{diag}(1,1,-1)$, $\epsilon^{abc}$ are the invariant metric and structure constants of $\mathrm{sl}(2)$ (we define $\epsilon^{123}=1$ etc.), respectively; indices are raised and lowered with $\eta_{ab}$. The unconventional choice of $\eta^{33}=-1$ will be convenient when gauging $J^3$ to go to the supercoset. Adding three Majorana fermions $\xi^a$ in the adjoint representation  of $\mathrm{sl}(2)$, with OPE $\xi^a(z)\xi^b(w)\sim\tfrac{\eta^{ab}}{z-w}$, one can define the following current
\begin{equation}\label{eq:combined_N=1_currents}
    J_a=j_a-\frac{\ii}{2}\epsilon_{abc}\xi^b\xi^c \ ,
\end{equation}
which can be shown to generate an affine $\mathrm{sl}(2)_\fk$ Kač-Moody algebra at level $\fk$. In other words, the current constructed out of the fermions generates an $\mathrm{sl}(2)_{-2}$ affine algebra and shifts the level. Moreover, this theory enjoys $\n=1$ supersymmetry, with the corresponding (holomorphic) supercurrent given by
\begin{equation}\label{eq:N=1_WZW_supercurrent}
    G=\sqrt{\frac{2}{\fk}}\Big(\eta_{ab}\xi^aj^b-\frac{\ii}{6}\epsilon_{abc}\xi^a\xi^b\xi^c\Big) \ .
\end{equation}
Essentially this is the only option for $G$, since we are looking for a dimension $\big(\tfrac{3}{2},0\big)$ operator that is also an $\mathrm{sl}(2)$ singlet, which should map the fermionic piece of the supermultiplet to the bosonic \eg $G(z)\xi^a(w)\sim\sqrt{\frac{2}{\fk}}\frac{J^a(w)}{z-w}$, and vice versa.

To go to the supercoset, the next step is to gauge the $\n=1$ $\mathrm{U}(1)$ supermultiplet $(J^3,\xi^3)$. Defining the complex currents and fermions as
\begin{equation}
    \xi^\pm=\frac{1}{\sqrt{2}}(\xi^1\pm \ii\xi^2), \quad j^\pm=j^1\pm \ii j^2 \ ,
\end{equation}
the $J^3$ current becomes
\begin{equation}\label{eq:J^3_expression}
    J^3=j^3-\ii\xi^1\xi^2=j^3+\xi^+\xi^-\equiv j^3+\ii\partial H_L \ ,
\end{equation}
where in the last equality we defined the bosonization of the fermionic current $\xi^+\xi^-=\ii\partial H_L$, where as usual $H_L$ is a compact boson. We can also bosonize the other abelian currents as
\begin{equation}\label{eq:current_bosomizations}
    \begin{aligned}
        J^3&=-\sqrt{\frac{\fk}{2}}\partial X^3 \ ,\\
        j^3&=-\sqrt{\frac{\fk+2}{2}}\partial x^3 \ ,
    \end{aligned}
\end{equation}
where the pre-factor conventions are chosen so that they agree with the current OPEs, while all chiral bosons $H_L,X^3,x^3$ have the standard OPE, \eg $H_L(z)H_L(w)\sim-\log{(z-w)}$. It is then straightforward to show that the $\mathrm{SL}(2)_\fk/\mathrm{U}(1)$ supercoset has $\n=2$ supersymmetry, with the supercurrents and $\mathrm R$-current given, respectively, by\footnote{One can of course construct these currents before gauging the $\mathrm{U}(1)$ multiplet, which further are $J^3$-neutral. Taking the $\mathrm{U}(1)$ coset simply isolates this $\n=2$ supersymmetry sector.}
\begin{equation}\label{eq:supercoset_currents}
    G^\pm = \sqrt{\frac{2}{\fk}}j^\mp \xi^\pm \ , \qquad J_R=\frac{2}{\fk}J^3+\ii\partial H_L    \ .
\end{equation}
Accordingly, bosonizing the $\mathrm R$-current as,
\begin{equation}\label{eq:R-symmetry_current_supercoset}
    J_R=\ii\sqrt{\frac{\fk+2}{\fk}}\partial X_R \ , \ \text{so that} \ J_R(z)J_R(w)\sim \frac{(\fk+2)/\fk}{(z-w)^2}\ ,
\end{equation}
the following relations between the bosonization scalars hold true,
\begin{equation}\label{eq:boson_relations}
    \begin{aligned}
        iH_L&=\sqrt{\frac{2}{\fk}}X^3+\ii\sqrt{\frac{\fk+2}{\fk}}X_R \ ,\\
        x^3&=\sqrt{\frac{\fk+2}{\fk}}X^3+\ii\sqrt{\frac{2}{\fk}}X_R \ .
    \end{aligned}
\end{equation}
These follow from the expressions~\eqref{eq:J^3_expression},~\eqref{eq:supercoset_currents} of the currents and their bosonizations.

In the above construction, we can trace the central charge to be
\begin{equation}\label{eq:supercoset_central_charge}
    c=\frac{3(\fk+2)}{(\fk+2)-2}+\frac{3}{2}-\frac{3}{2}=3\left(1+\frac{2}{\fk}\right) \ ,
\end{equation}
where, in the first equality, the first fraction is the central charge of the bosonic $\mathrm{SL}(2)_{\fk+2}$ WZW model, the second fraction is the contribution of the three Majorana fermions in the $\n=1$ WZW, and the last originates from gauging the $\mathrm{U}(1)$ supermultiplet (consisting of one boson and one Majorana fermion) to go to the supercoset. We notice that, due to the presence of the fermions, the level $\fk$ is shifted. As we will see in section \ref{sec:Liouville} this is the analogue, in this dual description, of a non-renormalization theorem in $\n=2$ Liouville that violates the $\fb$ to $\fb^{-1}$ symmetry. 

Turning to the primaries of the theory, the bosonic $\mathrm{SL}(2)_{\fk+2}$ WZW model primaries $\Phi_{j,m}$ are labeled (in the holomorphic sector) by $j,\,m$ which are, respectively, the eigenvalue of the $\mathrm{sl}(2,\mathbb{R})$ quadratic Casimir $\mathcal{C}_2=-j(j-1)$ and of $j^3_0$. Thus, the $\n=1$ $\mathrm{SL}(2)_\fk$ WZW primaries are simply the tensor product of the affine $\mathrm{sl}(2)_{\fk+2}$ representations and the Majorana fermions' ones, \ie using the bosonization of $\xi^+\xi^-$ they take the explicit form $\Phi_{j,m}\e^{\ii nH_L}\xi^3$. (We will discuss these representations in detail momentarily.) Then, we can separate these primaries into a vertex operator with definite $J^3_0$-charge and a neutral piece, namely
\begin{equation}\label{eq:building_primaries}
    \Phi_{j,m}\e^{\ii nH_L}\xi^3=V_{j,m}\e^{\sqrt{\frac{2}{\fk+2}}mx^3}\e^{\ii nH_L}\xi^3=V_{j,m}\e^{\ii\sqrt{\frac{\fk+2}{\fk}}\big(\frac{2m}{\fk+2}+n\big)X_R}\e^{\sqrt{\frac{2}{\fk}}(m+n)X^3}\xi^3 \ ,
\end{equation}
where $V_{jm}$ is a bosonic $\mathrm{SL}(2)_{\fk+2}/\mathrm{U}(1)$ primary and in the last equality we used eq.~\eqref{eq:boson_relations}. Quotienting out by the $\n=1$ $\mathrm{U}(1)$ supermultiplet  $(J^3,\xi^3)$, we arrive at the $\n=2$ supercoset primaries
\begin{equation}
    V^n_{j,m}\equiv V_{j,m}\e^{\ii\sqrt{\frac{\fk+2}{\fk}}\big(\frac{2m}{\fk+2}+n\big)X_R} \ .
\end{equation}
The dimension of these primaries can be obtained by appropriately combining the dimensions of the bosonic $\mathrm{SL}(2)_{\fk+2}$ primaries with the dimensions of the $X_R,\,X^3$ vertex operators in eq.~\eqref{eq:building_primaries}, respectively as 
\begin{equation}\label{eq: conformal dimension Vjmn}
      h(V^n_{j,m})=-\frac{j(j-1)}{\fk}+\frac{m^2}{\fk+2}+\frac{\fk+2}{2\fk}\left(\frac{2m}{\fk+2}+n\right)^2=-\frac{j(j-1)}{\fk}+\frac{(m+n)^2}{\fk}+\frac{n^2}{2} \ ,
\end{equation}
while the $\mathrm R$-charge can be deduced from the OPE with $J_R$, yielding
\begin{equation}\label{eq: gR Vjmn}
    J_R(z)V^n_{jm}(w)\sim \frac{q_R}{z-w}V^n_{jm}(w)  \ ,\quad q_R=\frac{2m}{\fk}+n\frac{\fk+2}{\fk} \ .
\end{equation}
Our discussion thus far has focused on the holomorphic sector. Similar expressions hold for the anti-holomorphic sector, where for later use we choose to bosonize the various currents with a relative sign compared to \eg \eqref{eq:current_bosomizations},~\eqref{eq:R-symmetry_current_supercoset}, culminating in the full supercoset primaries,
\begin{equation}\label{eq:full_supercoset_primaries}
    V^{n,\bar n}_{j,m,\bar m}= V_{j,m,\bar m}\e^{\ii\sqrt{\frac{\fk}{\fk+2}}q_RX_R-\ii\sqrt{\frac{\fk}{\fk+2}}\bar q_R\bar X_R} \ .
\end{equation}
The corresponding dimensions and R-charges take the following form ((\ref{eq: conformal dimension Vjmn}) and (\ref{eq: gR Vjmn}))
\begin{equation}\label{eq:full_supercoset_dimas_&_R_charges}
\begin{aligned}
     h&=-\frac{j(j-1)}{\fk}+\frac{(m+n)^2}{\fk}+\frac{n^2}{2} \ , \quad q_R =\frac{2m}{\fk}+n\frac{\fk+2}{\fk} \ ,\\
   \bar  h&=-\frac{j(j-1)}{\fk}+\frac{(\bar m+\bar n)^2}{\fk}+\frac{\bar n^2}{2} \ , \quad \bar q_R =\frac{2\bar m}{\fk}+\bar n\frac{\fk+2}{\fk} \ .
\end{aligned}     
\end{equation}
Note that the target space -- being the universal cover of $\mathrm{SL}(2,\mathbb{R})$ -- is uncompact, hence locality requires the eigenvalues of the left and right-moving Casimir to coincide, \ie $j=\bar{j\,}$.

\paragraph{Spectrum} We now discuss the supercoset primary representations; first, recall the bosonic $\mathrm{SL}(2)_{\fk+2}$ WZW primaries $\Phi_{j,m}$ labeled by $j,\,m$. For the supercoset cigar CFT we are interested in representations of the universal cover of $\mathrm{SL}(2,\mathbb R)$, since its Wick-rotation yields Euclidean hyperbolic space, \ie $H^+_3$. There are three representations of interest\footnote{There are also complementary representations in $\mathrm{sl}(2,\mathbb{R})$, which nevertheless we ignore since they are not normalizable in $\mathbb{L}^2\big(\mathrm{SL}(2,\mathbb{R})\big)$ and hence will not appear in standard quantizations of the theory.} for the universal cover of $\mathrm{SL}(2,\mathbb{R})$ (see \eg \cite{Dixon:1989cg,Maldacena:2000hw}). 
\begin{enumerate}
    \item {\bf Principal (continuous) representations $\mathcal{C}_j^\alpha$} where $j=\frac{1}{2}+\ii s$ and $m\in \mathbb{Z}+\alpha$, with $0\leq \alpha <1$, $s\in\mathbb{R}_+$.
    \item {\bf Discrete representations $\mathcal{D}^\pm_j$} where $\pm m-j=0,1,2,\dots$, with $j$ so that $\frac{1}{2}<j<\frac{\fk+1}{2}$. The lower bound is imposed by unitarity, while the upper from requiring smoothness of the supercoset two-point function, as we discuss later. In the supercoset, these will correspond to degenerate representations of the $\n=2$ superconformal algebra, see appendix~\ref{ap:degen_reps} for a review.
    \item {\bf Trivial representation} where $j=m=0$.
\end{enumerate}
As explained in~\cite{Maldacena:2000hw}, there are also spectrally flowed representations that one should consider. These arise from the following outer automorphism of the $\mathrm{sl}(2)_{\fk+2}$ algebra,\footnote{This should not be confused with the spectral flow automorphism of the $\n=2$ superconformal algebra~\eqref{eq:spectral flow1}. Note also that we are using a different convention for the spectral flow parameter $w$ than in \eg ref.~\cite{Maldacena:2000hw}, \ie $w_\text{here}=-w_\text{there}$, since that will be useful for the cigar discussion later.}
\begin{equation}\label{eq:currents_spectral-flow}
    j^3_n\mapsto\tilde j^3_n\equiv j^3_n+\frac{\fk+2}{2}w\delta_{n,0} \ , \quad j^\pm_n\mapsto \tilde j^\pm_n\equiv j^\pm_{n\mp w} \ ,
\end{equation}
and similarly for the anti-holomorphic generators, with locality of the theory in the universal cover requiring the same integer amount of holomorphic and anti-holomorphic spectral flow $w\in\mathbb Z$. Accordingly, the Sugawara modes of the stress-tensor transform as,
\begin{equation}\label{eq:bosonic_Sugawara_flowed_generatros}
    l_n\mapsto\tilde l_n\equiv l_n-w j^3_n-\frac{\fk+2}{4}w^2\delta_{n,0} \ ,
\end{equation}
obeying the Virasoro algebra with the same central charge $c=3(1+2\fk^{-1})$. The spectrally flowed primary states, which we denote as $\Phi^w_{j,m}$, have to be annihilated by the flowed generators $\tilde j^{\pm,3}_n$ for $n>0$ and correspond genuinely to novel not highest-weight representations. These do not live inside the unflowed representations $\mathcal C^\alpha_j,\mathcal D^\pm_j$, but can be obtained from them via spectral flow by $w$, yielding $\mathcal C^{\alpha,w}_j,\mathcal D^{\pm,w}_j$. As usual, to obtain the full affine modules $\hat{\mathcal C}^{\alpha,w}_j,\hat{\mathcal D}^{\pm,w}_j$ one has to act on the zero mode representations $\mathcal C^{\alpha,w}_j,\mathcal D^{\pm,w}_j$ with the raising operators $\tilde j^{\pm,3}_{-n},\,n>0$. 

In addition, for the $\n=1$ $\mathrm{SL}(2)_\fk$ WZW model we also have the three Majorana fermions $\xi^a$ that organize into the standard Fock space, which we denote by $\mathcal F$. More specifically, in the NS-sector the vacuum $|0\rangle_\text{NS}$ is defined as $\xi^{\pm,3}_r|0\rangle_\text{NS}=0,\,r\in\mathbb{Z}_{\geq 0}+\frac{1}{2}$ with the NS Fock space being,
\begin{equation}
    \mathcal{F}_\text{NS}=\text{span}\lbrace\xi^{a_1}_{-r_1}\dots\xi^{a_p}_{-r_p}|0\rangle_\text{NS}\rbrace \ , \ \text{for} \ r_i\in\mathbb{Z}_{\geq 0}+\frac{1}{2} \ \text{and} \ a_i=\pm,3 \ .
\end{equation}
In the R-sector there are three zero modes $\xi^{\pm,3}_0$ that generate a two-dimensional Clifford module $|\pm\rangle_\text{R}$, yielding the familiar double degeneracy of R-sector ground states. The corresponding Fock space reads,
\begin{equation}
    \mathcal{F}_\text{R}=\text{span}\lbrace\xi^{a_1}_{-r_1}\dots\xi^{a_p}_{-r_p}|\pm\rangle_\text{R}\rbrace \ , \ \text{for} \ r_i\in\mathbb Z_{> 0} \ \text{and} \ a_i=\pm,3 \ .
\end{equation}
The combined Fock space is simply $\mathcal{F}=\mathcal{F}_\text{NS}\oplus\mathcal{F}_\text{R}$. How do the fermions behave under spectral flow? To answer this, recall that the fermions live in the adjoint representation of $\mathrm{sl}(2)_\fk$, namely
\begin{equation}\label{eq:fermion_commutators}
    [J^a_n,\xi^b_r]=\ii \epsilon^{ab}_{\ \ c}\,\xi^c_{n+r} \  \ \Rightarrow  \ \ [J^\pm_n,\xi^3_r]=\mp\sqrt 2\xi^\pm_{n+r} \ , \ [J^3_n,\xi^\pm_r]=\pm\xi^\pm_{n+r} \ .
\end{equation}
Inspecting the spectral flow transformation of the $\mathrm{sl}(2)_\fk$ currents~\eqref{eq:currents_spectral-flow} (after shifting the level $\fk+2\to\fk$), motivates the following fermion spectral flow
\begin{equation}
    \xi^3_r\mapsto\tilde\xi^3_r\equiv\xi^3_r \ , \quad \xi^\pm_r\mapsto\tilde\xi^\pm_r\equiv\xi^\pm_{r\mp w} \ ,
\end{equation}
which is further consistent with the fermionic anticommutation relations. It is straightforward to verify that integer spectral flow maps $\mathcal F$ to itself. For instance, under positive spectral flow by $w\geq 1$ the NS vacuum $|0\rangle_\text{NS}$ gets mapped to $|0;w\geq1\rangle_\text{NS}\equiv \xi^+_{-w+\frac{1}{2}}\xi^+_{-w+\frac{3}{2}}\dots\xi^+_{-\frac{1}{2}} |0\rangle_\text{NS}$. Hence, one could choose to directly spectrally flow the combined $\n=1$ affine currents~\eqref{eq:combined_N=1_currents}. 

The above analysis results in the following spectrum for the $\n=1$ $\mathrm{SL}(2)_\fk$ WZW model,\footnote{We only include the lowest-weight $\hat{\mathcal D}^{+,w}_j$ discrete representations in $\mathcal H$, since the highest-weight ones are taken into account when $w\in\mathbb Z$ due to the identification $\hat{\mathcal D}^{-,w}_j\simeq \hat{\mathcal D}^{+,w+1}_{\frac{\fk+2}{2}-j}$.}
\begin{equation}\label{eq:N=1_spectrum}
   \mathcal{H}= \bigoplus_{w\in\mathbb Z} \left(\int_{\mathbb R_+}\d s\int_0^1\d\alpha\, \hat{\mathcal C}^{\alpha,w}_{\frac12 +\ii s}\otimes{\hat{\overline{{\mathcal C}}}}^{\alpha,w}_{\frac12 +\ii s}\oplus\int_\frac{1}{2}^\frac{\fk+1}{2}\d j\,\hat{\mathcal D}^{+,w}_j\otimes{\hat{\overline{\mathcal D}}}^{+,w}_j  \right)\otimes\mathcal F\otimes\overline{\mathcal{F}} \ .
\end{equation}
Therefore, recalling eq.~\eqref{eq:building_primaries}, the spectrum $\mathcal H$ can be thought of as containing flowed states of the form $\Phi_{j,m}^w\e^{\ii (n+w)H_L}$ which, with respect to the unflowed $L_0,J^3_0$ generators, have
\begin{equation}\label{eq:flowed_dims_&_ms}
\begin{aligned}
     h^w&\equiv \Big(-\frac{j(j-1)}{\fk}+wm-\frac{\fk+2}{4}w^2\Big)+\frac{(n+w)^2}{2} =-\frac{j(j-1)}{\fk}+\frac{n^2}{2}+w(m+n)-\frac{\fk}{4}w^2 \ ,\\
      m^w&\equiv\Big(m-\frac{\fk+2}{2}w\Big)+\left(n+w\right)= m+n-\frac{\fk w}{2}  \ ,
\end{aligned}
\end{equation}
with $j,m+n$ being the eigenvalues of $\Phi_{j,m}\e^{\ii nH_L}$ wrt the unflowed generators $\mathcal{C}_2,J^3_0$, taking the allowed values for the unflowed representations $\mathcal C^{\alpha}_j,\mathcal D^{\pm}_j$ described above.
The spectral flow transformation shifts the conformal weight by a term quadratic in the flow parameter, making them generally unbounded from below. However, in a fixed $J_0^3$ representation $L_0$ is bounded below.

Note that, in writing eq.~\eqref{eq:flowed_dims_&_ms} for the full spectrally flowed states, we used the fact that under $w$ spectral flow the $j^3_0,J^3_0$-charge of the bosonic $\Phi_{j,m}$ primary and fermion vertex operator $e^{\ii n H_L}$ change as $m\to m-\frac{\fk+2}{2}w$ and $n\to n+w$, respectively. The former follows from eq.~\eqref{eq:currents_spectral-flow}, while a simple way to see the latter is via noticing from eq.~\eqref{eq:fermion_commutators} that each $\xi^+_r$ mode increases the fermion charge by one unit. Since the flowed vacuum $|0;w\geq1\rangle_\text{NS}$ contains $w$ such modes the fermion charge gets shifted by $w$ units in the flowed sector. We lastly remark that if we were instead in the single cover of $\mathrm{SL}(2,\mathbb R)$ the difference would be that $m$ would now be discretized and different holomorphic $w_L$ and anti-holomorphic $w_R$ amounts of spectral flow would be allowed, with their difference $w_L-w_R$ being the string winding number along the compact timelike direction. 

We proceed with the observation that generically the $\n=1$ WZW spectrum $\mathcal H$~\eqref{eq:N=1_spectrum} contains negative norm states. For instance, given a primary $|j,m\rangle$ in a unitary representation of $\mathrm{sl}(2,\mathbb R)$, the descendant state $J^3_{-1}|j,m\rangle$ has norm $-\frac{\fk}{2}$ which is negative for $\fk>0$. This is not unexpected, since \eg the kinetic term in the action of the WZW model has indefinite sign.\footnote{Relatedly, the $H^+_3$ model whose correlation functions we will discuss later is not a unitary two-dimensional CFT.} Nonetheless, since these issues are mostly due to the timelike character of $J^3$, one might hope that by taking the $J^3$ coset these pathological states will be removed. Amusingly, this is indeed what happens. In particular, the states $|\psi\rangle\in\mathcal H$ that survive the axial $\mathrm{SL}(2)_\fk/\mathrm U(1)$ supercoset must satisfy
\begin{equation}\label{eq:supercoset_conditions}
    (J_0^3+\bar J_0^3)|\psi\rangle=0 \ , \quad J^3_n|\psi\rangle=\bar J^3_n|\psi\rangle=\xi^3_r|\psi\rangle=\bar \xi^3_r|\psi\rangle=0, \ n,r>0 \ ,
\end{equation}
and one can show that the second condition removes all negative norm states from the spectrum. This fact was shown in~\cite{Dixon:1989cg} for the unflowed representations and in~\cite{Maldacena:2000hw} for the flowed ones. This implies that both the supercoset, as well as its mirror $\n=2$ Liouville theory, are unitary two-dimensional SCFTs.

Accordingly, the supercoset spectrum is given by $\mathcal H$ quotiented by the conditions~\eqref{eq:full_supercoset_dimas_&_R_charges}. Affine primary states in the parent $\n=1$ WZW model obey by definition the second condition, while the first one is satisfied by $V^{n,\bar n}_{j,m,\bar m}$~\eqref{eq:full_supercoset_primaries} having factored out the $J^3_0$ charge in eq.~\eqref{eq:building_primaries}. To be concrete, the holomorphic supercoset dimensions are given by $h^w-h_{\mathrm{U}(1)}$; the Sugawara stress tensor for the $J^3$ currents reads $T_{\mathrm{U}(1)}=-\fk^{-1}(J^3)^2$ and taking its OPE with a primary $\mathcal{O}_{j,m^w}\in\mathcal H$ of $J^3_0$-charge $m^w$\eqref{eq:flowed_dims_&_ms} one learns that $h_{\mathrm{U}(1)}=-\fk^{-1}(m^w)^2$. Hence
\begin{equation}
    h^w- h_{\mathrm{U}(1)} = -\frac{j(j-1)}{\fk}+\frac{(m+n)^2}{\fk}+\frac{n^2}{2} ~,
\end{equation}
which agrees with the holomorphic dimension in (\ref{eq:full_supercoset_dimas_&_R_charges}).
Similarly, the R-charge can be obtained by the OPE of $\Phi_{j,m}^{w}\e^{\ii (n+w)H_L}$ with $J_R$~\eqref{eq:R-symmetry_current_supercoset}. Combining the above, the supercoset primaries are those of eq.~\eqref{eq:full_supercoset_primaries} belonging to the zero mode representations $\mathcal C^{\alpha,w}_j,\mathcal D^{\pm,w}_j$, with dimensions and R-charge as in eq.~\eqref{eq:full_supercoset_dimas_&_R_charges}. Put differently, from the total affine module~\eqref{eq:N=1_spectrum} the first condition in~\eqref{eq:supercoset_conditions} isolates the axial $\mathrm{U}(1)$-neutral representations that belong to the supercoset, with the role of the second condition being to remove the negative norm states amongst them -- yielding a unitary spectrum.

\subsection{$2$d fermionic black hole}\label{subsec:cigar_target}
Viewed as a non-linear sigma model (NLSM), the $\mathrm{SL}(2)_\fk/\mathrm{U}(1)$ supercoset has a $2$d target that corresponds to the fermionic Euclidean black hole. This is the supersymmetric generalization of the bosonic $2$d Euclidean black hole studied in~\cite{Elitzur:1990ubs,Mandal:1991tz,Witten:1991yr,Dijkgraaf:1991ba}. The line element of the target and the dilaton profile read,\footnote{We work in conventions where $\alpha'=2$.}
\begin{equation}\label{eq:cigar_metric}
    ds^2_{\text{cigar}}=2\fk\big(\d\rho^2 +\tanh^2{\rho}\,\d\theta^2\big) \ , \quad \Phi=-\log{\cosh{\rho}} \ ,
\end{equation}
where $\rho\in[0,\infty)$ and $\theta\sim\theta+2\pi$. This metric can be obtained by starting with the action of a NLSM with target $H^+_3 \cong \mathrm{SL}(2,\mathbb{C})/\mathrm{SU}(2)$ (\ie Euclidean AdS$_3$) and gauging the axial $\mathrm{U}(1)_{\mathrm{A}}$ action generated by $J^3$~\cite{Witten:1991yr}. We, therefore, have a NLSM with a cigar target, schematically: 
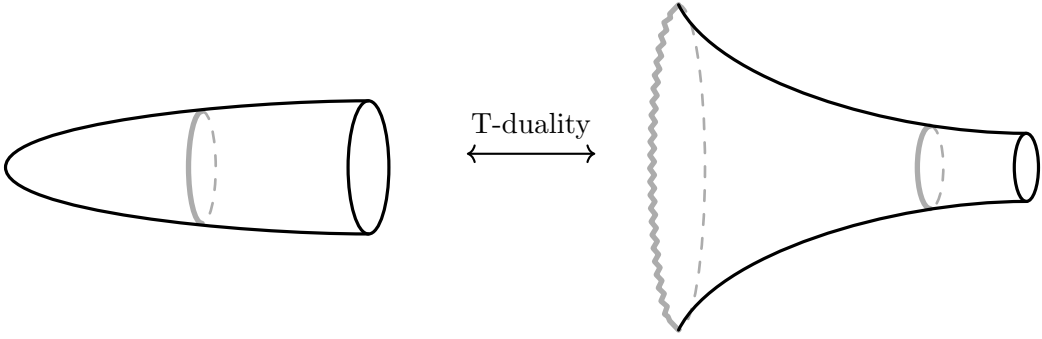
\begin{figure}[H]
    \begin{center}
\begin{tikzpicture}[
    line cap=round,
    line join=round,
    outline/.style={black, very thick},
    section/.style={gray!65, line width=2pt},
    hidden/.style={gray!65, line width=1pt, dashed, dash pattern=on 4pt off 5pt},
    singularity/.style={
        gray!65,
        line width=2pt,
        decorate,
        decoration={
            zigzag,
            amplitude=0.9pt,
            segment length=5pt,
            pre length=5pt,
            post length=5pt
        }
    }
]

\pgfmathsetmacro{\trumpetGrayX}{5.78}  
\pgfmathsetmacro{\trumpetGrayR}{0.545}       
\pgfmathsetmacro{\trumpetGraySquash}{0.315}  '
\pgfmathsetmacro{\trumpetGrayXR}{\trumpetGraySquash*\trumpetGrayR}


\draw[section]
    (-3.85,0.74)
    arc[start angle=90,end angle=270,x radius=0.18,y radius=0.74];

\draw[hidden]
    (-3.85,0.74)
    arc[start angle=90,end angle=-90,x radius=0.18,y radius=0.74];

\draw[outline]
    (-1.65,0.88)
    .. controls (-3.40,0.88) and (-6.45,0.58) .. (-6.45,0)
    .. controls (-6.45,-0.58) and (-3.40,-0.88) .. (-1.65,-0.88);

\draw[outline] (-1.65,0) ellipse [x radius=0.27, y radius=0.88];


\node[] at (0.50,0.55) {T-duality};
\draw[<->, thick] (-0.35,0.18) -- (1.35,0.18);


\draw[singularity]
    (2.45,2.15)
    arc[start angle=90,end angle=270,x radius=0.35,y radius=2.15];

\draw[hidden]
    (2.45,2.15)
    arc[start angle=90,end angle=-90,x radius=0.35,y radius=2.15];

\draw[section]
    (\trumpetGrayX,\trumpetGrayR)
    arc[
        start angle=90,
        end angle=270,
        x radius=\trumpetGrayXR,
        y radius=\trumpetGrayR
    ];

\draw[hidden]
    (\trumpetGrayX,\trumpetGrayR)
    arc[
        start angle=90,
        end angle=-90,
        x radius=\trumpetGrayXR,
        y radius=\trumpetGrayR
    ];

\draw[outline]
    (2.45,2.15)
    .. controls (2.95,1.15) and (5.25,0.45) .. (7.05,0.45);

\draw[outline]
    (2.45,-2.15)
    .. controls (2.95,-1.15) and (5.25,-0.45) .. (7.05,-0.45);

\draw[outline] (7.05,0) ellipse [x radius=0.16, y radius=0.45];

\end{tikzpicture}
    \end{center}
    \label{fig:placeholder}
\end{figure}
\noindent The alternative, namely taking a vector $\mathrm{U}(1)_{\mathrm{V}}$ quotient, yields a coset with a trumpet target that is T-dual to the cigar, as depicted above. This mirror target is related to $\n=2$ Liouville theory, as we discuss in section~\ref{sec:Liouville}.

In the large-$\fk$ limit, this NLSM corresponds to strings propagating in the Euclidean black hole background. From this perspective one expects that string states will be labeled by three quantum numbers; first, there will be the momentum of the string $\text{Im}\,j$ along the radial direction of the cigar, second the string momentum $\ell=m+n-\bar{m}-\bar n\in\mathbb Z$ along the angular cigar direction and, third, the winding number $w\fk=m+n+\bar{m}+\bar n,\,  w\in\mathbb{Z}$ of the string along the cigar circle.\footnote{We have expressed these quantum numbers also via the labels $j,m,n,\bar m,\bar n$ of the supercoset primaries. In particular, the relations for $\ell,w$ come from combining~\eqref{eq:flowed_dims_&_ms} with the locality condition $m^w-\bar m^w\in\mathbb Z$ and the first neutrality condition of eq.~\eqref{eq:supercoset_conditions}, respectively.} The string angular momentum is conserved, while the winding number is not; the string can simply unwind by sliding to the tip of the cigar. This will be important for our discussion of correlation functions later.

One can perform a semiclassical minisuperspace type of analysis on the cigar, see \eg refs.~\cite{Kazakov:2000pm,Ribault:2003ss} for further details. In that regime, strings can be thought of as point particles propagating in the black hole background. Semiclassical states are given by eigenfunctions of the Laplacian
\begin{equation}\label{eq:Laplacian_cigar}
    -\nabla^2=-\frac{1}{\e^{-2\Phi}\sqrt{g}}\partial_\mu\big({\e^{-2\Phi}}\sqrt{g}g^{\mu\nu}\partial_\nu\big)=-\frac{1}{2\fk}\big(\partial_\rho^2+(\coth{\rho}+\tanh{\rho})\partial_\rho+\coth^2{\rho}\partial^2_\theta\big) \ ,
\end{equation}
and take the form of hypergeometric functions, see \eg~\cite{Ribault:2003ss}. For large $\rho$, where the cigar asymptotes to the cylinder, eq.~\eqref{eq:Laplacian_cigar} becomes $-\nabla^2\simeq -(2\fk)^{-1}(\partial^2_\rho+2\partial_\rho+\partial^2_\theta)$, whose eigenfunctions with eigenvalue $\lambda=-\frac{4j(j-1)}{2\fk}+\frac{\ell^2}{2\fk}$ are of the form $\Psi(\rho)\e^{\ii \ell\theta}\simeq \e^{-2a\rho+\ii \ell\theta}$, when $a$ satisfies $a^2-a-j(j-1)=0$. Therefore, $a\in \{j,\,1-j\}$, leading to
\begin{equation}
    \Psi(\rho)\simeq \e^{-2(1-j)\rho}+R(j,m,\bar m)\e^{-2j\rho}\ ,
\end{equation}
where $R(j,m,\bar m)$ is the reflection coefficient, whose explicit form will be discussed later in eq.~\eqref{eq:supercoset_2pt_func}, and reads
\begin{equation}\label{eq:R(j)_relations_draft}
     R(j,m,\bar m)=\pi R(j)\frac{\Gamma(j-\bar m)\Gamma(j+m)\Gamma(1-2j)}{\Gamma(1-j-\bar m)\Gamma(1-j+m)\Gamma(2j)} \ ,\quad R(j)= 2\big( \fk^{-\fk}\gamma(\fk)\big)^\frac{2-2j}{\fk}\frac{\gamma\left(\frac{2j-1}{\fk}\right)}{\gamma(\fk^{-1})} ~.
\end{equation}
One can further argue from this analysis that $j=\frac{1}{2}+\ii s,\,s\in\mathbb{R}_+$ and $s$ can then be identified with the string radial momentum at infinity. These states correspond to principal (continuous) representations and, as concluded from this discussion, they are (string) scattering states. As explained in ref.~\cite{Dijkgraaf:1991ba} (see also~\cite{Jafferis:2021ywg}), the semiclassical analysis is agnostic to the discrete representations. These can be identified by noting that the reflection coefficient has poles at values of $j$ that correspond to the discrete $\mathrm{sl}(2,\mathbb{R})$ representations~\cite{Aharony:2004xn}; in that instance we have to keep the residue of $R(j,m,\bar{m})$ and thus $\Psi(\rho)\simeq \e^{-2j\rho}$, implying that these modes are localized near the tip of the cigar (recall that $j>1/2$ for discrete representations). This discussion also clarifies why continuous representations are delta-function normalizable, while discrete representations are normalizable. In particular, the semiclassical $\mathbb{L}^2$-norm in the target reads,
\begin{equation}
    \int_{\frac{\mathrm{SL}(2,\mathbb R)}{\mathrm U(1)}} \Big|\Psi(\rho)\e^{\ii \ell \theta}\Big|^2=\int \d\rho \d\theta \sqrt{g}\e^{-2\Phi}\Big|\Psi(\rho)\Big|^2=4\pi\fk\int_0^\infty \d\rho\,\sinh{\rho}\cosh{\rho}\,\e^{-4 \mathrm{Re}( a) \rho} \ .
\end{equation}
For large $\rho$, $\sinh{\rho}\cosh{\rho}\sim \e^{2\rho}$ and the $\rho$ integral converges for $a\in\mathbb{R}$ iff $a>1/2$. This corresponds to the discrete representations for $j>1/2$, with the normalizable mode being $\e^{-2j\rho}$. Principal representations where $j=\frac{1}{2}+\ii s$, on the other hand, result in a non-compact zero mode integral and are therefore only delta-function normalizable. Notice, also, that the $\lambda$-eigenvalues of the Laplacian above coincide with the CFT dimensions above, when $n=\bar n=0$ and up to an overall factor. To get the full states and dimensions when $n,\bar n\neq0$, we can simply tensor the Laplacian eigenfunctions with the vertex operator of the scalar that bosonizes the fermions. Meanwhile, in the large-$\fk$ and minisuperspace limit, strings behave as point-like particles and hence have trivial winding around the $S^1$, \ie $w=0$.

Finally, we can also express the conformal dimensions and R-charges (\ref{eq:full_supercoset_dimas_&_R_charges}) in terms of the winding number and angular momentum $w$ and $\ell$, which are integers, and the Casimir eigenvalue $j$. This implies 
\begin{equation}\label{eq:full_supercoset_dimas_&_R_charges_w}
\begin{aligned}
     h&=-\frac{j(j-1)}{\fk}+\frac{(\ell+w\fk)^2}{4\fk}+\frac{n^2}{2} \ , \quad q_R =\frac{\ell+w\fk}{\fk}+n \ ,\\
   \bar  h&=-\frac{j(j-1)}{\fk}+\frac{(\ell-w\fk )^2}{4\fk}+\frac{\bar n^2}{2} \ , \quad \bar q_R =\frac{w\fk -\ell}{\fk}+\bar n \ .
\end{aligned}     
\end{equation}
The winding number here is equivalent to the spectral flow parameter $w$ in the flowed representations of the parent theory (\ref{eq:flowed_dims_&_ms}). 
We observe that the spin of the supercoset primaries is quantized, namely
\begin{equation}
    \mathfrak{s}\equiv h-\bar{h} = \ell   w +\frac{n^2-\bar n^2}{2}~,\quad q_R - \bar{q}_R =  \frac{2\ell}{\fk}+n-\bar n~,
\end{equation}
where $n,\bar n\, \in\mathbb{Z},\mathbb{Z}+\frac{1}{2}$ in the NS and R-sector, respectively. In the NS-sector, locality of the bosonized vertex operators requires $n-\bar n\in\mathbb Z,n+\bar n\in2\mathbb Z,$ hence $\mathfrak{s}$ is always integer. Similarly, in the R-sector $(n-\bar n)(n+\bar n)$ is even since the sum of the two factors is odd, hence again $\mathfrak{s}\in\mathbb Z$.

\subsection{Supercoset structure constants}

We now turn to discuss the correlation functions of the supercoset model. We are aiming for the correlators of the suupercoset operators~\eqref{eq:full_supercoset_primaries}, which read
\begin{equation}\label{ew:wind_full_supercoset_correlators}
    \Big \langle\prod_{i=1}^N V^{n_i,\bar n_i}_{j_i,m_i,\bar m_i}(z_i,\bar z_i) \Big\rangle=\Big \langle\prod_{i=1}^N \Phi_{j_i,m_i,\bar m_i}(z_i,\bar z_i) \Big\rangle \frac{\Big \langle\prod_{i=1}^N \e^{\ii\sqrt{\frac{\fk}{\fk+2}}q_R^iX_R-\ii\sqrt{\frac{\fk}{\fk+2}}\bar q^i_R\bar X_R}  \Big\rangle}{\Big \langle\prod_{i=1}^N \e^{\sqrt{\frac{2}{\fk+2}}(m_ix^3-\bar m_i \bar x^3)}  \Big\rangle} \ ,
\end{equation}
where we recall from eq.~\eqref{eq:building_primaries} that $\Phi_{j,m,\bar m}=V_{j,m,\bar m}\e^{\sqrt{\frac{2}{\fk+2}}(mx^3-\bar m\bar x^3)}$.\footnote{Equivalently, one may start from the flowed parent primaries $\Phi^w_{j,m,\bar m}\e^{\ii(n+w)H_L-\ii(\bar n+w)H_R}$. As already anticipated from the discussion after eq.~\eqref{eq:supercoset_conditions}, the flowed bosonic primaries factorize into the same coset fields $V_{j,m,\bar m}$, but with shifted $x^3,\bar x^3$ charges. The extra $w$-dependent factors would be exactly canceled by the corresponding factors in the correlators of $X^3,\bar X^3$ vertex operators. Hence we would again land on~\eqref{ew:wind_full_supercoset_correlators}. The dependence on $w$ in the supercoset comes from the constraint $m+n+\bar m+\bar n=w\fk$.} The correlators of the vertex operators are straightforward to compute, yielding 
\begin{equation}\label{eq:XR_vertex_correlators}
    \Big \langle\prod_{i=1}^N \e^{\ii\sqrt{\frac{\fk}{\fk+2}}q_R^iX_R-\ii\sqrt{\frac{\fk}{\fk+2}}\bar q^i_R\bar X_R}  \Big\rangle=\delta_{\sum_{i=1}^N q_R^i,0}\delta_{\sum_{i=1}^N \bar q_R^i,0}\prod_{1\leq i<j\leq N}z_{ij}^{\frac{\fk}{\fk+2}q_R^iq_R^j}\bar z_{ij}^{\frac{\fk}{\fk+2}\bar q_R^i\bar q_R^j} \ ,
\end{equation}
for the numerator, and
\begin{equation}\label{eq:x3_vertex_correlators}
    \Big \langle\prod_{i=1}^N \e^{\sqrt{\frac{2}{\fk+2}}(m_ix^3-\bar m_i \bar x^3)}  \Big\rangle=\delta\Big(\sum_{i=1}^N m_i\Big)\delta\Big(\sum_{i=1}^N \bar m_i\Big)\prod_{1\leq i<j\leq N}z_{ij}^{-\frac{2}{\fk+2}m_im_j}\bar z_{ij}^{-\frac{2}{\fk+2}\bar m_i\bar m_j} \ ,
\end{equation}
for the denominator, where the Kronecker deltas and delta functions are coming from compact and non-compact zero mode integrals, respectively, and in eq.~\eqref{eq:x3_vertex_correlators} we had to rotate the contour to make sense of the integral. The remaining step is to evaluate the correlation functions of the bosonic $\mathrm{SL}(2)_{\fk+2}$ WZW model primaries $\Phi_{j,m,\bar m}$. These can be readily deduced by leveraging the correlators of the $H^+_3$ WZNW model derived by Teschner~\cite{Teschner:1997ft,Teschner:1999ug,Teschner:2001gi}, which we review in appendix~\ref{ap:H^+_3_model}. 

This is a good point to make an important observation. In the $\mathrm{SL}(2)_{\fk+2}$ parent theory, and correspondingly in the supercoset, spectral flow $w$ is not a conserved quantum number. In the language of the cigar target, the integer $w$ becomes the string winding number along the angular circle, which need not be conserved  since strings can unwind at the tip of the cigar. However, only winding number preserving correlators can be computed through the simple relation~\eqref{ew:wind_full_supercoset_correlators}. Indeed, on the support of the delta function constraints in~\eqref{eq:XR_vertex_correlators},~\eqref{eq:x3_vertex_correlators}, the correlator~\eqref{ew:wind_full_supercoset_correlators} is non-vanishing only when $\sum_{i=1}^Nw_i=0$. Both in the parent theory and in the supercoset there exist more general correlation functions that violate spectral flow conservation, whose treatment requires a slightly more involved analysis. We will come back to such correlators shortly.

In appendix~\ref{ap:H^+_3_model} the correlation functions of the bosonic $\mathrm{SL}(2)_{\fk+2}$ WZW model are given in the basis $\Phi_j(x,\bar x;z,\bar z)$ of $\mathrm{sl}(2)$ primaries. For the application to the supercoset, it will be useful to express these in a basis where the generators $j^3,\bar j^3$ are diagonal. This can be achieved through the following complex Mellin transform,\footnote{Here $(z,\bar z)$ are the worldsheet coordinates, while $(x,\bar x)$ are auxiliary complex coordinates. Under the action of $\mathrm{SL}(2,\mathbb{C})$ we have $\Phi_j(x,\bar{x}; z, \bar{z})\mapsto(cx+d)^{2j}(\bar{c}\bar{x}+\bar{d})^{2j}\Phi_j(x,\bar{x}; z, \bar{z})$. In discussions of strings in AdS$_3$, the latter can be interpreted as parameterizing the conformal boundary of AdS. Pulled back to the boundary, $j^3,\bar j^3$ act as dilatations and this explains why a Mellin transform of the form~\eqref{eq:complex_Mellin} is required to diagonalize them.}
\begin{equation}\label{eq:complex_Mellin}
    \Phi_{j,m,\bar{m}}(z,\bar z)=\int_\mathbb{C}\d^2x\,x^{j+m-1}\bar{x}^{j+\bar{m}-1}\Phi_j(x,\bar x;z,\bar z) \ ,
\end{equation}
where henceforth the integration measures in the complex plane are defined as $d^2x=(2\pi)^{-2}\d x\d\bar{x}$. Eq.~\eqref{eq:complex_Mellin} is a combination of an ordinary Mellin and a Fourier transform, which hence requires $m-\bar m\in\mathbb{Z}$, \ie the angular momentum of the string in the cigar is quantized. We are now ready to discuss the two and three-point functions of the supercoset, in the winding number preserving and violating sectors, in turn.

\subsubsection{Winding number preserving correlators}

In the winding number preserving (WNP) sector, supercoset correlation functions can be obtained via substituting the $H^+_3$ model correlators (see appendix~\ref{ap:H^+_3_model}) into eq.~\eqref{ew:wind_full_supercoset_correlators}.

\paragraph{WNP supercoset two-point functions} Given the two-point correlation function of the $\Phi_j$ primaries~\eqref{eq:H3+_2pt-func}, we can express it in the $\Phi_{j,m,\bar{m}}$ basis by performing the transform~\eqref{eq:complex_Mellin}. More specifically,\footnote{We leave the dependence of $\Phi_j$ on the anti-holomorphic coordinates implicit.}
\begin{equation}
\big\langle \Phi_{j_1,m_1,\bar m_1}(z_1)\Phi_{j_2,m_2,\bar m_2}(z_2)\big\rangle=\int_{\mathbb{C}^2} \d^2x_1\d^2x_2\,
x_1^{a_1}\bar x_1^{\bar a_1}
x_2^{a_2}\bar x_2^{\bar a_2}
\big\langle \Phi_{j_1}(x_1;z_1)\Phi_{j_2}(x_2;z_2)\big\rangle \ ,
\end{equation}
where $a_i\equiv j_i+m_i-1,\, \bar a_i\equiv j_i+\bar m_i-1$. We will perform this integral explicitly here to give a taste of what goes into such manipulations, but will defer the evaluation of similar integrals encountered later to appendix~\ref{app:bestiary}. Splitting the two-point function as 
\begin{equation}
\begin{aligned}
    \big\langle \Phi_{j_1}(x_1;z_1)\Phi_{j_2}(x_2;z_2)\big\rangle&=\frac{I_{\rm a}+I_{\rm b}}{|z_{12}|^{4\hat h_{1}}} \ ,\\
    \ I_\text{a}=T(\fk)\delta^{(2)}(x_{12})\delta(j_1+j_2-1) \ &, \ I_\text{b}=\frac{B(j_1)}{|x_{12}|^{4j_1}}\delta(j_1-j_2) \ ,
\end{aligned}
\end{equation}
where the dimensions read $\hat h_i=-\frac{j_i(j_i-1)}{\fk}$, we have the transform, for $I_\text{a}$,
\begin{equation}\label{eq:Ia}
\begin{aligned}
I_\text{a}&=T(\fk)\int_{\mathbb{C}^2} \d^2x_1\d^2x_2\,x_1^{a_1}\bar x_1^{\bar a_1}x_2^{a_2}\bar x_2^{\bar a_2}\delta^{(2)}(x_{12})\delta(j_1+j_2-1)\\
   &= T(\fk)\delta(j_1+j_2-1)
\int_\mathbb{C} \d^2x\, x^{M_{12}-1}\bar x^{\overline{M}_{12}-1}\\
&=  T(\fk)\delta(j_1+j_2-1)\tilde{\delta\,}(M_{12}) \ ,
\end{aligned}
\end{equation}
where in the last step we used the complex Mellin representation of a delta function and $M_{12}\equiv m_1+m_2$.\footnote{Explicitly, we have
\begin{equation}\label{eq:delta_Mellin}
    \int_\mathbb{C}\frac{\d x}{2\pi}\frac{\d\bar x}{2\pi}x^{a-1}\bar x^{\bar a-1}\overset{\ x=r\e^{i\theta}}{=}(2\pi)^{-2}\int_0^\infty \d r\,r^{a+\bar a-1}
\int_0^{2\pi}\d\theta\,\e^{i\theta(a-\bar a)}=\delta_{a-\bar a,0}\delta(a+\bar a)\equiv \tilde{\delta\,}(a) \ ,
\end{equation} 
where in the $r$ integral we rotated the contour of integration along the imaginary axis.} The compact notation $\tilde{\delta\,}(a)$ denotes a Kronecker delta for $a-\bar a$ and a delta function for $a+\bar a$. Similarly, the transform of $I_\text{b}$ reads,
\begin{equation}
   \begin{aligned}
        &B(j_1)\delta(j_1-j_2)
\int_{\mathbb{C}^2} \d^2x_1\d^2x_2\,
x_1^{a_1}\bar x_1^{\bar a_1}
x_2^{a_2}\bar x_2^{\bar a_2}
|x_{12}|^{-4j_1}\\
&\ =  B(j_1)\delta(j_1-j_2)\int_\mathbb{C} \d^2u\d^2y\,|y|^2
(uy)^{a_1}(\bar u\bar y)^{\bar a_1}
y^{a_2}\bar y^{\bar a_2}
|y|^{-4j_1}|u-1|^{-4j_1}\\
&\ =B(j_1)\delta(j_1-j_2)
\left[\int_\mathbb{C} \d^2y\,y^{M_{12}-1}\bar y^{\overline{M}_{12}-1}\right]
\left[\int_\mathbb{C} \d^2u\,u^{j-1+m_1}\bar u^{j-1+\bar m_1}|1-u|^{-4j}\right]\\
&\ =B(j_1)\delta(j_1-j_2)\,\tilde{\delta\,}(M_{12})\,\pi
\frac{\Gamma(j+m)\Gamma(1-2j)\Gamma(j-\bar m)}
{\Gamma(1-j-\bar m)\Gamma(2j)\Gamma(1-j+m)} \ ,
   \end{aligned}
\end{equation}
where, in the second line we implemented the change of variables $x_1=ux_2,\,x_2=y,\,\d^2x_1\d^2x_2=|y|^2\d^2u\d^2y$, and in the last line we again used the integral representation of $\delta^{(2)}(a)$, as well as the complex Beta integral
\begin{equation}\label{eq:complex_Beta_int}
    \int_{\mathbb{C}} \d^2u\,
u^{\alpha-1}\bar u^{\bar\alpha-1}
(1-u)^{\beta-1}(1-\bar u)^{\bar\beta-1}
=\pi
\frac{\Gamma(\alpha)\Gamma(\beta)\Gamma(1-\bar\alpha-\bar\beta)}
{\Gamma(1-\bar\alpha)\Gamma(1-\bar\beta)\Gamma(\alpha+\beta)} \ ,
\end{equation}
for $\alpha=j+m,\,\bar\alpha=j+\bar m,\,\beta=\bar \beta=1-2j$. Putting everything together, we arrive at
\begin{align}\label{eq:2pt_func_Phi-jmm}
&\big\langle \Phi_{j_1,m_1,\bar m_1}(z_1,\bar z_1)\Phi_{j_2,m_2,\bar m_2}(z_2,\bar z_2)\big\rangle
=\frac{\tilde{\delta\,}(M_{12})}{|z_{12}|^{4\hat h_{1}}}\mathsf T(\fk)
\Bigg[\delta(j_1+j_2-1)+\cr
&\qquad\qquad \quad  +\pi R(j_1)\delta(j_1-j_2) \frac{\Gamma(j_1-\bar m_1)\Gamma(j_1+m_1)\Gamma(1-2j_1)}{\Gamma(1-j_1-\bar m_1)\Gamma(1-j_1+m_1)\Gamma(2j_1)} \Bigg] \ ,
\end{align}
where the explicit expression for $R(j_1)$ is given in eq.~\eqref{eq:R(j)_relations}. 

Then, eqs.~\eqref{ew:wind_full_supercoset_correlators},~\eqref{eq:XR_vertex_correlators},~\eqref{eq:x3_vertex_correlators},~\eqref{eq:2pt_func_Phi-jmm} imply the following winding number preserving two-point correlation function for the supercoset  
\begin{equation}\label{eq:supercoset_2pt_func}
\begin{aligned}
&\Big \langle  V^{n_1,\bar n_1}_{j_1,m_1,\bar m_1}(z_1,\bar z_1)V^{n_2,\bar n_2}_{j_2,m_2,\bar m_2}(z_2,\bar z_2) \Big\rangle = \frac{\delta\Big(\sum_{i=1}^2 w_i\Big)\delta_{\sum_{i=1}^2  q_R^i,0}\delta_{\sum_{i=1}^2  \bar q_R^i,0}}{z_{12}^{2 h_{1}}\bar z_{12}^{2\bar h_{1}}}\\
& \ \times
\mathsf T(\fk)\bigg[\delta(j_1+j_2-1)+ R(j_1) \frac{\pi\Gamma(j_1-\bar m_1)\Gamma(j_1+m_1)\Gamma(1-2j_1)}{\Gamma(1-j_1-\bar m_1)\Gamma(1-j_1+m_1)\Gamma(2j_1)} \delta(j_1-j_2)\bigg] \ ,
\end{aligned}
\end{equation}
Note that unitarity of the discrete series representations of $\mathrm{sl}(2)_{\fk+2}$ demands $\frac{1}{2}<j<\frac{\fk+2}{2}$. The lower bound ensures that the second term in the correlator~\eqref{eq:supercoset_2pt_func} does not have a pole. By further inspecting the second term of that correlator, one is required to impose the stronger bound $j<\frac{\fk+1}{2}$ to avoid additional poles. In addition, the fraction of gamma functions in the second term has LSZ-type poles that correspond precisely to the discrete representations $|m|-j\in\mathbb{N}$ of $\mathrm{sl}(2)$~\cite{Aharony:2004xn}. When one scatters strings of principal representations in the cigar target, the physical meaning of these poles is that they correspond to bound states localized near the tip~\cite{Dijkgraaf:1991ba}.

\paragraph{WNP supercoset three-point functions} We now wish to transform to the $\Phi_{j,m,\bar m}$ basis the correlator~\eqref{eq:Phij_3-point_func}, \ie
\begin{equation}\label{eq:Phij_3-point_func}
\big\langle
\Phi_{j_1}(x_1;z_1)\Phi_{j_2}(x_2,;z_2)\Phi_{j_3}(x_3;z_3)
\big\rangle=\frac{D(j_1,j_2,j_3)}{
|z_{12}|^{2\hat h_{123}}
|z_{23}|^{2\hat h_{231}}
|z_{13}|^{2\hat h_{312}}
|x_{12}|^{2j_{123}}
|x_{23}|^{2j_{231}}
|x_{13}|^{2j_{312}}} \ ,
\end{equation}
where the corresponding $H^+_3$ model structure constants are~
\begin{equation}\label{eq:H3+ structure constant}
   D(j_1,j_2,j_3)=\mathrm A(b^{-1}) \big(b^{2b^{-2}-2}\gamma(b^{-2})\big)^{b^2(1-\tilde j)}\frac{\Upsilon_{b}(b)\prod_{i=1}^3\Upsilon_{b}(2bj_i)}{\Upsilon_{b}\big(b(\tilde{j \,}-1)\big)\prod_{\text{cyc}}\Upsilon_{b}(bj_{ijk})} \ ,
\end{equation} 
where $\mathrm A(b)$ denotes a function of $b$ only, which cannot be fixed unambiguously.\footnote{In bootstrap language, one can always multiply the structure constant by a momenta-independent function and the shift equations will continue to be obeyed.} Here we introduced $b\equiv \fk^{-1/2}$, $\tilde{j \,}=j_1+j_2+j_3$, $j_{ijk}=j_i+j_j-j_k$ and the cyclic product is taken over cyclic permutations of $i,j,k\in\lbrace{1,2,3\rbrace}$. To do so we implement the complex Mellin transform~\eqref{eq:complex_Mellin}, \ie 
\begin{equation}
\begin{aligned}
\left\langle \prod_{i=1}^3 \Phi_{j_i,m_i,\bar m_i}(z_i,\bar z_i)\right\rangle
&=
\int_{\mathbb{C}^3} \prod_{i=1}^3 \d^2x_i\,
x_i^{j_i+m_i-1}\bar x_i^{j_i+\bar m_i-1}
\left\langle \prod_{i=1}^3 \Phi_{j_i}(x_i,\bar x_i;z_i,\bar z_i)\right\rangle \\
&=
D(j_1,j_2,j_3)\,\mathcal J(j_i,m_i,\bar m_i)
\prod_{\mathrm{cyc}}|z_{ab}|^{-2\hat h_{abc}}
 \ ,
\end{aligned}
\end{equation}
where $i, a,b,c\in \{1,2,3\}$; We also use $\hat h_a=-\frac{j_a(j_a-1)}{\fk}$ and $\hat h_{abc}\equiv \hat h_a +\hat h_b- \hat h_c$ and
\begin{equation}\label{eq:J_integral}
\mathrm{WNP}_{\mathcal{J}}:\quad \mathcal J_{\mathrm{VNP}}(j_i,m_i,\bar m_i)
\equiv
\int \prod_{i=1}^3 \d^2x_i\,
x_i^{j_i+m_i-1}\bar x_i^{j_i+\bar m_i-1}\,
|x_{12}|^{-2j_{123}}|x_{23}|^{-2j_{231}}|x_{13}|^{-2j_{312}} \ .
\end{equation}
The integral $\mathrm{WNP}_{\mathcal{J}}$ is evaluated in appendix \ref{app:bestiary} and yields 
\begin{equation}
\mathcal J_{\mathrm{VNP}}(j_i,m_i,\bar m_i)
=
\tilde{\delta\,}\Big(\sum_{i=1}^3m_i\Big)\,
\mathcal W(j_i,m_i,\bar m_i) \ .
\end{equation}
The delta function imposes both angular momentum $\sum_{i=1}^3 (m_i-\bar m_i)=0$ and winding $\sum_{i=1}^3 (m_i+\bar m_i)=0$ conservation. Overall, the three-point function of the bosonic $\mathrm{SL}(2)_{\fk+2}$ model in the $\Phi_{j,m,\bar m}$ basis, reads 
\begin{equation}\label{eq:3pt_func_Phi-jmm}
    \left\langle \prod_{i=1}^3 \Phi_{j_i,m_i,\bar m_i}(z_i,\bar z_i)\right\rangle=\delta\Big(\sum_{i=1}^3m_i\Big)\delta\Big(\sum_{i=1}^3\bar m_i\Big)\,
C(j_i,m_i,\bar m_i)
\prod_{\mathrm{cyc}}|z_{ab}|^{-2\hat h_{abc}}
 \ ,
\end{equation}
where 
\begin{equation}
    C(j_i,m_i,\bar m_i) \equiv D(j_1,j_2,j_3)\mathcal{W}(j_i,m_i,\bar m_i) \ ,
\end{equation}
and we defined $\hat{h}_i$ below equation (\ref{eq:Ia}). 
Combining eqs.~\eqref{eq:XR_vertex_correlators},~\eqref{eq:x3_vertex_correlators},~\eqref{eq:3pt_func_Phi-jmm} in eq.~\eqref{ew:wind_full_supercoset_correlators}, the supercoset three-point functions in the winding number preserving sector, read~
\begin{equation}\label{eq:supercoset_3pt_funcs}
    \begin{aligned}
        \Big \langle \prod_{i=1}^3 V^{n_i,\bar n_i}_{j_i,m_i,\bar m_i}(z_i,\bar z_i) \Big\rangle &= \delta\Big(\sum_{i=1}^3 w_i\Big)\delta_{\sum_{i=1}^3q^i_R,0}\delta_{\sum_{i=1}^3\bar q_R^i,0} \, C(j_i,m_i,\bar m_i)
\prod_{\mathrm{cyc}}z_{ab}^{-h_{abc}}\bar z_{ab}^{-\bar h_{abc}} \ ,
    \end{aligned}
\end{equation}
where the dimensions and R-charges are as in eq.~\eqref{eq:full_supercoset_dimas_&_R_charges}, the $H_3^+$ structure constant is in (\ref{eq:H3+ structure constant}) and 
\begin{equation}
    \mathcal W(j_i,m_i,\bar m_i)
=
\int_{\mathbb{C}^2} \d^2y \d^2z\,
y^{a_2}\bar y^{\bar a_2}
z^{a_3}\bar z^{\bar a_3}
|1-y|^{-2j_{123}}|1-z|^{-2j_{312}}|y-z|^{-2j_{231}} \ .
\end{equation}
We recall that $\tilde{j \ }=j_1+j_2+j_3,\,j_{abc}=j_a+j_b-j_c,\,a_i=j_i+m_i-1,\,\bar a_i=j_i+\bar m_i-1$.
We now summarize a few important points about this structure constant, from the supercoset point of view:
\begin{itemize}
    \item The apparent asymmetry in the $\mathcal W$ integral is an artifact of the particular choice of variables~\eqref{eq:J_integral_variables_change}, since the original $\mathcal J$ integral in~\eqref{eq:J_integral} exhibits manifest symmetry under the exchange of $j_i,m_i,\bar m_i$. In the support of the delta functions, the full supercoset structure constant~\eqref{eq:H3+ structure constant} is symmetric in the external states.
    \item A form of the $\mathcal W$ integral was evaluated in ref.~\cite{Fukuda:2001jd}, however the explicit closed form expression is rather complicated and unilluminating, hence we do not present it here. On the other hand, when evaluating particular correlation function below we will be able to perform the integral in special cases explicitly.
    \item Apart from shifting the level from $\fk\to\fk+2$, the role of the fermions is minor, since they only enter in~\eqref{eq:supercoset_3pt_funcs} through the conservation of the R-charges. In particular, in the winding number preserving sector one is forced to have the special constraints $\sum_i n_i=\sum_i \bar n_i=0$ for the fermionic vertex operators $\e^{\ii (n H_L-\bar n H_R)}$ in eq.~\eqref{eq:full_supercoset_primaries}. 

\item There also exists a formula, originally due to Ribault and Teschner~\cite{Ribault:2005wp}, relating arbitrary primary $N$-point correlators of the $H^+_3$ model to $(2N-2)$-point bosonic Liouville theory correlators on the sphere (this formula was later generalized to Riemann surfaces of arbitrary genus~\cite{Hikida:2007tq}). This formula, implies the following expression for the winding number preserving $N$-point correlation functions of the supercoset primaries 
\begin{equation}\label{eq:wind_final_supercoset_corellator}
    \begin{aligned}
        &\Big \langle\prod_{i=1}^N V^{n_i,\bar n_i}_{j_i,m_i,\bar m_i}(z_i,\bar z_i) \Big\rangle_{\n=2\,\frac{\mathrm{SL(2)_{k}}}{\mathrm{U}(1)}}=\frac{2\pi^{3-2N}}{(N-2)!\sqrt{\fk}}
\delta\Big(\sum_{i=1}^N w_i\Big)\delta_{\sum_{i=1}^Nq^i_R,0}\delta_{\sum_{i=1}^N\bar q_R^i,0}
\\
&\quad\times
\prod_{1\leq i<j\le N}
z_{ij}^{n_in_j+m_i+m_j+\frac{2}{\fk}(n_i+m_i)(n_j+m_j)+\frac{\fk}{2}+1}
\bar z_{ij}^{\bar n_i\bar n_j+\bar m_i+\bar m_j+\frac{2}{\fk}(\bar n_i+\bar m_i)(\bar n_j+\bar m_j)+\frac{\fk}{2}+1}\\
&\quad\times\prod_{i=1}^N
\pi^{j_i} \big(b^{2b^{-2}-2}\gamma(b^{-2})\big)^{-b^2j_i}b^{1-2j_i+2b^2(1-j_i)}\frac{\gamma\big(b^2(2j_i-1)\big)\Gamma(j_i+m_i)}{\gamma(b^2)^{j_i}\Gamma(1-j_i-\bar m_i)}
\\
&\quad\times
\int_{\mathbb C}
\prod_{i=1}^{N-2} \d^2y_i
\prod_{1\le i<j\le N-2}|y_{ij}|^{\fk+2}\prod_{i=1}^N\prod_{j=1}^{N-2}
(z_i-y_j)^{-m_i-\frac{\fk}{2}-1}
(\bar z_i-\bar y_j)^{-\bar m_i-\frac{\fk}{2}-1}
\\
&\quad\times\left\langle
\prod_{i=1}^N
V_{b(1-j_i)+\frac{1}{2b}}(z_i,\bar z_i)
\prod_{j=1}^{N-2}
V_{-\frac{1}{2b}}(y_j, \bar y_j)
\right\rangle_{\n=0\,\text{Liouville}} \  .
    \end{aligned}
\end{equation}
where we dropped an overall $j$-independent factor. This expression can be obtained by combining eq.~$(3.36)$ in~\cite{Ribault:2005wp} (in our chosen operator normalization) with eq.~\eqref{ew:wind_full_supercoset_correlators} here. A similar relation is true also for the winding number violating correlation functions~\cite{Ribault:2005ms}, which we will now describe.

\end{itemize}

\subsubsection{Winding number violating correlators}

In the winding number violating (WNV) sector, a given $N$-point function can violate winding number up to $N-2$ units. This was noted in the original work on FZZ duality, and was shown in~\cite{Maldacena:2001km} to be forced by the $\mathrm{sl}(2)_{\fk+2}$ representation theory (see appendix~\ref{ap:H^+_3_model} for a review of the argument).

\paragraph{WNV supercoset two-point functions} The above statement implies that only two-point functions that preserve winding number exist in the supercoset. Consequently, the corresponding two-point function is as in eq.~\eqref{eq:supercoset_2pt_func}.

\paragraph{WNV supercoset three-point functions} In the supercoset, apart from the WNP three-point function of primaries~\eqref{eq:supercoset_3pt_funcs}, there exists a three-point function that violates winding by $\pm1$ unit. As we explained in the paragraph after eq.~\eqref{eq:x3_vertex_correlators}, with a direct application of the relation~\eqref{ew:wind_full_supercoset_correlators} one is unable to obtain these correlators -- a more elaborate manipulation is required. Such a manipulation was proposed by Fateev-Zamolodchikov-Zamolodchikov in their work on the homonymous duality.\footnote{We are very grateful to Al. Zamolodchikov for sharing with us his unpublished notes on winding number violating three-point functions of the bosonic $\mathrm{SL}(2)_\fk/\mathrm{U}(1)$ coset.} The application of this argument in our context proceeds as follows; consider the primary $\mathrm V_\sigma\equiv V^{-\sigma,-\sigma}_{\frac{\fk+2}{2},\sigma \frac{\fk+2}{2},\sigma \frac{\fk+2}{2}}$, where $\sigma=\pm1$. From~\eqref{eq:full_supercoset_dimas_&_R_charges}, this operator has $h=q_R=0$ and since the supercoset is a unitary SCFT it can be identified with the identity operator, \ie $\mathrm V_\sigma\sim \mathbbm{1}_\text{supercoset}$. Additionally, recalling eq.~\eqref{eq:building_primaries}, this operator can be decomposed as $\mathrm V_\sigma=\Phi_{\frac{\fk+2}{2},\sigma \frac{\fk+2}{2},\sigma \frac{\fk+2}{2}}\e^{-\sigma\sqrt{\frac{\fk+2}{2}}(x^3-\bar x^3)}$, it corresponds to a degenerate representation (since $j=\pm m=\pm\bar m$) and further carries winding number $w=\sigma$. Suppose, now, that we attempt to compute a four-point function of the form $\big\langle\mathrm V_\sigma(z_2)\prod_{i=1,3,4}V^{n_i,\bar n_i}_{j_i,m_i,\bar m_i}(z_i)\big\rangle\equiv \big\langle\prod_{i=1,3,4}V^{n_i,\bar n_i}_{j_i,m_i,\bar m_i}(z_i)\big\rangle$ through eq.~\eqref{ew:wind_full_supercoset_correlators}. Due to the fact that $\mathrm V_\sigma$ has $q_R=\bar q_R=0$, the vertex operator correlator in the numerator of~\eqref{ew:wind_full_supercoset_correlators} just imposes R-charge conservation for the remaining three-point function. The correlator in the denominator yields the constraints,
\begin{equation}
    \sum_{i=1,3,4} m_i+\sigma \frac{\fk+2}{2}=\sum_{i=1,3,4} \bar m_i+\sigma \frac{\fk+2}{2}=0 \ ,
\end{equation}
whereas R-charge conservation further implies,
\begin{equation}
    0=\sum_{i=1,3,4}q^i_R=\sum_{i=1,3,4}\Big(\frac{2m_i}{k}+n_i\frac{k+2}{k}\Big)\Rightarrow \sum_{i=1,3,4}n_i=\sum_{i=1,3,4}\bar n_i = \sigma \ .
\end{equation}
Combining these two, the winding number for the remaining supercoset three-point function becomes
\begin{equation}\label{eq:wnv_winding_no}
    \sum_{i=1,3,4} w_i \fk=\sum_{i=1,3,4}(m_i+n_i+\bar m_i+\bar n_i)=-\sigma \fk \ ,
\end{equation}
\ie the resulting supercoset three-point function violates winding by $\pm 1$ unit. The conclusion is that the WNV three-point functions of supercoset primaries can be obtained from eq.~\eqref{ew:wind_full_supercoset_correlators}, via evaluating the four-point function $\big\langle\mathrm \Phi_{\frac{\fk+2}{2},\sigma\frac{\fk+2}{2},\sigma\frac{\fk+2}{2}}(z_2)\prod_{i=1,3,4}\Phi_{j_i,m_i,\bar m_i}(z_i)\big\rangle$ of bosonic $\mathrm{sl}(2)_{\fk+2}$ primaries. In particular, because $\Phi_{\frac{\fk+2}{2},\sigma\frac{\fk+2}{2},\sigma\frac{\fk+2}{2}}$ is a degenerate $\mathrm{sl}(2)_{\fk+2}$ field, this four-point function can be determined exactly, as we now turn to explain. As before, we will first compute the correlator in the $x$-basis and then transform it to the $m$-basis through~\eqref{eq:complex_Mellin}. This procedure was also outlined in~\cite{Maldacena:2001km}, but we review it here because many of the intermediate steps were omitted in that reference.

Starting from the $x$-basis field $\Phi_{\frac{\fk+2}{2}}(x;z)$, we note that it corresponds to a $\mathrm{sl}(2)_{\fk+2}$ degenerate representation, whose null state condition at level-$1$ reads
\begin{equation}
    (j^+_{-1}-2xj^0_{-1}+x^2j^-_{-1})\Phi_{\frac{\fk+2}{2}}(x;z) = 0 \ .
\end{equation}
For details on the explicit derivation of this and of the following relations, see appendix~\ref{ap:H^+_3_model}. Using the current Ward identity~\eqref{eq:current_Ward_identity} this yields the following equation for the four-point function,
\begin{equation}
    \sum_{i=1,3,4}\frac{(x_i-x_2)^2\partial_{x_i}+2j_i(x_i-x_2)}{z_i-z_2}\big\langle\Phi_{\frac{\fk+2}{2}}(x_2;z_2)\prod_{i=1,3,4}\Phi_{j_i}(x_i;z_i)\big\rangle=0 \ .
\end{equation}
Parameterizing the four-point function like eq.~\eqref{eq:G-4-pt-func}, the above relation further implies a differential equation for the holomorphic conformal block $\mathcal F(x,z)$, \ie
\begin{equation}\label{eq:null_state_eq}
    x(x-1)(x-z)\partial_x\mathcal F=\Big[d(x^2-2xz+z)+2j_1x(1-z)+2j_3z(1-x)\Big]\mathcal F \ .
\end{equation}
Moreover, the Knizhnik–Zamolodchikov (KZ) equation for this degenerate field takes the form,
\begin{equation}
    \partial_z\Phi_{\frac{\fk+2}{2}}(x;z)=(j^0_{-1}-xj^-_{-1})\Phi_{\frac{\fk+2}{2}}(x;z) \ .
\end{equation}
or, equivalently, in terms of the four-point function,
\begin{equation}
    \partial_{z_2}\big\langle\Phi_{\frac{\fk+2}{2}}(x_2;z_2)\prod_{i=1,3,4}\Phi_{j_i}(x_i;z_i)\big\rangle=\sum_{i=1,3,4}\frac{(x_i-x_2)\partial_{x_i}+j_i}{z_i-z_2}\big\langle\Phi_{\frac{\fk+2}{2}}(x_2;z_2)\prod_{i=1,3,4}\Phi_{j_i}(x_i;z_i)\big\rangle \ ,
\end{equation}
with the corresponding differential equation for the conformal block being
\begin{equation}\label{eq:degen_KZ_eq}
    \Big[z(z-1)\partial_z+x(x-1)\partial_x\Big]\mathcal F=\Big[dx+j_1-z\Big(j_4-\frac{\fk+2}{2}\Big)\Big]\mathcal F \ .
\end{equation}
As detailed in appendix~\ref{ap:H^+_3_model}, it is straightforward to integrate the system of equations~\eqref{eq:null_state_eq}, \eqref{eq:degen_KZ_eq}, uniquely determining the conformal block (up to an overall multiplicative constant), to be
\begin{equation}\label{eq:main-text_degen_conformal_block_F}
    \mathcal F(x,z)=z^{j_1}(z-1)^{j_3}x^{2j_3+d}(x-1)^{2j_1+d}(x-z)^{r} \ ,
\end{equation}
where
\begin{equation}\label{eq:d and r}
    d=j_4-j_1-\frac{\fk+2}{2}-j_3 \ , \ r=\frac{\fk+2}{2} -j_1-j_3-j_4 \ .
\end{equation}
The obtained conformal block has the expected behavior when $x,z\to\lbrace 0,1\rbrace$, while an interpretation of the branch cut as $x\to z$ was proposed in~\cite{Maldacena:2001km} in terms of worldsheet instantons (see appendix~\ref{ap:H^+_3_model} for further comments). Moreover, as is well known, conformal invariance fixes the small-$z$ expansion of the s-channel conformal block to be of the form
\begin{equation}
    \mathcal{F}(x,z)=z^{\hat h_s-\hat h_1-\hat h_2}\big(1+\dots\big) \ .
\end{equation}
The leading contribution of our conformal block~\eqref{eq:main-text_degen_conformal_block_F} in the small-$z$ limit is $z^{j_1}$. Hence, we obtain
\begin{equation}\label{eq:j_s-solns}
    \hat h_s-\hat h_1-\hat h_2=j_1\Rightarrow-\frac{j_s(j_s-1)-j_1(j_1-1)-\frac{\fk+2}{2}\big(\frac{\fk+2}{2} -1\big)}{\fk}=j_1\Rightarrow j_s=\begin{cases}
        \frac{\fk+2}{2} -j_1 \ , \\
        j_1+1-\frac{\fk+2}{2} \ ,
    \end{cases}
\end{equation}
which implies the following fusion rule for our degenerate field\footnote{The alternative solution for $j_s=j_1+1-\frac{\fk+2}{2}$ in~\eqref{eq:j_s-solns} corresponds to the reflected primary under $j\mapsto 1-j$.}
\begin{equation}
    \Phi_{\frac{\fk+2}{2}}\times\Phi_{j_1}\sim \Phi_{\frac{\fk+2}{2} -  j_1} \ ,
\end{equation}
which is expected since the space of solutions of the system~\eqref{eq:null_state_eq},~\eqref{eq:degen_KZ_eq} is one-dimensional and therefore there ought to be a single fusion channel, as anticipated in writing~\eqref{eq:G-4-pt-func}.

The remaining step is to evaluate the coefficient $\tilde D(j_1,j_3,j_4)$, substituting $j_s=\frac{\fk+2}{2} -j_1$. Explicitly, after some massaging, we find 
\begin{equation}
    D\Big(j_1,\frac{\fk+2}{2},j_s\Big)= N(\fk)\delta\Big(j_s+j_1-\frac{\fk+2}{2}\Big) \ ,
\end{equation}
where 
\begin{equation}\label{eq:N(k)}
    N(\fk) = {A}(\fk^{\frac{1}{2}})\fk^{-2}\big(\fk^{-\fk}\gamma(\fk)\big)^{-\frac{1}{\fk}-1} \gamma(\fk^{-1})~.
\end{equation}
We therefore conclude that,
\begin{equation}\label{eq:interm_D}
    \tilde D(j_1,j_3,j_4)=\frac{D\big(j_1,\frac{\fk+2}{2},\frac{\fk+2}{2} -j_1\big)D(j_s,j_3,j_4)}{B(j_s)}=\frac{N(\fk)}{B\big(\frac{\fk+2}{2}-j_1\big)}D\Big(\frac{\fk+2}{2} -j_1,j_3,j_4\Big) \ ,
\end{equation}
for $B(j)$ the two-point function normalization~\eqref{eq:H3+_2pt-func}. This discussion has completely determined the four-point function~\eqref{eq:G-4-pt-func}, to be
\begin{equation}\label{eq:full_4pt_G}
    \big\langle\Phi_{\frac{\fk+2}{2}}(x_2;z_2)\prod_{i=1,3,4}\Phi_{j_i}(x_i;z_i)\big\rangle=\tilde D(j_1,j_3,j_4)\big|P_z(z_i)P_x(x_i)\mathcal F(x,z)\big|^2 \ ,
\end{equation}
where,
\begin{equation}
    \begin{aligned}
        &P_z(z_i)=z_{42}^{-2\hat h_2}z_{41}^{\hat h_2+\hat h_3-\hat h_1-\hat h_4}z_{43}^{\hat h_1+\hat h_2-\hat h_3-\hat h_4}z_{31}^{\hat h_4-\hat h_1-\hat h_2-\hat h_3} \ ,\\
        &P_x(x_i)=x_{42}^{-2j_2}x_{41}^{j_2+j_3-j_1-j_4}x_{43}^{j_1+j_2-j_3-j_4}x_{31}^{j_4-j_1-j_2-j_3} \ ,\\
        &\mathcal F(x,z)=z^{j_1}(z-1)^{j_3}x^{2j_3+d}(x-1)^{2j_1+d}(x-z)^{r} \ ,\\
        &\hat h_i=-\frac{j_i(j_i-1)}{\fk} \ , \ d=j_4-j_1-j_3-\frac{\fk+2}{2} \ , \ r=\frac{\fk+2}{2} -j_1-j_3-j_4 \ ,
    \end{aligned}
\end{equation}
and similarly for the anti-holomorphic expressions. As usual, when $(z_1,z_2,z_3,z_4)\mapsto(0,z,1,\infty)$ (and similarly for $x_i$) the $P_z,P_x$ factors go to one.

Next, we will transform this correlator into the $m$-basis, in which we have primary representations of definite spectral flow. The first thing to note is that the Mellin transform~\eqref{eq:complex_Mellin} for the degenerate field $\Phi_{\frac{\fk+2}{2}}$ diverges, as
\begin{equation}
   \Phi_{\frac{\fk+2}{2},\sigma\frac{\fk+2}{2},\sigma\frac{\fk+2}{2}}(z)=\begin{cases}
       +: \int_\mathbb{C}\d^2x\,|x|^{2(\fk +1)}\Phi_{\frac{\fk+2}{2}}(x;z) \ ,\\
       -:\int_\mathbb{C}\d^2x\,|x|^{-2}\Phi_{\frac{\fk+2}{2}}(x;z) \ .
   \end{cases} 
\end{equation}
Nonetheless, we can extract the corresponding residues via the limits
\begin{equation}\label{eq:res_lims}
    \begin{cases}
        \Phi_{\frac{\fk+2}{2},\frac{\fk+2}{2},\frac{\fk+2}{2}}(z)=\lim_{x\to\infty}|x|^{2(\fk+2)}\Phi_{\frac{\fk+2}{2}}(x;z) \ ,\\
        \Phi_{\frac{\fk+2}{2},-\frac{\fk+2}{2},-\frac{\fk+2}{2}}(z)=\lim_{x,\bar x\to 0}\Phi_{\frac{\fk+2}{2}}(x;z) \ .
    \end{cases}
\end{equation}
The complex Mellin transform for the other three insertions in~\eqref{eq:full_4pt_G} is as usual, namely via eq.~\eqref{eq:complex_Mellin}. With the above in mind, we will transform the degenerate four-point function~\eqref{eq:full_4pt_G} to the $m$-basis in each case separately. As an important note, since the three-point function can violate winding up to $\pm1$ units and the degenerate primary $\Phi_{\frac{\fk+2}{2},\sigma\frac{\fk+2}{2},\sigma\frac{\fk+2}{2}}$ carries $w=\pm1$, we can accommodate the four-point function~\eqref{eq:full_4pt_G} to preserve winding number. As was shown in ref.~\cite{Ribault:2005ms}, in that instance the spectrally flowed KZ equation coincides with the unflowed one and hence the structure constants of the $H^+_3$ model~\cite{Teschner:1997ft} can be used (hence we used $D(j_s,j_3,j_4)$ in eq.~\eqref{eq:interm_D}).

\subparagraph{$m$-basis flowed three-point function with $\sum_iw_i=-1$} We have to evaluate the expression
\begin{equation}\label{eq:def_wnv_-1}
   \mathcal{G}^{(-1)}(z_i)= \int_{\mathbb C^3}\prod_{i=1,3,4}\d^2x_i\,x_i^{j_i+m_i-1}\bar x_i^{j_i+\bar m_i-1}\lim_{x_2\to\infty}|x_2|^{2(\fk+2)} G(x_i,z_i) \ ,
\end{equation}
with $G(x_i,z_i)$ given by eq.~\eqref{eq:full_4pt_G}. Let us do the limit first, with
\begin{equation}
    \lim_{x_2\to\infty}x=-\frac{x_{43}}{x_{31}} \ , \ \lim_{x_2\to\infty}(x-1)=-\frac{x_{41}}{x_{31}} \ , \ \lim_{x_2\to\infty}(x-z)=-\frac{x_{43}+zx_{31}}{x_{31}} \ ,
\end{equation}
which, after combining with the $x_2^{\fk+2}$ pre-factor of the limit, yield
\begin{equation}
    \begin{aligned}
        &\lim_{x_2\to\infty}x^{\fk+2}_2P_x(x_i)x^{2j_3+d}(x-1)^{2j_1+d}(x-z)^r=(-1)^d(x_{43}+zx_{31})^r \ ,
    \end{aligned}
\end{equation}
where $d$ and $r$ are defined in (\ref{eq:d and r}).
We remark that, from now on, we will ignore the various branch phases, like $(-1)^d$ above, which can be absorbed into an overall rescaling of the correlation function. Therefore, the holomorphic piece surviving the limit reads
\begin{equation}
    P_z(z_i)z^{j_1}(z-1)^{j_3}\big(x_4+x_3(z-1)-zx_1\big)^r \ .
\end{equation}
From~\eqref{eq:def_wnv_-1} we obtain,
\begin{equation}\label{eq:main-text_WNV_int}
    \begin{aligned}
        \mathcal{G}^{(-1)}(z_i) &=\tilde D(j_1,j_3,j_4)\big|P_zz^{j_1}(z-1)^{j_3}\big|^2 \mathcal{J}(j_i,m_i,\bar m_i)~,\cr
    \hspace{-.6cm}\mathrm{WNV}^{-1}_{\mathcal{J}}:~ \mathcal{J}_{\mathrm{WNV}}^{-1}(j_i,m_i,\bar m_i) &= \int_{\mathbb C^3}\prod_{i=1,3,4}\d^2x_i\,x_i^{j_i+m_i-1}\bar x_i^{j_i+\bar m_i-1}(x_{43}+zx_{31})^r(\bar x_{43}+\bar z\bar x_{31})^r \ .
    \end{aligned}
\end{equation}
We evaluate this integral in appendix \ref{app:bestiary}. It yields
\begin{equation}
 \mathcal{J}_{\mathrm{WNV}}(j_i,m_i,\bar m_i)=  \tilde\delta\Big(m_1+m_3+m_4+\frac{\fk+2}{2} \Big)\pi^2 \frac{1}{\gamma\big(j_1+j_3+j_4-\frac{\fk+2}{2} \big)}\prod_{i=1,3,4}\frac{\Gamma(j_i+m_i)}{\Gamma(1-j_i-\bar m_i)}~.  
\end{equation}
Lastly, the leftover $z$-dependence of the four-point function can be massaged to give
\begin{equation}
    \begin{aligned}
         f_m(z_i)&\equiv P_zz^{j_1}(z-1)^{j_3}z^{-j_1-m_1}(z-1)^{-j_3-m_3}\\
         &= z_{21}^{-m_1}z_{23}^{-m_3}z_{42}^{-m_4}z_{43}^{\hat h_1+\hat h_2-\hat h_3-\hat h_4-m_1}z_{41}^{\hat h_2+\hat h_3-\hat h_1-\hat h_4-m_3}z_{31}^{\hat h_2+\hat h_4-\hat h_1-\hat h_3-m_4} \ .
    \end{aligned}
\end{equation}
Combining the above results, the four-point function in the $m$-basis reads,
\begin{equation}\label{eq:ful_m_basis_wnv-1_sl2}
    \begin{aligned}
        &\big\langle\Phi_{j_1,m_1,\bar m_1}(z_1)\Phi_{\frac{\fk+2}{2},\frac{\fk+2}{2},\frac{\fk+2}{2}}(z_2)\Phi_{j_3,m_3,\bar m_3}(z_3)\Phi_{j_4,m_4,\bar m_4}(z_4)\big\rangle&\\
        & \ =\tilde\delta\Big(m_1+m_3+m_4+\frac{\fk+2}{2}\Big)\frac{\pi^2 N(\fk)}{B\big(\frac{\fk+2}{2}-j_1\big)}\prod_{i=1,3,4}\frac{\Gamma(j_i+m_i)}{\Gamma(1-j_i-\bar m_i)} \frac{D\Big(\frac{\fk+2}{2} -j_1,j_3,j_4\Big)}{\gamma\big(j_1+j_3+j_4-\frac{\fk+2}{2} \big)}|f_{m}(z_i)|^2 \ .
    \end{aligned}
\end{equation}

\subparagraph{$m$-basis flowed three-point function with $\sum_iw_i=1$} The aim is to compute the following integral transform,
\begin{equation}\label{eq:def_wnv_+1}
   \mathcal{G}^{(+1)}(z_i)= \int_{\mathbb C^3}\prod_{i=1,3,4}\d^2x_i\,x_i^{j_i+m_i-1}\bar x_i^{j_i+\bar m_i-1}\lim_{x_2,\bar x_2\to 0} G(x_i,z_i) \ .
\end{equation}
This can be achieved via very similar manipulations to the ones we just discussed. The details are presented in appendix \ref{app:bestiary}. The full $m$-basis correlator in the case $\sum_i \omega_i =1$ is then
\begin{equation}\label{eq:ful_m_basis_wnv+1_sl2}
    \begin{aligned}
        &\big\langle\Phi_{j_1,m_1,\bar m_1}(z_1)\Phi_{\frac{\fk+2}{2},-\frac{\fk+2}{2},-\frac{\fk+2}{2}}(z_2)\Phi_{j_3,m_3,\bar m_3}(z_3)\Phi_{j_4,m_4,\bar m_4}(z_4)\big\rangle&\\
        & \ =\tilde\delta\Big(m_1+m_3+m_4-\frac{\fk+2}{2}\Big)\frac{\pi^2 N(\fk)}{B\big(\frac{\fk+2}{2}-j_1\big)}\prod_{i=1,3,4}\frac{\Gamma(j_i-m_i)}{\Gamma(1-j_i+\bar m_i)} \frac{D\Big(\frac{\fk+2}{2} -j_1,j_3,j_4\Big)}{\gamma\big(j_1+j_3+j_4-\frac{\fk+2}{2} \big)}|f_{-m}(z_i)|^2 \ ,
    \end{aligned}
\end{equation}
where 
\begin{equation}
     f_{-m}(z_i)= z_{21}^{m_1}z_{23}^{m_3}z_{42}^{m_4}z_{43}^{\hat h_1+\hat h_2-\hat h_3-\hat h_4+m_1}z_{41}^{\hat h_2+\hat h_3-\hat h_1-\hat h_4+m_3}z_{31}^{\hat h_2+\hat h_4-\hat h_1-\hat h_3+m_4} \ .
\end{equation}

\subparagraph{WNV supercoset primary structure constants} Up to this point we have determined the following $m$-basis four-point function of the $\mathrm{SL}(2)_{\fk+2}$ WZW model,
\begin{equation}\label{eq:ful_m_basis_wnv_sl2}
    \begin{aligned}
        &\big\langle\Phi_{j_1,m_1,\bar m_1}(z_1)\Phi_{\frac{\fk+2}{2},\sigma\frac{\fk+2}{2},\sigma\frac{\fk+2}{2}}(z_2)\Phi_{j_3,m_3,\bar m_3}(z_3)\Phi_{j_4,m_4,\bar m_4}(z_4)\big\rangle\\
        &\qquad=\delta\Big(\sum_{i=1,3,4}m_i+\sigma\frac{\fk+2}{2}\Big)\delta\Big(\sum_{i=1,3,4}\bar m_i+\sigma\frac{\fk+2}{2}\Big)C^{w}(j_i,m_i,\bar m_i)|f_{\sigma m}(z_i)|^2 \ ,
    \end{aligned}
\end{equation}
with $\sigma =\pm$ and
\begin{equation}\label{eq:f_sigmam_etc}
\begin{aligned}
    &C^w(j_i,m_i,\bar m_i)=\frac{\pi^2 N(\fk)}{B\big(\frac{\fk+2}{2}-j_1\big)}\prod_{i=1,3,4}\frac{\Gamma(j_i+\sigma m_i)}{\Gamma(1-j_i-\sigma\bar m_i)} \frac{D\Big(\frac{\fk+2}{2} -j_1,j_3,j_4\Big)}{\gamma\big(j_1+j_3+j_4-\frac{\fk+2}{2} \big)} \ , \\
    &f_{\sigma m}(z_i)= z_{21}^{-\sigma m_1}z_{23}^{-\sigma m_3}z_{42}^{-\sigma m_4}z_{43}^{\hat h_1+\hat h_2-\hat h_3-\hat h_4-\sigma m_1}z_{41}^{\hat h_2+\hat h_3-\hat h_1-\hat h_4-\sigma m_3}z_{31}^{\hat h_2+\hat h_4-\hat h_1-\hat h_3-\sigma m_4} \ .
\end{aligned}
\end{equation}
As explained around eq.~\eqref{eq:wnv_winding_no}, the remaining step for evaluating the supercoset WNV three-point functions is to substitute~\eqref{eq:ful_m_basis_wnv_sl2} in~\eqref{ew:wind_full_supercoset_correlators}. The holomorphic dependence from the vertex operator correlators in~\eqref{ew:wind_full_supercoset_correlators} reduces to
\begin{equation}\label{eq:first_factor}
    \prod_{i<j}z_{ij}^{\frac{1}{\fk+2}(\fk q_R^iq_R^j+2m_im_j)}\prod_{i=1,3,4} z_{i2}^{\sigma m_i} \ .
\end{equation}
Combining this with the holomorphic expression coming from $f_{\sigma m}(z_i)$~\eqref{eq:f_sigmam_etc}, we observe that the $z_2$-dependence disappears, as expected from the fact that the operator insertion at $z_2$ corresponds to the supercoset identity operator. It is straightforward to massage the remaining $z$-dependence into the standard three-point function form dictated by conformal invariance. 

Therefore, we have arrived at the supercoset primary three-point functions, when the winding number is violated by $\sigma=\pm 1$ units, 
\begin{equation}\label{eq:WNV structure constant}
    \begin{aligned}
        &\Big\langle\prod_{i=1}^3V^{n_i,\bar n_i}_{j_i,m_i,\bar m_i}(z_i)\Big\rangle_{\sum_{i=1}^3w_i\!=-\sigma}=\delta\Big(\sum_{i=1}^3w_i+\sigma\Big)\delta_{\sum_{i=1}^3q^i_R,0}\delta_{\sum_{i=1}^3\bar q_R^i,0} \frac{C^{w}(j_i,m_i,\bar m_i)}{\prod_{\mathrm{cyc}}z_{ab}^{h_{abc}}\bar z_{ab}^{\bar h_{abc}}} \ ,
    \end{aligned}
\end{equation}
where the corresponding structure constants take the form,
\begin{equation}\label{eq:WNV structure constants coset}
   C^{w}(j_i,m_i,\bar m_i)=\frac{\pi^2 N(\fk)}{B\big(\frac{\fk+2}{2}-j_1\big)}\prod_{i=1}^3\frac{\Gamma(j_i+\sigma m_i)}{\Gamma(1-j_i-\sigma\bar m_i)} \frac{D\Big(\frac{\fk+2}{2} -j_1,j_2,j_3\Big)}{\gamma\big(j_1+j_2+j_3-\frac{\fk+2}{2} \big)}  \ ,
\end{equation}
with $N(\fk)$ defined in~\eqref{eq:N(k)}. Notice that the expression for these WNV structure constants is somewhat simpler than the WNP ones~\eqref{eq:supercoset_3pt_funcs}, since in this case we were able to explicitly perform the integrals coming from the complex Mellin transforms. This was possible in the WNV sector due to the limits involved in extracting the residues in eq.~\eqref{eq:res_lims}, that significantly simplified the resulting $x$-integrals~\eqref{eq:def_wnv_-1},~\eqref{eq:def_wnv_+1}.
Using the explicit expressions for $D(j_1,j_2,j_3)$ (\ref{eq:H3+ structure constant}) and $B(j_1)$ (\ref{eq:B(j),T(k)_relations}) we obtain 
\begin{equation}
    \frac{D\Big(\frac{\fk+2}{2} -j_1,j_2,j_3\Big)}{B\big(\frac{\fk+2}{2}-j_1\big)} = \frac{1}{2\pi \fk^{\frac{1}{2}}} \big(\fk^{1-\fk}\gamma(\fk)\big)^{\frac{1}{2}+\frac{1}{\fk}(1-\tilde j)} \frac{\Upsilon_{\sqrt{\fk}}(\fk^{\frac{1}{2}})\prod_i \Upsilon_{\sqrt{\fk}}(2\fk^{-\frac{1}{2}}j_i)}{\Upsilon_{\sqrt{\fk}}(\fk^{-\frac{1}{2}}(\tilde j-1) -\frac{1}{2}\fk^{\frac{1}{2}}) \prod_{\mathrm{cyc}} \Upsilon_{\sqrt{\fk}} (\frac{1}{2}\fk^{\frac{1}{2}} +\fk^{-\frac{1}{2}}j_{ijk})}~.
\end{equation}
To summarize, we have obtained the three-point functions for the supercoset in the winding number preserving (WNP) and violating (WNV) case in eq.~(\ref{eq:supercoset_3pt_funcs}) and (\ref{eq:WNV structure constant}), respectively. We emphasize that in all supercoset correlators angular momentum is always preserved. We will now turn to discuss $\n=2$ Liouville theory and express them in Liouville parameters.

\section{$\mathcal{N}=2$ Liouville theory}\label{sec:Liouville}

In this section, we introduce $\n=2$ Liouville theory, discuss its mirror duality with the supercoset model explored in the previous section and present its structure constants in the winding number preserving and violating sectors, which on the Liouville side translate to the angular momentum preserving and violating cases.

\subsection{Basics of $\n=2$ Liouville theory}

We focus our discussion on $\mathcal{N}=2$ Liouville theory placed on a sphere topology, with its action given by~\cite{Anninos:2023exn}
\begin{equation}\label{eq:N=2_Liouville_action}
\begin{aligned}
    S_{\text{L}}^{\mathcal{N}=2} = \frac{1}{4\pi} \int_{S^2} \d^2z \, \sqrt{\tilde{g}} \Big(  \partial_\mu \tilde{\varphi} \partial^\mu \varphi &+ \frac{\fQ\widetilde{R}}{2} \left( \varphi + \tilde{\varphi} \right) -F\tilde F+\mu \e^{\fb\varphi}F+\mu^\ast \e^{\fb\tilde \varphi}\tilde F \\
    &\qquad - \ii \bar{\tilde\psi} \slashed{D}\psi
     - \frac{\ii}{2}\mu \fb \e^{\fb\varphi} \bar{\psi}\psi - \frac{\ii}{2}  \mu^* \fb \e^{\fb\tilde{\varphi}} \bar{\tilde\psi}\tilde{\psi} \Big) ~ ,
\end{aligned}  
\end{equation}
where all indices are raised and lowered with the background metric $\tilde{g}_{\mu\nu},\,\tilde{g}^{\mu\nu}$, which is the standard round metric on $S^2$, its Ricci scalar reads $\widetilde{R} =\tfrac{2}{r^2}$ and the bar over the fermions stands for $\bar \psi\equiv\psi^\mathrm{T}C$, where $C$ is the charge conjugation matrix.\footnote{ We pick the following representation for the gamma matrices,
\begin{equation}\label{eq:conventions SUSY}
    \gamma^1=\begin{pmatrix}
        0 &1\\
        1&0
    \end{pmatrix} \ , \quad  \gamma^2=\begin{pmatrix}
        0 &-i\\
        i&0
    \end{pmatrix} \ ,\quad   \gamma_*\equiv-i\gamma^1\gamma^2=\begin{pmatrix}
        1 &0\\
        0&-1
    \end{pmatrix} \ , \quad C\equiv \gamma^2=\begin{pmatrix}
        0 &-i\\
        i&0
    \end{pmatrix} \ .
\end{equation}
The chiral gamma matrices in $(z,\bar z)$ coordinates, then, read
\begin{equation}\label{eq:gammachiral}
        \gamma^z\equiv\frac{1}{2}\big(\gamma^1+i\gamma^2\big)=\begin{pmatrix}
            0&1\\0&0
        \end{pmatrix} \ , \quad
        \gamma^{\bar z}\equiv\frac{1}{2}\big(\gamma^1-i\gamma^2\big)=\begin{pmatrix}
            0&0\\1&0
        \end{pmatrix} \ .
\end{equation}} The above action can be obtained from the $\n=(2,2)$ superspace via the standard procedure (see \eg ref.~\cite{Hori:2003ic}) and describes the interaction of a chiral multiplet, consisting of a complex scalar $\varphi$, a Dirac spinor $\psi$ and an auxiliary complex scalar $F$ and a corresponding anti-chiral multiplet $(\tilde{\varphi},\tilde{\psi}, \widetilde{F})$, in the exponential superpotentials $W(\varphi) = \mu\fb^{-1}\e^{\fb\varphi},\,\widetilde{W}(\tilde\varphi) = \mu^*\fb^{-1}\e^{\fb\tilde{\varphi}}$. Furthermore, $\mathcal{N}=2$ Liouville theory is a $\mathcal{N}=2$ superconformal field theory with central charge $c= 3+6\mathsf{Q}^2$, where $\fQ=\fb^{-1}$. Notably, unlike the $\mathcal{N}=0,1$ versions of the theory, for $\mathcal{N}=2$ Liouville the central charge is subject to a non-renormalization theorem and does not receive quantum corrections, \ie $\fQ= \fb^{-1}$ \cite{Distler:1989nt}. This can also be deduced from the OPEs of the $\n=2$ superconformal algebra, given the explicit form of the supercurrents and $\rm R$-current that we present in this section. From the dual supercoset perspective, this is related to the presence of the fermions that shift the level from $\fk\to\fk+2$, as discussed around eq.~\eqref{eq:supercoset_central_charge}.

\paragraph{Symmetries} On the sphere the action~\eqref{eq:N=2_Liouville_action} is invariant under the following supersymmetry transformations
\begin{equation}
\label{eq:SuperConformalTransformations_Timelike}
\begin{aligned}
\delta \varphi &= \btepsilon \psi~, & \quad \delta \tvarphi &=  \overline{\epsilon} \tpsi~, \\
\delta \psi &=  \ii \slashed{\partial}\varphi \, \epsilon -\frac{1}{\mathsf{b}r}\epsilon  -\ii F \, \tepsilon  ~, &\quad  \delta \tpsi &= \ii \slashed{\partial}\tvarphi \,  \tepsilon -\frac{1}{\mathsf{b}r}\tepsilon  - \ii \tF \, \epsilon~, \\
\delta F &= - \overline{\epsilon} \slashed{D}\psi ~, &\quad \delta \tF &=- \btepsilon \slashed{D} \tpsi ~,
\end{aligned}
\end{equation}
where $\epsilon$ and $\tilde{\epsilon}$ are positive Killing spinors on the two-sphere, 
\begin{equation}
    \nabla_\mu \epsilon = \frac{\ii}{2r}\gamma_\mu \epsilon~,\quad \nabla_\mu \tilde{\epsilon} = \frac{\ii}{2r} \gamma_\mu \tilde{\epsilon} \ .
\end{equation}
The Noether currents deduced from the above supersymmetry transformations are the supercurrents $G^\pm$, and take the form
\begin{subequations}
\label{eq:G_Liouville_supercurrents}
\begin{align}
        G^+(z)=\ii\sqrt{2}\big(\tilde\psi_2\partial\varphi_L-\fQ\partial\tilde\psi_2\big) \ , \quad \bar G^+(\bar z)=\ii\sqrt{2}\big(\tilde\psi_1\bar\partial\varphi_R-\fQ\bar\partial\tilde\psi_1\big) \ ,\\
        G^-(z)=\ii\sqrt{2}\big(\psi_2\partial\tilde\varphi_L-\fQ\partial\psi_2\big) \ , \quad \bar G^-(\bar z)=\ii\sqrt{2}\big(\psi_1\bar\partial\tilde\varphi_R-\fQ\bar\partial\psi_1\big) \ ,
\end{align}
\end{subequations}
where we choose the background metric $\tilde g_{\mu\nu}$ to be flat for simplicity, as we can go back to the round metric on $S^2$ via a simple rescaling and the rescaling by $\ii\sqrt{2}$ will be useful later. We indicate by a $\mathrm{L},\mathrm{R}$ subscript the left and right-moving components of the fields in the free-field limit, as we shortly explain, and further decompose the Dirac fermions as
\begin{equation}\label{eq:Liouville_fermions}
    \psi=\begin{pmatrix}
        \psi_1(\bar z)\\
        \psi_2(z)
    \end{pmatrix} \ , \quad  \tilde\psi=\begin{pmatrix}
        \tilde\psi_1(\bar z)\\
        \tilde\psi_2(z) 
    \end{pmatrix} \ ,
\end{equation}
In addition, the theory enjoys a global $\mathrm{U}(1)_\mathrm{A}\times \mathrm{U}(1)_{\mathrm{V}}$ symmetry 
\begin{equation}
\label{eq:N2_RSymmetry}
\begin{aligned}
&\mathrm{U}(1)_{\mathrm{V}} &\, &: &\, \varphi &\to \varphi + 2  \ii \mathsf{Q} \lambda~,  &\ \tvarphi &\to \tvarphi - 2  \ii \mathsf{Q}  \lambda~, &\  \psi &\to \e^{ -\ii\lambda } \psi~, &\  \tpsi &\to \e^{\ii\lambda} \tpsi ~,  \\ 
& & & &  F &\to \e^{ -2\ii \lambda } F~, &\  \tF &\to \e^{2\ii \lambda} \tF ~, \\[5pt]
&\mathrm{U}(1)_{\mathrm{A}} &\, &: &\, \psi &\to \e^{\ii \alpha \gamma_*} \psi~,  &\  \tpsi &\to \e^{-\ii \alpha \gamma_*} \tpsi ~,
\end{aligned}
\end{equation}
where $\lambda$ is a constant and $\gamma_*=-\ii \gamma^1\gamma^2$ is the analogue of $\gamma^5$ in two dimensions (\ref{eq:gammachiral}). The corresponding $\mathrm{U}(1)_{\mathrm{V}}$ and $\mathrm{U}(1)_{\mathrm{A}}$ currents obtained via the Noether procedure are, \begin{subequations}
\label{eq:N2_RSymmetry}
\begin{align}
j^\mu_{\mathrm{U}(1)_\mathrm{V}}&=2\ii\fQ\big(\partial^\mu\tvarphi-\partial^\mu\varphi\big)-\bar{\tilde{\psi}}\gamma^\mu\psi~,\\
j^\mu_{\mathrm{U}(1)_\mathrm{A}}&= \bar{\tilde{\psi}}\gamma^\mu \gamma_*\psi ~.
\end{align}
\end{subequations}
The currents (\ref{eq:N2_RSymmetry}) are non-chiral, but are as usual related to the left (L) and right (R) chiral $\rm R$-currents through a simple linear transformation. More specifically,
the holomorphic and anti-holomorphic $R$-currents, respectively, read
\begin{subequations}\label{eq:L,R_R-currents}
    \begin{align}
        J_R(z)=\frac{1}{2}\big(J^z_{\mathrm{U}(1)_V}+J^z_{\mathrm{U}(1)_A}\big) =\fQ\big(\partial\varphi_L-\partial\tvarphi_L\big)+\tilde\psi_2\psi_2\ ,\\
        \bar J_R(\bar z)=\frac{1}{2}\big(J^{\bar z}_{\mathrm{U}(1)_V}-J^{\bar z}_{\mathrm{U}(1)_A}\big) = \fQ\big(\bar\partial\varphi_R-\bar\partial\tilde\varphi_R\big)-\tilde\psi_1\psi_1 \ .
    \end{align}
\end{subequations}
where, 
\begin{subequations}
    \begin{align}
        &J^z_{\mathrm{U}(1)_{\mathrm{V}}} \equiv \ii j^z_{\mathrm{U}(1)_{\mathrm{V}}} = 2\mathsf{Q}(\partial\varphi - \partial \tvarphi)+ \tpsi_2 \psi_2~,\qquad J^z_{\mathrm{U}(1)_{\mathrm{A}}} \equiv \ii j^z_{\mathrm{U}(1)_{\mathrm{A}}} = \tpsi_2 \psi_2 \ ,\\
        &J^{\bar{z}}_{\mathrm{U}(1)_{\mathrm{V}}}  \equiv \ii j^{\bar{z}}_{\mathrm{U}(1)_{\mathrm{V}}}  = 2\mathsf{Q}(\bar\partial\varphi - \bar\partial \tvarphi)- \tpsi_1 \psi_1~,\qquad J^{\bar{z}}_{\mathrm{U}(1)_{\mathrm{A}}} \equiv \ii j^{\bar{z}}_{\mathrm{U}(1)_{\mathrm{A}}}= \tpsi_1 \psi_1 \ ,
    \end{align}
\end{subequations}
As a simple check, using free-field OPEs and eqs.~\eqref{eq:G_Liouville_supercurrents},~\eqref{eq:L,R_R-currents} it is easy to verify that $J_R(z)G^\pm(w)\sim\pm\frac{G^\pm(w)}{z-w}$, and similarly for the anti-holomorphic components.~From the action~\eqref{eq:N=2_Liouville_action} the stress-tensor can also be read-off, where for a flat background metric $\tilde g_{\mu\nu}=\delta_{\mu\nu}$, we obtain
\begin{subequations}
\label{eq:T_Liouville}
\begin{align}
    T(z)=-\partial\tilde\varphi_L\partial\varphi_L-\dfrac{1}{2}\big(\tpsi_2\partial \psi_2+\psi_2\partial\tpsi_2\big)+\dfrac{\fQ}{2}\big(\partial^2\varphi_L+\partial^2\tilde\varphi_L\big) \ ,\\
    \bar T(\bar z)=-\bar\partial\tilde\varphi_R\bar\partial\varphi_R+\dfrac{1}{2}\big(\tpsi_1\bar\partial \psi_1+\psi_1\bar\partial\tpsi_1\big)+\dfrac{\fQ}{2}\big(\bar\partial^2\varphi_R+\bar\partial^2\tilde\varphi_R\big) \ .
\end{align}    
\end{subequations}
Bosonizing the fermions as $\tpsi_2\psi_2=\ii\partial K_L,\,\tpsi_1 \psi_1=-\ii\partial K_R$ (where \eg $K_L(z)K_L(w)\sim-\log{(z-w)}$) eqs.~\eqref{eq:G_Liouville_supercurrents},~\eqref{eq:L,R_R-currents},~\eqref{eq:T_Liouville} become,
\begin{equation}\label{eq:Liouville stress tensor and currents}
    \begin{aligned}
        &T(z)=-\partial\tilde\varphi_L\partial\varphi_L-\dfrac{1}{2}\big(\partial K_L\big)^2+\dfrac{\fQ}{2}\big(\partial^2\varphi_L+\partial^2\tilde\varphi_L\big) \ ,\\
        &G^+(z)=\ii\sqrt{2}\big(\tilde\psi_2\partial\varphi_L-\fQ\partial\tilde\psi_2\big) \ ,\\
        &G^-(z)=\ii\sqrt{2}\big(\psi_2\partial\tilde\varphi_L-\fQ\partial\psi_2 \big)\ , \\
        &J_R(z) = \fQ(\partial \varphi_L-\partial \tilde\varphi_L)+i\partial K_L \ ,
    \end{aligned}
\end{equation}
and similarly for their anti-holomorphic counterparts. When treated as operators in the quantum theory, all the expressions above are understood as normal ordered. Using the expressions in eq.~\eqref{eq:Liouville stress tensor and currents} and the standard free-field OPEs, one can verify the $\n=2$ superconformal OPEs and algebra~\eqref{eq:N=2-SCA} with the specific $\n=2$ Liouville central charge mentioned above (we rescaled the supercurrents in~\eqref{eq:G_Liouville_supercurrents} by the factor of $\ii\sqrt{2}$ to get the standard convention for the normalization of the $G^+G^-$ OPE here).

\paragraph{Vertex operators}
Let us turn to the discussion of the spectrum of the theory. To do that, we will resort to the regime where $\varphi,\, \tvarphi \rightarrow -\infty$, where the effect of the exponential superpotentials in~\eqref{eq:N=2_Liouville_action} is small and a free-field realization can be implemented. In this regime, all fields can be decomposed into holomorphic (L) and anti-holomorphic (R) components, \eg $\varphi(z,\bar z)=\varphi_L(z)+\varphi_R(\bar z)$, as was already hinted in the expressions for the stress-tensor and (super)currents earlier. Then, the complete set of left-moving primary operators for $\mathcal{N}=2$ Liouville theory is
\begin{equation}\label{eq:primariy_holo_Liouville}
    V_{\alpha,\tilde{\alpha}}^\eta(z)= \e^{\alpha\varphi_L + \tilde{\alpha}\tvarphi_L + \ii \eta K_L}~,
\end{equation}
where $\eta$ is related to the contribution of the fermions; in particular, $\eta\in \mathbb{Z}$ in the NS-sector, $\eta \in \mathbb{Z}+\tfrac{1}{2}$ in the Ramond sector. From the OPE of the operators~\eqref{eq:primariy_holo_Liouville} with the stress-tensor and the R-current~\eqref{eq:Liouville stress tensor and currents}, as well as the $\mathrm{U}(1)_{\mathrm{V}}$ and $\mathrm{U}(1)_{\mathrm{A}}$ currents~(\ref{eq:N2_RSymmetry}),
we infer the holomorphic dimension $h_{\alpha,\tilde{\alpha}}^\eta$, R-charge $q^\eta_{\alpha ,\tilde{\alpha}}$, vector $v^\eta_{\alpha,\tilde{\alpha}}$ and axial $a^\eta_{\alpha,\tilde{\alpha}}$ charge of the $\mathcal{N}=2$ Liouville primaries,
\begin{subequations}\label{eq:Liouville_dimas_&_R_charges}
\begin{align}\label{subeq:h}
    h_{\alpha,\tilde{\alpha}}^\eta &= -\alpha \tilde{\alpha} + \frac{\mathsf{Q}}{2}(\alpha + \tilde{\alpha}) +\frac{\eta^2}{2}~,\\ \label{subeq:q}
    q^\eta_{\alpha,\tilde{\alpha}} &= \mathsf{Q}(\alpha - \tilde{\alpha}) +\eta~,\\
    v^\eta_{\alpha,\tilde{\alpha}}&= 2\mathsf{Q}(\alpha - \tilde{\alpha})+\eta \ ,\\
    a^\eta_{\alpha,\tilde{\alpha}}&=\eta~.
    \end{align}
\end{subequations}
Similar formulae hold for the right-moving sector, labeled by $\bar\alpha,\bar{\tilde{\alpha}},\bar\eta$. Note that $\alpha+\tilde\alpha=\bar\alpha+\bar{\tilde{\alpha}}$, since the radial direction in the target of the theory $\varphi+\tilde\varphi$ is non-compact (this point will become clearer when discussing the supercoset mirror below).~On the other hand, the differences $\alpha-\tilde\alpha,\,\bar\alpha-\bar{\tilde{\alpha}}$ are in general independent.

\paragraph{$\n=2$ Liouville theory as $2$d supergravity} As a side comment, note that $\n=2$ Liouville theory -- similarly to its $\n=0,1$ counterparts -- can also be thought of as two-dimensional $\n=2$ supergravity coupled to a $\n=2$ SCFT, in the super-Weyl gauge. This point was explained in~\cite{Antoniadis:1990mx,Brink:1976vg,Anninos:2023exn}. In particular, the physical supermultiplet consists of the zweibein $\e^a_\mu$, a $\mathrm{U}(1)$ gauge field $A_\mu$, a Dirac gravitino $\chi_\mu$ and a complex scalar $B$, being related to the Liouville fields appearing in the action~\eqref{eq:N=2_Liouville_action}, as~\cite{Anninos:2023exn}
\begin{equation}\label{eq:SUGRA}
\begin{aligned}
    \e^a_\mu=\e^{\frac{1}{2}\fb(\varphi+\tilde\varphi)}\tilde \e^a_\mu \ , \ A_\mu=&-\frac{\ii}{2}\epsilon_{\mu\nu}\partial^\nu(\varphi-\tilde\varphi) \ , \ \chi_\mu=\e^{\frac{1}{2}\fb\varphi}\gamma_\mu\psi \ , \ \chi_\mu^\ast=\e^{\frac{1}{2}\fb\tilde\varphi}\gamma_\mu\tilde\psi \ , \\
    &\ B=\e^{-\fb\varphi}F \ , \ B^\ast=\e^{-\fb\tilde\varphi}\widetilde F \ ,
\end{aligned}    
\end{equation} 
where the overhead tilde in $\tilde \e^a_\mu$ denotes the zweibein of the round two-sphere. The central charge of the SCFT $c_m$ is related to the Liouville parameters as $\fQ=\fb^{-1}=\sqrt{\frac{3-c_m}{6}}$. See~\cite{Anninos:2023exn} for further discussion.

\subsection{$\n=2$ Liouville $-$ $\mathrm{SL}(2)_\fk/\mathrm{U}(1)$ dictionary}

The main tool that goes into our derivation of the $\n=2$ Liouville structure constants is the duality of the theory with the $\mathrm{SL}(2)_\fk/\mathrm{U}(1)$ supercoset model, some aspects of which we now discuss in more detail.

The aforementioned duality was shown in~\cite{Hori:2001ax} to be an instance of mirror symmetry~\cite{Hori:2000kt}. Roughly speaking, the idea of ref.~\cite{Hori:2001ax} was to construct a gauged linear sigma model (GLSM) that flows to the $\mathrm{SL}(2)_\fk/\mathrm{U}(1)$ supercoset in the IR. Implementing mirror symmetry, one dualizes the GLSM to a theory that can be argued to flow to $\n=2$ Liouville in the IR. Therefore, mirror symmetry provides an infrared duality according to which,
\begin{equation*}
    \n =2~\mathrm{Liouville~theory} \quad \overset{\text{Mirror}}{\longleftrightarrow} \quad \n=2~\frac{\mathrm{SL}(2,\mathbb{R})_k}{\mathrm{U}(1)}~\mathrm{supercoset} \ .
\end{equation*}
This is the supersymmetric analogue of FZZ duality, where the bosonic $\mathrm{SL}(2)_k/\mathrm{U}(1)$ coset (whose target is the $2$d Euclidean cigar) is dual to (bosonic) sine-Liouville theory.\footnote{In fact, FZZ duality follows from this supersymmetric duality after gauging the $\mathrm{U}(1)$ R-symmetry. This can be seen by \eg inspecting the explicit form of the primary operators~\eqref{eq:full_supercoset_primaries},~\eqref{eq:primariy_holo_Liouville} or by appropriately massaging the actions of the supersymmetric theories.} Moreover, in ref.~\cite{Creutzig:2010bt} the equivalence of correlation functions between the two sides of the duality was demonstrated, although the authors there did not explicitly compute any correlators.

From the target space point of view, the equivalence of the two theories follows from T-duality. Recall from our discussion in subsection~\ref{subsec:cigar_target} that the supercoset has a cigar target. T-dualizing the cigar, one obtains a trumpet geometry with a singularity at the position of the cigar's tip. One can deduce this trumpet target from the Liouville side, via redefining the complex scalars as 
\begin{equation}\label{eq:rho,Y_def}
    \varphi \equiv \frac{1}{\sqrt{2}}\big(\rho+iY\big)~,\quad \quad \tvarphi \equiv \frac{1}{\sqrt{2}}\big(\rho-iY\big)~,
\end{equation}
and inspecting the exponential superpotentials in the action~\eqref{eq:N=2_Liouville_action}, which become
\begin{equation}\label{eq:T-dual-story_superpotentials}
    \e^{\fb \varphi}=\e^{\frac{\fb}{\sqrt 2}\rho}\e^{\ii\frac{\fb}{\sqrt 2}Y} \ , \quad \e^{\fb \tilde\varphi}=\e^{\frac{\fb}{\sqrt 2}\rho}\e^{-\ii\frac{\fb}{\sqrt 2}Y} \ .
\end{equation}
From these expressions, one observes that $Y$ corresponds to an angular coordinate in the target parameterizing an $S^1$, whereas $\rho$ is a radial coordinate. Because of the presence of the Liouville wall $\e^{\fb(\varphi+\tilde\varphi)}=\e^{\sqrt 2\fb \rho}$, the target manifold has a singularity when $\rho\to\infty$. From the line element~\eqref{eq:cigar_metric} we learn that the asymptotic radius of the cigar is $\sqrt{2\fk}$, while from the superpotentials~\eqref{eq:T-dual-story_superpotentials} one reads-off the asymptotic radius of the trumpet to be $\sqrt{2/ \fk}$, as expected.

Let us now present the dictionary between the two sides of the duality. Choosing to parameterize
\begin{subequations}
    \begin{align}
        &\alpha=\fQ(j+m+n) \ , \qquad \tilde{\alpha}=\fQ(j-m-n) \ ,\qquad \eta =n~,\\
        &\bar\alpha=\fQ(j+\bar m+\bar n) \ , \qquad \bar{\tilde{\alpha}}=\fQ(j-\bar m-\bar n) \ ,\qquad \bar \eta =\bar n~,
    \end{align}
\end{subequations}
the dimensions~(\ref{subeq:h}) and (\ref{subeq:q}), along with their anti-holomorphic counterparts, become
\begin{subequations}
    \begin{align}
        h_{\alpha,\tilde{\alpha}}^\eta&= -\mathsf{Q}^2j(j-1)+\mathsf{Q}^2(m+n)^2+\dfrac{n^2}{2}~, \quad q^\eta_{\alpha,\tilde{\alpha}}= 2\mathsf{Q}^2m+n\big(1+2\mathsf{Q}^2\big) \ ,\\
        h_{\bar\alpha,\bar{\tilde{\alpha}}}^{\bar\eta}&= -\mathsf{Q}^2j(j-1)+\mathsf{Q}^2(\bar m+\bar n)^2+\dfrac{\bar n^2}{2} \ , \quad q^{\bar\eta}_{\bar\alpha,\bar{\tilde{\alpha}}}= 2\mathsf{Q}^2\bar m+\bar n\big(1+2\mathsf{Q}^2\big)~.
    \end{align}
\end{subequations}
Under the identification $\mathsf{Q}^{-2}=\mathsf{b}^2\equiv\mathsf{k}$, which follows by comparing the central charges of the two theories, these coincide with the supercoset dimensions and $\rm R$-charges of eq.~\eqref{eq:full_supercoset_dimas_&_R_charges}. Notice that in this dictionary the well-known mapping $\bar q_R
\to-\bar q_R$ induced by mirror symmetry\footnote{See appendix~\ref{ap:N=2_SCFTs_Basics} for a brief review of the mirror automorphism of the $\n=2$ algebra.} is absent. This is because both anti-holomorphic R-charges have been deduced in the two theories separately. If we were, instead, to evaluate the OPE of \eg a supercoset primary with the Liouville anti-holomorphic R-current~\eqref{eq:L,R_R-currents} (expressed in T-dual variables) we would get the required overall minus sign. Lastly, it is straightforward to check that the above mapping is $1$-$1$.\footnote{Up to the reflection symmetry $j\mapsto 1-j$ obeyed by the primaries.} 

Let us make a couple of comments; first, from this mapping to the supercoset, it also follows that it must be $\alpha+\talpha=\bar\alpha+\bar\talpha$, as discussed earlier. Second, the $\n=2$ algebra has an outer automorphism parameterized by $\beta\in\mathbb{R}$, commonly referred to as \textit{spectral flow}~\cite{Schwimmer:1986mf}. After spectral flow (\ref{eq:spectral flow1}) by $\beta$, an operator with holomorphic dimension $h$ and $\rm R$-charge $q_R$ is mapped to an operator with (\ref{eq:spectral flow})
\begin{equation}
    h^\beta\equiv h+\beta q_R+\beta^2\frac{c}{6} \ , \quad q^\beta_R \equiv q_R+\beta\frac{c}{3} \ ,
\end{equation}
and similarly for the anti-holomorphic superconformal algebra. For instance, under half-integer spectral flow NS-sector primaries are mapped to R-sector primaries, and vice versa. A beautiful feature of the theories we are discussing is that the supercoset primaries $V^{n,\bar n}_{j,m,\bar m}$ of eq.~\eqref{eq:full_supercoset_primaries} can be obtained from the primaries $V^{0,0}_{j,m,\bar m}$, via spectral flow (\ref{eq:spectral flow}) by $\beta=n,\,\bar\beta=\bar n$. In particular, the operators with $\beta=\bar\beta=\pm\frac{1}{2}$ belong to the R-sector, with the vertex operators $\e^{\pm\frac{\ii}{2}\sqrt{\frac{\fk+2}{\fk}}\big(X_R-\bar X_R\big)}$ being the spin fields $\sigma^\pm$.\footnote{As usual, the $\pm$ index corresponds to the double degeneracy of R-sector states.} As is typical in $\n=0,1$ Liouville theories, we can also define the Liouville momentum $P \in \mathbb{R}$ through the relation $\alpha + \tilde{\alpha} ={\mathsf{Q}} + 2\ii P$, which implies for the dimensions~\eqref{subeq:h}, 
\begin{equation}
    h^{\eta}_{\alpha,\tilde{\alpha}} = \frac{\mathsf{Q}^2}{4} +P^2 +\frac{(q^{\eta}_{\alpha,\tilde{\alpha}}-\eta)^2}{4\mathsf{Q}^2}+\frac{\eta^2}{2} \geq \frac{c-3}{24}~.
\end{equation}
 From this, we observe that when $c>3$ the dimensions are positive, while they are negative when $c<3$. In the former case -- which is the main focus of this paper -- the theory is unitary and typically referred to as \textit{spacelike} $\n=2$ Liouville theory, while in the latter case the theory is non-unitary and usually called \textit{timelike} $\n=2$ Liouville theory. Due to the fact that $\fQ=\fb^{-1}$ does not get renormalized, unlike the $\n=0,1$ instances of the theory, in the $\n=2$ case there is no intermediate central charge regime between the spacelike and the timelike theories.

We now summarize the dictionary between the parameters of the supercoset model and of $\n=2$ Liouville, with the identifications
\begin{equation}\label{eq:Liouville-supercoset_dictionary}
\begin{cases}
    j=\dfrac{\alpha+\tilde\alpha}{2\fQ}=\dfrac{\bar\alpha+\bar{\tilde{\alpha}}}{2\fQ} \ , \\[2mm]
    m=\dfrac{\alpha-\tilde\alpha}{2\fQ}-\eta \ , \  \bar m=\dfrac{\bar\alpha-\bar{\tilde{\alpha}}}{2\fQ}-\bar\eta \ , \\
    n=\eta, \  \bar n=\bar \eta \ ,
\end{cases}
\leftrightarrow \quad 
\begin{cases}
    \alpha = \fQ(j+m+n) \ , \  \bar\alpha = \fQ(j+\bar m+\bar n) \ ,\\[2mm]
    \tilde\alpha = \fQ(j-m-n) \ , \  \bar{\tilde\alpha} = \fQ(j-\bar m-\bar n)  \ ,\\
    \eta=n,\ \bar\eta=\bar n \ ,
\end{cases}
\end{equation}
in Table \ref{tab: dictionary}.~To match the full spectrum in both sides of the duality one needs the above identifications, along with the primaries~\eqref{eq:full_supercoset_primaries},~\eqref{eq:primariy_holo_Liouville} and the dimensions and R-charges~\eqref{eq:full_supercoset_dimas_&_R_charges}, \eqref{eq:Liouville_dimas_&_R_charges}. 
\begin{table}[ht]
\centering
\small
\setlength{\arrayrulewidth}{.7pt}
\renewcommand{\arraystretch}{1.75}

\begin{tabular}{|c!{\vrule width 1.3pt}c|c!{\vrule width 1.3pt}c}
\hline
\rule{0pt}{3.5ex} &$\frac{\mathrm{SL}(2,\mathbb{R})_\fk}{\mathrm{U}(1)}$ supercoset
& $\mathcal{N}=2$ Liouville \\

\hhline{|=|=|=|}

\rule{0pt}{3.5ex}
Central charge &
$c=3+\frac{6}{\fk}$
& $c=3+\frac{6}{\fb^2}$  \\

\hhline{|=|=|=|}

\rule{0pt}{3.5ex}
Vertex~operators &
$V_{j,m}^n(z) = V_{j,m}\e^{\ii \sqrt{\frac{\mathsf{k}}{\mathsf{k}+2}} q_R X_R}$
& $V_{\alpha,\tilde{\alpha}}^\eta(z) = \e^{\alpha \varphi_L + \tilde{\alpha}\tvarphi_L + \ii \eta K_L}$  \\

\hhline{|=|=|=|}

\rule{0pt}{3.5ex}
Conformal dimension &
$h^n_{j,m}= -\frac{j(j-1)}{\mathsf{k}} +\frac{(m+n)^2}{\mathsf{k}} + \frac{n^2}{2}$
& $h_{\alpha,\tilde{\alpha}}^\eta = -\alpha \tilde{\alpha} + \frac{\mathsf{Q}}{2}(\alpha + \tilde{\alpha}) +\frac{\eta^2}{2}$  \\

\hhline{|=|=|=|}

\rule{0pt}{3.5ex}
R-charge &
$q_R = \frac{2m}{\mathsf{k}} + n\frac{\fk+2}{\mathsf{k}}$
& $q^\eta_{\alpha \tilde{\alpha}} = \mathsf{Q}(\alpha - \tilde{\alpha}) +\eta$  \\

\hline
\end{tabular}

\caption{Comparison of $\tfrac{\mathrm{SL}(2,\mathbb{R})_\fk}{\mathrm{U}(1)}$ supercoset and $\mathcal{N}=2$ Liouville, in the holomorphic sector. Similar formulae hold true in the anti-holomorphic sector.}
\label{tab: dictionary}

\end{table}

\subsection{Liouville structure constants}

Building on the duality with the supercoset and the explicit dictionary presented in the previous subsection, we can now proceed to obtain the winding number preserving (WNP) and violating (WNV) structure constants of primary operators in $\n=2$ Liouville theory.

As an important note, mirror symmetry (or T-duality) trades winding number for angular momentum, and vice versa, as we will quickly review. 
The $\mathcal{N}=2$ Liouville vertex operators are
\begin{equation}\label{eq: full Liouville vertex operator}
    V = \e^{\frac{\alpha+\tilde\alpha}{\sqrt{2}} (\rho_L +\rho_R) + \ii \mathsf{P}_L Y_L + \ii \mathsf{P}_RY_R +i\eta K_L + i\bar\eta K_R}~,\quad \mathsf{P}_L\equiv   \frac{\alpha -\tilde\alpha}{\sqrt{2}}~,\quad \mathsf{P}_R\equiv \frac{\bar \alpha -\bar{\tilde{\alpha}}}{\sqrt{2}}
\end{equation}
where we used (\ref{eq:primariy_holo_Liouville}) and (\ref{eq:rho,Y_def}). 
$Y_L$, $Y_R$ are compact, and we infer the radius $\mathsf{R} = \tfrac{\sqrt{2}}{\fb}$ from (\ref{eq:T-dual-story_superpotentials}). Since we work with $\alpha'=2$, the dual radii satisfy $\mathsf{R}R= 2$,
where for the supercoset we have $R = \sqrt{2\fk}$ (\ref{eq:cigar_metric}). From $V$ we now easily infer the winding number $\mathsf{w}$ and angular momentum $\lambda$ on the Liouville side, to be
\begin{subequations}\label{eq:Liouville_winding_and_momentum}
    \begin{align}
        \mathsf{w} &= \frac{1}{\mathsf{R}}(\mathsf{P}_L- \mathsf{P}_R) = \frac{\fb}{2}(\alpha -\tilde\alpha -(\bar \alpha - \bar{\tilde{\alpha}}))~,\\
        \lambda &= \frac{\mathsf{R}}{2}(\mathsf{P}_L+ \mathsf{P}_R)= \frac{1}{2\fb}(\alpha -\tilde\alpha +\bar \alpha - \bar{\tilde{\alpha}})~.
    \end{align}
\end{subequations}
Using the dictionary (\ref{eq:full_supercoset_dimas_&_R_charges_w}) we hence confirm that mirror symmetry flips winding number and angular momentum. Therefore, winding number preserving/violating correlators in the supercoset get mapped to angular momentum preserving/violating correlators on the Liouville side, and we will present them explicitly now. We highlight that for both the AMP and the AMV case the winding number is always conserved, \ie $\sum_i \mathsf{w}_i=0$.

\subsubsection{Angular momentum preserving correlators}

More specifically, starting with the AMP case, all we have to do is to express eq.~\eqref{eq:supercoset_3pt_funcs} in the Liouville parameters using Table~\ref{tab: dictionary}. Doing so, we arrive at the following expression for the angular momentum preserving three-point functions of $\n=2$ Liouville theory, 
\begin{equation}\label{eq:N=2-Liouville_3pt_funcs}
    \begin{aligned}
        \Big \langle \prod_{i=1}^3 V^{\eta_i,\bar \eta_i}_{\alpha_i,\tilde\alpha_i,\bar \alpha_i,\bar{\tilde{\alpha}}_i}(z_i,\bar z_i) \Big\rangle &= \delta\Big(\sum_i \lambda_i\Big)\delta_{\sum_{i=1}^3  q^{\eta_i}_{\alpha_i,\tilde\alpha_i},0}\delta_{\sum_{i=1}^3  \bar q^{\bar\eta_i}_{\bar\alpha_i,\bar{\tilde{\alpha}}_i},0}\\
        &\qquad\times\, C(\alpha_i,\tilde{\alpha}_i;\eta_i)
\prod_{\mathrm{cyc}}z_{ab}^{-h_{abc}}\bar z_{ab}^{-\bar h_{abc}}  \ ,
    \end{aligned}
\end{equation}
where $C(\alpha_i,\tilde{\alpha}_i;\eta_i) =  D(\alpha_1,\alpha_2,\alpha_3;\tilde\alpha_1,\tilde\alpha_2,\tilde\alpha_3)\mathcal W(\alpha_i,\tilde{\alpha}_i,\eta_i) $. We will use the shorthand notation $D(\alpha_1,\alpha_2,\alpha_3;\tilde\alpha_1,\tilde\alpha_2,\tilde\alpha_3) \equiv D(\alpha_i,\tilde\alpha_i)$ leading to  \cite{Teschner:1997ft, Ribault:2014hia},
\begin{equation}\label{eq:N=2-Liouville_structure_constant}
    \begin{aligned} 
       &\ \ D(\alpha_i,\tilde \alpha_i)={A}(\fb)\bigg(\frac{|\mu|^2}{4} \fb^{2-2\fb^2}\gamma(\fb^2)\bigg)^{\frac{\fQ-\frac{1}{2}\sum_{i=1}^3(\alpha_i+\talpha_i)}{\fb}}\frac{\Upsilon_\fb(\fb)\prod_{i=1}^3\Upsilon_\fb(\alpha_i+\talpha_i)}{\Upsilon_\fb\big(\frac{1}{2}\sum_{i=1}^3(\alpha_i+\talpha_i)-\fQ\big)\prod_{\text{cyc}}\Upsilon_\fb\big(\frac{1}{2}(\alpha_{ijk}+\talpha_{ijk})\big)} \ ,\\
&\mathcal W(\alpha_i,\tilde{\alpha}_i,\eta_i)
=
\int_{\mathbb{C}^2} \d^2y \d^2z\,
y^{a_2}\bar y^{\bar a_2}
z^{a_3}\bar z^{\bar a_3}
|1-y|^{-\fb(\alpha_{123}+\talpha_{123})}|1-z|^{-\fb(\alpha_{312}+\talpha_{312})}|y-z|^{-\fb(\alpha_{231}+\talpha_{231})} \ .
    \end{aligned}
\end{equation}
As in the supercoset model we denote by $\alpha_{abc}=\alpha_a+\alpha_b-\alpha_c,\,\tilde\alpha_{abc}=\tilde\alpha_a+\tilde\alpha_b-\tilde\alpha_c,\,a_i=\fb \alpha_i-\eta_i-1,\,\bar a_i=\fb \bar{\alpha}_i-\bar{\eta}_i-1$, the $\fQ$ fraction in the pre-factor comes from translating the supercoset delta functions to Liouville parameters and we have also leveraged the self-dual symmetry of the Ypsilon function $\Upsilon_{\fb^{-1}}(w)=\Upsilon_\fb(w)$ (see appendix~\ref{ap:special_functions}). Note that in translating to Liouville parameters above we had to use the identification of the supercoset level, \ie $\sqrt \fk=\fb$. Notice, further, that in $D(\alpha_i,\talpha_i)$ in eq.~\eqref{eq:N=2-Liouville_structure_constant} the Liouville parameters appear always in combinations of the form $\alpha+\talpha\equiv\bar\alpha+\bar{\tilde{\alpha}}$, hence the structure constant exhibits symmetry between the holomorphic and anti-holomorphic sectors, while on the support of the delta functions the $\mathcal W$ integral is also symmetric. Moreover, because $\fQ=\fb^{-1}$ we observe that the Ypsilon function fraction in the structure constant~\eqref{eq:N=2-Liouville_structure_constant} does not enjoy the self-dual $\fb\leftrightarrow\fb^{-1}$ symmetry, as anticipated.\footnote{The overall change in the $\fb$-dependent pre-factor under $\fb\to\fb^{-1}$ could be absorbed by transforming the cosmological constants $\mu,\mu^*$ to their duals, as in the $\n=0,1$ theories.} Finally we included the dependence on the $\mathcal{N}=2$ Liouville cosmological constants $\mu$ and $\mu^*$, whose precise form will be justified below.~
\begin{itemize}
\item \textit{Liouville momentum} The $\alpha_i$, $\tilde{\alpha}_i$ are related to the Liouville momentum by
\begin{equation}\label{eq:Liouville momentum}
    \alpha_i+ \tilde\alpha_i = \fQ +2i P~,\quad P \in \mathbb{R}~.
\end{equation}
Conformal dimension and R-charge are summarized in table \ref{tab: dictionary}.
\item \textit{Angular momentum} The correlation function (\ref{eq:N=2-Liouville_3pt_funcs}) is valid in the angular momentum preserving case
\begin{equation}
    \sum_i \lambda_i =0 ~.
\end{equation}
Below we explore instances where either $\alpha_i= \tilde{\alpha}_i$, $\forall i$, or where the sum over the $\alpha_i$ and $\tilde{\alpha}_i$ individually vanishes.  
\item \textit{Poles and zeros} From (\ref{eq:Y-zeros}) we infer that $D(\alpha_i,\tilde\alpha_i)$ has zeros for
\begin{equation}\label{eq: zeros AMP}
    \alpha_i +\tilde\alpha_i \in -r\fb-s\fb^{-1}~,\quad \alpha_i+\tilde\alpha_i \in (r+1)\fb + (s+1)\fb^{-1}~.
\end{equation}
Similarly we find poles for 
\begin{equation}\label{eq:poles 1 AMP}
    \frac 12 \sum_{i=1}^3(\alpha_i+\talpha_i) -\fQ\in -r\fb-s\fb^{-1} ~,\quad \frac 12 \sum_{i=1}^3(\alpha_i+\talpha_i) -\fQ  \in (r+1)\fb + (s+1)\fb^{-1} ~,
\end{equation}
and 
\begin{align}
    \alpha_{ijk} + \tilde{\alpha}_{ijk} &= -2r\fb-2s\fb^{-1}~,\quad \alpha_{ijk} + \tilde{\alpha}_{ijk} =2(r+1)\fb + 2(s+1)\fb^{-1}~,
\end{align}
where $r,s\in \mathbb{Z}_{\geq 0}$. 
Without a closed form for the $\mathcal W$ integral it is difficult to anticipate whether it contributes any zeros. In the AMV case instead discussed below, we can explicitly evaluate the analogous $\mathcal{W}$ integral and discuss the poles/zeros generally. 
\end{itemize}
\paragraph{2-point function}
The corresponding two-point function can be derived from the above structure constant by taking appropriate limits, as in the instance of the $H_3^+$ model discussed around eq.~\eqref{eq:H3+_2pt-func}. 
The limits correspond to taking one of the operators to be the identity. Although it is not a normalizable operator of the theory, the identity operator can be obtained via analytic continuation in $\alpha+\talpha$. From the three-point function
(\ref{eq:N=2-Liouville_structure_constant}) we then obtain the two point function 
\begin{subequations}\label{eq:Liouville_2-pt-function}
\begin{align}
\lim_{(\alpha_3+\tilde\alpha_3)\rightarrow 0} D(\alpha_i,\tilde\alpha_i) &= \frac{4\pi}{\fb}\, {A}(\fb) \gamma\left(\frac{\alpha_1+\tilde\alpha_1}{\fb}-\frac{1}{\fb^2}\right) \left(\frac{|\mu|^2}{4\fb^2}f(\fb)\right)^{\frac{1}{\fb^2}-\frac{\alpha_1+\tilde\alpha_1}{\fb}}\!\!\!\delta(\alpha_{12}+\tilde\alpha_{12}) ~,\\
     \lim_{(\alpha_3+\tilde\alpha_3)\rightarrow 2\fQ}D(\alpha_i,\tilde\alpha_i) &=\frac{2\pi}{\fb}  {A}(\fb)\gamma(\fQ^2) \left(\frac{|\mu|^2\fb^2}{4}f(\fb)\right)^{-\frac{1}{\fb^2}}\delta(\alpha_1+\tilde\alpha_1+\alpha_2+\tilde\alpha_2-2\fQ)~,
\end{align}    
\end{subequations}
where $\alpha_{12}\equiv \alpha_1- \alpha_2,\,\talpha_{12}=\talpha_1-\talpha_2$ and $f(\fb)=\fb^{2-2\fb^2}\gamma(\fb^2)$.

\subsubsection{Angular momentum violating correlators}

Using the dictionary (\ref{eq:Liouville-supercoset_dictionary}) we can write the structure constants (\ref{eq:WNV structure constants coset}) in terms of Liouville parameters. Explicitly, we find 
\begin{equation}\label{eq:WNV structure constant L}
    \begin{aligned}
        &\Big\langle\prod_{i=1}^3V^{\eta_i,\bar \eta_i}_{\alpha_i,\tilde\alpha_i, \bar \alpha_i, \bar{\tilde{\alpha}}_i}(z_i)\Big\rangle_{\sum_{i=1}^3\lambda_i=-\sigma}=\delta\Big(\sum_i \lambda_i+\sigma\Big)\delta_{\sum_{i=1}^3  q^{\eta_i}_{\alpha_i,\tilde\alpha_i},0}\delta_{\sum_{i=1}^3  \bar q^{\bar\eta_i}_{\bar\alpha_i,\bar{\tilde{\alpha}}_i},0}\\
&\qquad\qquad\qquad\qquad\qquad\qquad\quad\times C^{\lambda}(\alpha_i,\tilde\alpha_i,\eta_i)\prod_{\mathrm{cyc}}z_{ab}^{-h_{abc}}\bar z_{ab}^{-\bar h_{abc}} \ ,
    \end{aligned}
\end{equation}
where the corresponding structure constants take the form 
\begin{align}\label{eq:Liouville_AMV_structure_constant}
   C^{\lambda}(\alpha_i,\tilde\alpha_i;\eta_i)&=\left(\frac{\mu}{2}\right)^{\frac{\fQ}{\fb}-\frac{1}{2\fb}\sum_{i=1}^3(\alpha_i+\bar\alpha_i)} \left(\frac{\mu^*}{2}\right)^{\frac{\fQ}{\fb}-\frac{1}{2\fb}\sum_{i=1}^3(\tilde{\alpha}_i+\bar{\talpha}_i)}\frac{\mathsf{a}(\fb)\big(\fb^{2-2\fb^2}\gamma(\fb^2)\big)^{\frac{\fQ-\frac{1}{2}\sum_i(\alpha_i+\tilde\alpha_i)}{\fb} +\frac{1}{2}}}{\gamma(\frac{\fb}{2}\sum_i(\alpha_i +\tilde\alpha_i) -\frac{\fb^2+2}{2})}\cr
&\quad \times\begin{cases}
    \prod_{i=1}^3\frac{\Gamma(\fb\alpha_i -\eta_i)}{\Gamma(1-\fb \bar{{\alpha}}_i+\bar\eta_i)}~,~\sigma=+1\\
\prod_{i=1}^3\frac{\Gamma(\fb\tilde\alpha_i +\eta_i)}{\Gamma(1-\fb \bar{\tilde{\alpha}}_i -\bar\eta_i)}~,~\sigma=-1 \end{cases} \cr
&\quad \times \frac{\Upsilon_\fb(\fb)\prod_{i}\Upsilon_\fb(\alpha_i+\tilde\alpha_i)}{\Upsilon_\fb(\frac{1}{2}\sum_i(\alpha_i+\tilde{{\alpha}}_i) -\frac{\fb}{2}-\fQ)\prod_{\mathrm{cyc}}\Upsilon_\fb(\frac{\fb}{2} +\frac{1}{2}(\alpha_{ijk}+ \tilde\alpha_{ijk}))}~.
\end{align}
where we absorbed purely $\fb$ dependence into $\mathsf{a}(\fb)$.
We list a few properties
\begin{itemize}
    \item \textit{Angular momentum} The correlation function (\ref{eq:Liouville_AMV_structure_constant}) is valid in the angular momentum violating case
\begin{equation}
    \sum_i \lambda_i =\pm 1~.
\end{equation}
\item \textit{Poles and zeros} From (\ref{eq:Y-zeros}) we easily infer that the ratio of Ypsilon functions has zeros for 
\begin{equation}
    \alpha_i +\tilde\alpha_i \in -r\fb-s\fb^{-1}~,\quad \alpha_i+\tilde\alpha_i \in (r+1)\fb + (s+1)\fb^{-1}~.
\end{equation}
which coincide with the zeros of the AMP structure constant (\ref{eq: zeros AMP}). Similarly combining the $\gamma$-function and the first Ypsilon function in the denominator we obtain the poles
\begin{equation}\label{eq: poles 1 AMV}
    \frac{1}{2}\sum_i(\alpha_i +\tilde\alpha_i)-\fQ \in -\left(r+\frac{1}{2}\right)\fb - s \fb^{-1}~,\quad \frac{1}{2}\sum_i(\alpha_i +\tilde\alpha_i)-\fQ \in \left(r+\frac{1}{2}\right)\fb + (s+1) \fb^{-1}
\end{equation}
and 
\begin{equation}
    \frac{1}{2}(\alpha_{ijk} +\tilde\alpha_{ijk}) \in -\left(r+\frac{1}{2}\right)\fb -s\fb^{-1}~,\quad \frac{1}{2}(\alpha_{ijk} +\tilde\alpha_{ijk}) \in \left(r+\frac{1}{2}\right)\fb +(s+1)\fb^{-1}~.
\end{equation}
Furthermore the explicit $\Gamma$-function ratios contribute
\begin{equation}
    \sigma= +1:\quad \fb \alpha_i - \eta_i \in -\mathbb{N}_0~,\quad \quad \sigma=-1:\quad \sigma= +1:\quad \fb \tilde\alpha_i + \eta_i \in -\mathbb{N}_0~,
\end{equation}
and zeros for
\begin{equation}
    \sigma=+1:\quad \fb \bar \alpha_i +\bar \eta_i \in 1+\mathbb{N}_0~,\quad \quad \sigma=-1:\quad \fb \bar{\tilde{\alpha}}_i -\bar\eta_i \in 1+ \mathbb{N}_0~.
\end{equation}
\end{itemize}

\section{Semiclassical analysis of $\n=2$ Liouville theory}\label{sec:Liouville_semiclassical}

The purpose of this section is to test the structure constants of $\n=2$ Liouville theory~\eqref{eq:N=2-Liouville_structure_constant} against the semiclassical expansion of the corresponding Liouville path integral based on the action~\eqref{eq:N=2_Liouville_action}, following the classic work~\cite{Zamolodchikov:1995aa} in bosonic Liouville theory (see also~\cite{Harlow:2011ny}). The parameter that controls this asymptotic expansion is $\fb$, with the semiclassical limit being $\fb\to 0$.

\paragraph{Path integral and correlation functions}
Given the action~\eqref{eq:N=2_Liouville_action}, $N$-point correlation functions of the exponential primaries~\eqref{eq:primariy_holo_Liouville} on the sphere in $\n=2$ Liouville theory are defined semiclassically through the following path integral,
\begin{equation}\label{eq:Liouville_correlators_def}
    \Big\langle\prod_{i=1}^NV^{\eta_i,\bar\eta_i}_{\alpha_i,\bar{\alpha_i},\talpha_i,\bar{\talpha}_i}(z_i,\bar z_i)\Big\rangle\equiv\int [\mathcal D X] \e^{-S_{\rm L}^{\mathcal{N}=2}} \prod_{i=1}^NV^{\eta_i,\bar\eta_i}_{\alpha_i,\bar{\alpha_i},\talpha_i,\bar{\talpha}_i}(z_i,\bar z_i) \ ,
\end{equation}
where all the operators are normal ordered and $X$ collectively denotes the measure over all the fields $\varphi,\tvarphi,\psi,\tpsi,F,\tF$.~For clarity, we will leave implicit the anti-holomorphic dependence of operators and correlators in what follows.

As in the $\n=0,1$ instances of the theory, we can determine the cosmological constant dependence of the $\n=2$ Liouville correlators via performing the following KPZ scaling~\cite{Knizhnik:1988ak}~
\begin{equation}
    \varphi \rightarrow \varphi - \frac{1}{\fb}\log \mu~,\quad \tilde{\varphi} \rightarrow \tilde{\varphi} - \frac{1}{\fb}\log \mu^*~.
\end{equation}
In particular, on the two-sphere this leads to
\begin{equation}\label{eq:Liouville_mu,mu*-dependence}
    G(z_1,\ldots ,z_N)\equiv\Big\langle\prod_{i=1}^NV^{\eta_i}_{\alpha_i,\talpha_i}\Big\rangle   =(\mu\mu^*)^{\frac{\fQ}{\fb}}\mu^{-\frac{1}{2\fb}\sum_{i=1}^N(\alpha_i+\bar\alpha_i)} (\mu^*)^{-\frac{1}{2\fb}\sum_{i=1}^N(\tilde{\alpha}_i+\bar{\talpha}_i)} F(z_1,\ldots , z_N)~,
\end{equation}
where the function $F$ is independent of the cosmological constants $\mu,\mu^*$. This last expression fixes the dependence of correlation functions on the cosmological constants $\mu,\mu^*$. In the winding number preserving sector, the delta functions of eq.~\eqref{eq:N=2-Liouville_structure_constant} make the $\mu,\mu^*$ exponents in~\eqref{eq:Liouville_mu,mu*-dependence} equal, yielding
\begin{equation}\label{eq:wnp_KPZ}
    G_\text{wnp}(z_1,\ldots ,z_N)  =(\mu\mu^*)^{\frac{\fQ-\frac{1}{2}\sum_{i=1}^N(\alpha_i+\talpha_i)}{\fb}}F_\text{wnp}(z_1,\ldots , z_N)~,
\end{equation}
where recall that $\alpha_i+\talpha_i=\bar\alpha_i+\bar\talpha_i$.

Furthermore, as a path integral check, by implementing the various symmetry transformations, \eg eq.~\eqref{eq:N2_RSymmetry}, on the correlation functions $G(x_1,\ldots , x_N)$ we get the conservation of the corresponding charges~\eqref{eq:Liouville_dimas_&_R_charges} on the sphere. For instance, from the $\mathrm{U}(1)_{\rm V}$ symmetry we infer that $G(x_1,\ldots , x_N)$ does not vanish iff 
\begin{equation}
    \sum_{i=1}^N \big(2\mathsf{Q}(\alpha_i-\tilde{\alpha}_i) + \eta_i\big)=0~,
\end{equation}
from the $\rm U(1)$ R-symmetries we obtain the conservation of the R-charges, and so on.

\subsection{Perturbative poles of correlation functions}
Following the analysis for bosonic and $\mathcal{N}=1$ Liouville theory~\cite{Goulian:1990qr,Rashkov:1996np}, we start by discussing the analytic structure of Liouville correlation functions from a path integral perspective. To discuss that, as well as the saddle point expansion later, it will be convenient to work with the alternative field variables $\rho,Y$ defined in~\eqref{eq:rho,Y_def}, instead of the complex scalars $\varphi,\tvarphi$. In the supergravity language presented around eq.~\eqref{eq:SUGRA}, $\rho$ plays the role of the Weyl factor while $Y$ controls the $\mathrm{U}(1)$ gauge field. Then, after integrating out the auxiliary scalars $F$ and $\tF$, the $\mathcal{N}=2$ action~\eqref{eq:N=2_Liouville_action} takes the form
\begin{equation}\label{eq:rho_Y_Liouville_action}
\begin{aligned}
S_{\mathrm{L}}^{\mathcal{N}=2}
&= \frac{1}{4\pi}\int_{S^2}\d^2z\sqrt{\tilde g}\,
\Bigg(
\begin{aligned}[t]
&\frac12 \partial_\mu\rho\partial^\mu\rho
+\frac12 \partial_\mu Y\partial^\mu Y
+\frac{\mathsf Q\widetilde R}{\sqrt2}\rho
+|\mu|^2\e^{\sqrt2\fb\rho}
-\ii\btpsi\slashed D\psi
\\
&\qquad\qquad
-\frac{\ii}{2}\mu\fb\,
\e^{\frac{\fb}{\sqrt2}(\rho+\ii Y)}\bpsi\psi
-\frac{\ii}{2}\mu^*\fb
\e^{\frac{\fb}{\sqrt2}(\rho-\ii Y)}\btpsi\tpsi\Bigg)
\end{aligned}
\\
&= \frac{1}{4\pi}\int_{S^2}\d^2z\sqrt{\tilde g}\,
\Bigg(
\begin{aligned}[t]
&\frac12 \partial_\mu\rho\partial^\mu\rho
+\frac12 \partial_\mu Y\partial^\mu Y
+\frac12 \partial_\mu K\partial^\mu K
+\frac{\mathsf Q\widetilde R}{\sqrt2}\rho
\\
&\quad
+|\mu|^2\e^{\sqrt2\fb\rho}-\mu\fb
\e^{\frac{\fb}{\sqrt2}(\rho+\ii Y)-\ii K}
-\mu^*\fb
\e^{\frac{\fb}{\sqrt2}(\rho-\ii Y)+\ii K}\Bigg) \  ,
\end{aligned}
\end{aligned}
\end{equation}
where, in the second equality, we also bosonized the fermions~\eqref{eq:Liouville_fermions} as,
\begin{equation}
    \psi_2\sim\e^{-\ii K_L} \ , \ \tpsi_2\sim\e^{\ii K_L} \ , \ \psi_1\sim\e^{-\ii K_R} \ , \ \tpsi_1\sim\e^{\ii K_R} \ .
\end{equation}
We also recall the explicit form of the Liouville vertex operators from eq.~\eqref{eq: full Liouville vertex operator}. 

For the zero mode treatment, we further split the fields as
\begin{equation}\label{eq:zero-mode_splitting}
   \rho = \rho_0+ \hat{\rho} \ ,\ Y= Y_0 +\widehat{Y} \ , \ K=K_0+\widehat K \ ,
\end{equation}
where $\rho_0,\,Y_0,\, K_0$ are the constant zero modes and $\hat{\rho},\,\widehat{Y},\,\widehat K$ are fields orthogonal to them, respectively. For the compact scalars, these zero modes additionally split into left and right-moving parts. Furthermore, we denote the various screening operators as
\begin{equation}\label{eq:screening_ops}
    \Lambda_0 =\e^{\sqrt{2}\fb {\rho}} \ ,  \ 
    \Lambda_+ =  \e^{\frac{\fb}{\sqrt{2}}(\rho + \ii Y)-\ii K} \ ,\   \Lambda_- =\e^{\frac{\fb}{\sqrt{2}}(\rho - \ii Y)+\ii K} \ .    
\end{equation}
With this parametrization, the $N$-point correlation function~\eqref{eq:Liouville_mu,mu*-dependence} takes the form
\begin{equation}\label{eq:zero-mode_G}
    G(z_i)=\sum_{p,q\geq 0}\frac{\lambda^p\lambda^q_*}{p!q!}\int_{S^2} \prod_{r=1}^p\d^2y_r\prod_{l=1}^q\d^2w_l\Big\langle\prod_{i=1}^N V_i(z_i)\prod_{r=1}^p\Lambda_+(y_r)\prod_{l=1}^q\Lambda_-(w_l)\Big\rangle_0 \ ,
\end{equation}
with $\lambda=\frac{\mu\fb}{4\pi},\,\lambda_*=\frac{\mu^*\fb}{4\pi}$ and $V_i(z_i)$ is given in (\ref{eq: full Liouville vertex operator}). The subscript $0$ on the rhs implies that the correlator is evaluated against an action that does not have interactions carrying angular momentum. The strategy in spacelike Liouville theory, inherited from the $\n=0,1$ cases~\cite{Goulian:1990qr,Rashkov:1996np}, is to split the path integration measure according to~\eqref{eq:zero-mode_splitting} and then do the zero mode integral first. Accordingly, the integral of the holomorphic compact zero modes in the correlator on the rhs of~\eqref{eq:zero-mode_G} becomes
\begin{equation}
    \int_{(S^1)^2} \d Y_{L,0}\d K_{L,0} \,\e^{\sum_{i=1}^N\big( \ii \mathsf{P}^i_L Y_{L,0}  +i\eta_i K_{L,0} \big)} \e^{\ii p\big(\frac{\fb}{\sqrt 2}Y_{L,0}- K_{L,0}\big)}\e^{-\ii q\big(\frac{\fb}{\sqrt 2}Y_{L,0}- K_{L,0}\big)} \ ,
\end{equation}
and similarly for the anti-holomorphic zero modes $Y_{R,0},\,K_{R,0}$; $P_L^i$ and $P_R^i$ are defined in (\ref{eq: full Liouville vertex operator}). These integrals yield the following delta constraints,
\begin{subequations}\label{eq:compact_zero-mode_constraints}
    \begin{align}
        &\sum_{i=1}^N(\alpha_i-\talpha_i)+\fb\big(p-q\big)=\sum_{i=1}^N(\bar\alpha_i-\bar\talpha_i)+\fb\big(p-q\big)=0 \ ,\\
        &\sum_{i=1}^N\eta_i+q-p=\sum_{i=1}^N\bar\eta_i+q-p=0 \ .
    \end{align}
\end{subequations}
Together, these imply R-charge~\eqref{subeq:q} conservation, for all screening numbers $p,q\in\mathbb N$. The remaining $\rho_0$ zero mode integral, then, reads\footnote{The integral~\eqref{eq:rho_zero-mode} converges along the chosen $\mathbb R$ contour when $s_{p,q}<0$ (assuming $b,\hat\Lambda_0>0$). In more general cases it can be evaluated via an analytically continued contour, as we explain later.}
\begin{equation}\label{eq:rho_zero-mode}
    \int_{\mathbb{R}} \d \rho_0 \, \e^{-\sqrt 2 \fQ \rho_0-\hat\Lambda_0\e^{\sqrt 2 \fb \rho_0}+\frac{\rho_0}{\sqrt2}\big(\sum_{i=1}^N(\alpha_i+\talpha_i)+b(p+q)\big)}=\frac{1}{\sqrt 2 \fb} \hat\Lambda_0^{s_{p,q}}\Gamma\big(-s_{p,q}\big) \ ,
\end{equation}
where $\hat\Lambda_0=\frac{|\mu|^2}{4\pi}\int_{S^2}\d^2z\sqrt{\tilde g}\,\e^{\sqrt 2 \fb \hat\rho}$ and,
\begin{equation}\label{eq:spq_def}
    s_{p,q}\equiv \frac{\fQ-\frac 12 \sum_{i=1}^N(\alpha_i+\talpha_i)}{\fb} -\frac{p+q}{2} \ .
\end{equation}
The above perturbative analysis yields the expectation that $\n=2$ Liouville correlation functions shall exhibit poles when $s_{p,q}\in\mathbb N$. It appears, however, that the zero mode analysis is unable to dictate any constraints on angular momentum violation. We expect that these constraints arise when evaluating the remaining Coulomb gas integrals (something along these lines was also discussed in~\cite{Hosomichi:2004ph}). With this in mind, we proceed to discuss in more detail these perturbative poles for the three-point function, in the instances where angular momentum is preserved or violated by $\pm1$ unit.

\paragraph{AMP perturbative poles} When angular momentum is preserved, the relations~\eqref{eq:Liouville_winding_and_momentum} imply
\begin{equation}
    \sum_i(\alpha_i-\talpha_i)=\sum_i(\bar\alpha_i-\bar\talpha_i)=0 \ ,
\end{equation}
and therefore, from eqs.~\eqref{eq:compact_zero-mode_constraints} it must be $p=q$. This is just the statement that, since the screening operators $\Lambda_{\pm}$ carry $\pm1$ unit of angular momentum, they have to pair up so that the correlator in the rhs of~\eqref{eq:zero-mode_G} preserves angular momentum.\footnote{The latter statement is true due to the fact that the path integral of that correlator is evaluated with an action whose only interaction term is $\Lambda_0$~\eqref{eq:screening_ops}. This screening operator does not carry angular momentum and hence the $\langle\dots\rangle_0$ correlators have to preserve it.} Thus, the three-point functions' perturbative poles~\eqref{eq:spq_def} in the AMP sector occur when
\begin{equation}\label{eq:AMP_GL_poles}
    \frac{\fQ-\frac 12 \sum_{i=1}^3(\alpha_i+\talpha_i)}{\fb} \in \mathbb N \ .
\end{equation}
The AMP structure constant~\eqref{eq:N=2-Liouville_structure_constant} contains $\Upsilon_\fb\big(\frac 12 \sum_{i=1}^3(\alpha_i+\talpha_i)-\fQ\big)$ in the denominator, and given that the Ypsilon function is an entire function with zeros when~\eqref{eq:Y-zeros}, taking $r\in\mathbb N,\,s=0$ (in the first family of zeros) the exact answer (\ref{eq:poles 1 AMP}) reproduces the perturbative poles~\eqref{eq:AMP_GL_poles}.

\paragraph{AMV perturbative poles} In the instance where the three-point function violates angular momentum conservation by $\pm1$ unit, \ie $\sum_{i=1}^3\lambda_i=-\sigma$ where again $\sigma=\pm1$, by a similar reasoning as above the number of screening operators must satisfy $p=q+\sigma$. This implies that the AMV three-point function exhibits perturbative poles at
\begin{equation}\label{eq:AMV_GL_poles}
        \frac{\fQ-\frac 12 \sum_{i=1}^3(\alpha_i+\talpha_i)}{\fb}-\frac{\sigma}{2} \in \mathbb N \ .
\end{equation}
Since the proposed AMV structure constant~\eqref{eq:Liouville_AMV_structure_constant} contains $\Upsilon_\fb\big(\frac 12 \sum_{i=1}^3(\alpha_i+\talpha_i)-\fQ-\frac \fb 2\big)$ in the denominator, for $r=\mathsf{N}+\frac{\sigma+1}{2},\,\mathsf{N}\in\mathbb N$ and $s=0$ (\ref{eq: poles 1 AMV}) the perturbative poles~\eqref{eq:AMV_GL_poles} are again reproduced.\\
We conclude that the perturbative Goulian-Li type poles \cite{Goulian:1990qr} are reproduced by the proposed structure constants of $\n=2$ Liouville theory~\eqref{eq:N=2-Liouville_structure_constant},~\eqref{eq:Liouville_AMV_structure_constant}. As is the case for the bosonic and $\n=1$ (spacelike) versions of the theory, the exact structure constants exhibit much more additional poles.

Having set up the stage, we can now discuss the semiclassical expansion of various correlators around a particular sphere saddle which we shortly specify. More specifically, in the $\fb\to 0$ limit, the correlators~\eqref{eq:Liouville_correlators_def} admit a saddle point expansion with quantum corrections that can be systematically evaluated order-by-order in perturbation theory. In that vein, we start by exploring AMP and AMV non-BPS correlators and, subsequently, discuss $\frac 12$-BPS correlation functions in the next section.

\subsection{Semiclassical AMP path integral} 
We are now interested in calculating a AMP three-point function from a path integral perspective. In particular we will consider
\begin{equation}\label{eq: example AMP PI correlator}
 \big\langle V^0_{\fb,\fb}(0)V^0_{\fb,\fb}(1)V^0_{\fb,\fb}(\infty)\big\rangle  = \langle \e^{\fb(\varphi + \tilde\varphi)} \e^{\fb(\varphi + \tilde\varphi)}\e^{\fb(\varphi + \tilde\varphi)}\rangle = \langle \e^{\sqrt{2}\fb \rho} \e^{\sqrt{2}\fb \rho}\e^{\sqrt{2}\fb \rho}\rangle ~,
\end{equation}
where in the last equality we used the change of variables (\ref{eq:rho,Y_def}). The correlator has $\alpha_i=\tilde\alpha_i=\fb$ and $\eta_i =\bar \eta_i =0$ for $i=1,2,3$ and hence $\sum_i \lambda_i=0$.

\paragraph{Sphere saddles} We start by discussing the semiclassical expansion of the $\n=2$ Liouville path integral~\eqref{eq:Liouville_correlators_def}, expressed in the bosonic variables $\rho,Y$ of eq.~\eqref{eq:rho,Y_def}.\footnote{In the $\fb\to 0$ limit, $V_{\alpha,\alpha}^0=\e^{\sqrt 2 \fb\rho}$ is a light operator and only disturbs the saddles~\eqref{eq:complex_saddles} at order $\mathcal{O}(\fb^2)$.} The equation of motion for the Weyl factor $\rho$ stemming from the action~\eqref{eq:rho_Y_Liouville_action}, reads
\begin{equation}\label{eq: eom varphis}
\begin{aligned}
&-\widetilde{\nabla}^2 \rho + \frac{\mathsf{Q}\widetilde{R}}{\sqrt{2}} + \sqrt{2}|\mu|^2 \mathsf{b}\e^{\sqrt{2}\mathsf{b}\rho} -\frac{\ii}{2\sqrt{2}}\mu \mathsf{b}^2\e^{\frac{\mathsf{b}}{\sqrt{2}}(\rho + \ii Y)}\bpsi\psi -\frac{\ii}{2\sqrt{2}}\mu^* \mathsf{b}^2\e^{\frac{\mathsf{b}}{\sqrt{2}}(\rho -\ii Y)}\btpsi\tpsi &=0 ~.
\end{aligned}    
\end{equation}
There exists a simple constant saddle point solution, that can be deduced as follows. To satisfy the equations of motion for the fermions, we take $\psi_*= \tpsi_*=0$ on the saddle.~Combining this with eq. (\ref{eq: eom varphis}) and further setting $Y_{*} =0$,\footnote{In fact, the most general solution consistent with the fermions vanishing on the saddle is $Y_*=\text{const.}$. The other constant solutions are obtained by acting on $Y_*=0$ with the generator of $\mathrm U(1)\subset\mathrm{OSp}(2|2,\mathbb C)$.} we obtain a one-parameter family of constant complex saddles on the sphere
\begin{equation}\label{eq:complex_saddles}
    \rho_{*}^{(s)} = \frac{1}{\sqrt{2}\mathsf{b}}\log \frac{1}{|\mu|^2\mathsf{b}^2r^2}+\frac{\ii\pi}{\sqrt{2}\mathsf{b}}(2s+1)~,\quad \quad s\in \mathbb{Z}~.
\end{equation}
From the supergravity perspective, all these saddles correspond to the same complex metric on the two-sphere $ds^2_*=-\frac{1}{|\mu|^2\fb^2}d\Omega^2_2$, with the fermions and the gauge field being classically trivial. The above constant solutions break the global $\n=2$ superconformal group $\mathrm{OSp}(2|2,\mathbb C)$ into a subgroup. In particular, the bosonic part inside $\mathrm{OSp}(2|2,\mathbb C)$ is $\mathrm{PSL}(2,\mathbb{C})\times \mathrm U(1)$, with the constant solutions breaking it to $\mathrm{PSU}(2)$ and yielding $4$ Goldstone modes. Similarly, there will be $4$ Goldstone modes from the fermionic part. We will identify these zero modes in our one-loop analysis below.

If instead we consider Liouville theory as two-dimensional supergravity, the above interpretation gets altered. Whereas, naively, the constant solutions break the global $\n=2$ superconformal group $\mathrm{OSp}(2|2,\mathbb C)$, the theory (\ref{eq:rho_Y_Liouville_action}) has a $\mathrm{OSp}(2|2,\mathbb C)$ family worth of solutions which are gauge equivalent from the perspective of the underlying $\mathcal{N}=2$ supergravity and hence the symmetry is not broken. From that point of view, to evaluate correlation functions from the path integral of $\mathcal{N}=2$ Liouville supergravity coupled to matter, we have to sum over all the constant saddles and then divide by the volume of $\mathrm{OSp}(2|2,\mathbb C)$.

Which of the above saddles contribute to the $\n=2$ Liouville path integral? Normally, one has to evaluate the fluctuations around each saddle and then address the question of which to sum over in the path integral. As we will illustrate later, however, in our case the one-loop corrections do not depend on $s$ and hence we can sum over saddles at tree-level. To address the question of which saddles contribute we resort for inspiration to the minisuperspace truncation of the Liouville zero mode $\rho_0$ path integral, as in bosonic Liouville theory~\cite{Seiberg:1990eb,Polchinski:1990mh}, which after a rescaling takes the form,
\begin{equation}\label{eq:zero-mode_int}
    \int_{\cal C\subset\mathbb{C}} \d z\,\e^{-\frac{z}{\fb^2}-c\e^{z}} \ ,
\end{equation}
where $z=\sqrt{2}\fb\rho_0,c=|\mu|^2r^2$ and $\cal C$ is the yet unspecified integration contour. This integral is closely related to the integral representation of the Gamma function, and one can perform a standard Picard-Lefschetz type analysis of its saddles and corresponding steepest descent/ascent contours, the results of which are summarized in the following graph:
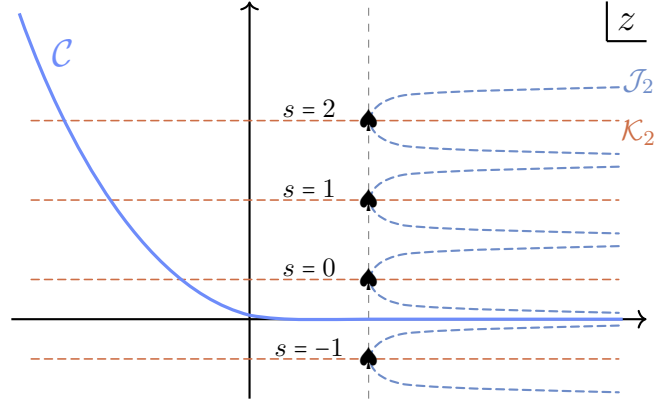
\begin{center}
\begin{tikzpicture}[scale=1.05]
    \definecolor{mygreen}{RGB}{110,140,250}   
    \definecolor{myred}{RGB}{110,140,200}    
    \definecolor{myorange}{RGB}{205,110,70}  
    \draw[->,thick] (0,-1) -- (0,4);
    \draw[->,thick] (-3,0) -- (5,0);
    \draw[dashed,gray] (1.5,-1) -- (1.5,4);
    \draw[thick] (4.5,3.5) -- (5,3.5);
    \draw[thick] (4.5,3.5) -- (4.5,4);
    \node at (4.75,3.75) {\Large $z$};
    \foreach \y/\s in {-0.5/-1,0.5/0,1.5/1,2.5/2} {

        \draw[myred,thick,dash pattern=on 3pt off 2pt,line cap=round]
            (4.65,\y+0.42)
            .. controls (3.95,\y+0.40) and (2.45,\y+0.38) .. (1.95,\y+0.32)
            .. controls (1.62,\y+0.26) and (1.52,\y+0.12) .. (1.52,\y)
            .. controls (1.52,\y-0.12) and (1.62,\y-0.26) .. (1.95,\y-0.32)
            .. controls (2.45,\y-0.38) and (3.95,\y-0.40) .. (4.65,\y-0.42);

    }
    \node[myred] at (4.9,3) {$\mathcal J_{2}$};
    \foreach \y/\s in {-0.5/-1,0.5/0,1.5/1,2.5/2} {
        \draw[myorange,semithick,dash pattern=on 3pt off 2pt,line cap=round]
            (-2.75,\y) -- (4.65,\y);
    }
     \node[myorange] at (4.9,2.35) {$\mathcal K_{2}$};
    \foreach \y/\s in {-0.5/-1,0.5/0,1.5/1,2.5/2} {
        \node at (1.5,\y) {\Large $\boldsymbol\spadesuit$};
        \node at (0.75,\y+.15) {\small $s=\s$};
    }
    \draw[very thick,mygreen]
        (-2.9,3.85)
        .. controls (-2.55,2.85) and (-1.55,0.45) .. (0,0.05)
        .. controls (0.45,-0.03) and (0.95,0.00) .. (1.5,0.00)
        -- (4.70,0.00);
    \node[mygreen] at (-2.35,3.35) {\Large $\mathcal C$};
\end{tikzpicture}
\end{center}
The integral~\eqref{eq:zero-mode_int} converges along the real axis for $\fb^2<0$, yielding $c^{\fb^{-2}}\Gamma(-\fb^{-2})$, and can be defined for the case of interest $\fb^2>0$ via analytic continuation. In particular, in the graph we have represented the complex saddle points of~\eqref{eq:zero-mode_int} (which coincide with the Liouville saddles~\eqref{eq:complex_saddles}) with the symbol $\Large\boldsymbol\spadesuit$, with the gray and orange dashed curves corresponding to the downward ($\mathcal J_s$) and upward ($\mathcal K_s$) Lefschetz thimbles of each saddle, respectively. The light-blue solid line is one possible defining contour $\cal C$ along which the integral converges when $\fb^{2}>0$. This contour can be further decomposed as $\mathcal{C} = \sum_{s\geq 0} \mathcal{J}_{s}$, which reproduces the analytic continuation of the Gamma function $\Gamma(-\fb^{-2})$ to positive values of $\fb^{-2}$. This line of reasoning instructs us to sum over all the $s\geq 0$ saddles~\eqref{eq:complex_saddles} in the $\n=2$ Liouville path integral.\footnote{Note, in passing, that since all these saddles yield the same 2$d$ metric with two negative eigenvalues, the Kontsevich-Segal-Witten criterion~\cite{Kontsevich:2021dmb,Witten:2021nzp} would render them non-allowable. Nonetheless, as we shall see, summing over the $s\geq0$ saddles in the path integral is important for recovering the correct semiclassical limit of the exact structure constants~\eqref{eq:N=2-Liouville_structure_constant},~\eqref{eq:Liouville_AMV_structure_constant}.} Doing so we get the following tree-level expression for (\ref{eq: example AMP PI correlator}) 
\begin{equation}\label{eq:corr_tree}
    \begin{aligned}
        &\big\langle V^0_{\fb,\fb}(0)V^0_{\fb,\fb}(1)V^0_{\fb,\fb}(\infty)\big\rangle \simeq \sum_{s\geq 0}\Big(\e^{\sqrt 2 \fb \rho_*^{(s)}}\Big)^3\e^{-S^{\n=2}_\text{L}[\rho_*^{(s)}]}\\
        &\qquad\simeq-(|\mu|^2\fb^2 r^2)^{-3}(|\mu|^2 \mathsf{b}^2 r^2 \e)^{\frac{1}{\mathsf{b}^2}} \e^{-\frac{\ii \pi}{\mathsf{b}^2}}\sum_{s\geq 0}\e^{-\frac{2 \pi\ii}{\mathsf{b}^2}s}\simeq\frac{\ii}{2}(|\mu|^2\fb^2r^2)^{\frac{1}{\fb^2}-3}\e^\frac{1}{\fb^2}\frac{1}{\sin{\big(\frac{\pi}{\fb^{2}}\big)}}~,
    \end{aligned}
\end{equation}
where we chose $\alpha=\talpha=\bar\alpha=\bar\talpha=\fb,\,\eta=\bar\eta=0$ and dropped the anti-holomorphic labels of the primaries for simplicity. We remark that the imaginary factor in the above expression is an artifact of just keeping the tree-level contributions and will be lifted once we incorporate one-loop corrections below. Note that we could also have considered spectrally flowed states~\eqref{eq:primariy_holo_Liouville} so that $\sum_i\eta_i=\sum_i\bar\eta_i=0$, \eg $V^\eta_{\fb,\fb}=\e^{\sqrt 2\fb\rho+\ii\eta K}$ with $\eta=\bar\eta\sim\mathcal{O}(1)$, for which the leading behavior of the three-point function would be the same as in eq.~\eqref{eq:corr_tree}. We will discuss spectrally flowed correlators more in the AMV case below.

The strategy followed above is equivalent to that of bosonic Liouville theory on the sphere \cite{Harlow:2011ny,Mahajan:2021nsd, Muhlmann:2021clm}. As a matter of fact, when setting $\psi=\tpsi=Y=0$ the action~\eqref{eq:rho_Y_Liouville_action} becomes,
\begin{equation}
     \frac{1}{4\pi} \int_{S^2} \d^2 z \sqrt{\tilde{g}}\, \bigg(\partial_\mu\rho\partial^\mu\rho +\fQ\tilde R \rho+|\mu|^2\e^{2\fb\rho}\bigg) \ ,
\end{equation}
which is the action of bosonic Liouville with background charge $\fQ=\fb^{-1}$ and cosmological constant $|\mu|^2/4\pi$. Moreover, the form of the Ypsilon function fraction in the AMP structure constant~\eqref{eq:N=2-Liouville_structure_constant} closely resembles the DOZZ formula when $\alpha_i=\talpha_i$ and $\fQ=\fb^{-1}$. Hence, we expect the various tree-level tests of DOZZ in bosonic Liouville, as performed \eg in refs.~\cite{Zamolodchikov:1995aa,Harlow:2011ny}, to be applicable to the $\n=2$ case for correlators of primaries of the form $V^0_{\alpha,\alpha}=\e^{2\alpha\rho}$. The main difference from the bosonic case arises for more general correlators that we will explore later, as well as when higher-loop contributions are taken into account -- which we now turn to discuss.

\paragraph{One-loop corrections} 

The one and higher-loop computations follow closely the treatment for $\n=2$ timelike Liouville theory given in~\cite{Anninos:2023exn}, hence our presentation here will be relatively brief. We will, however, treat carefully the fermionic zero modes. Adding fluctuations to the complex saddle point solutions, as
\begin{equation}
    \rho=\rho_*^{(s)}+\delta\rho \ , \ Y=\delta Y \ , \ \psi=\delta\psi \ , \ \tpsi=\delta\tpsi \ ,
\end{equation}
and expanding the action~\eqref{eq:rho_Y_Liouville_action} to quadratic order, we arrive at
\begin{equation}\label{eq:S2}
 \begin{aligned}
     S^{(2)}_\text{L}&= \frac{1}{4\pi} \int_{S^2} \d^2 z \sqrt{\tilde{g}}\, \bigg(\frac{1}{2}\delta \rho (-\widetilde{\nabla}^2 -\frac{2}{r^2})\delta \rho +\frac{1}{2}\delta Y (-\widetilde{\nabla}^2)\delta Y \bigg)\cr
     &\quad \quad +\frac{1}{4\pi} \int_{S^2} \d^2 z \sqrt{\tilde{g}}\,\bigg(-\ii\delta\btpsi \slashed{D}\delta \psi +\frac{(-1)^s}{2r}\sqrt{\frac{\mu}{\mu^*}} \delta \bpsi \delta \psi +\frac{(-1)^s}{2r} \sqrt{\frac{\mu^*}{\mu}} \delta \btpsi\delta \tpsi\bigg) \ .
 \end{aligned}
\end{equation}
We can decompose the bosonic fluctuations $\delta\rho$ and $\delta Y$ into eigenfunctions of the Laplacian on the two-sphere $-\widetilde{\nabla}^2 Y_{lm} = \frac{l(l+1)}{r^2}Y_{lm}$, \eg
\begin{equation}\label{eq:spherical_harm_deco}
    \delta\rho=\sum_{l\geq0}\sum_{m=-l}^l\delta\rho_{lm}Y_{lm} \ ,
\end{equation}
from which we infer that (\ref{eq:S2}) has $3$ ($l=1$ for $\delta\rho$) $+1$ ($l=0$ for $\delta Y$) zero modes. These are the $4$ Goldstone modes anticipated from the saddles breaking the bosonic symmetry $\mathrm{PSL}(2,\mathbb C)\times \mathrm U(1)\subset \mathrm{OSp(2|2,\mathbb{C})}$, and will be treated separately in computing the bosonic determinant below. Furthermore, the $l=0$ mode for $\delta \rho$ leads to a Gaussian unsuppressed term in~\eqref{eq:S2}, which we cure by Wick rotating $\delta\rho_{00}\rightarrow \pm\ii \delta\rho_{00}$~\cite{Gibbons:1978ac,Polchinski:1988ua}. Defining the bosonic measures as~\cite{Anninos:2023exn}
\begin{equation}
    [\mathcal{D}\delta\rho]=\prod_{l=0}^\infty\prod_{m=-l}^l \left(\frac{r^2\Lambda_\text{uv}}{\pi}\right)^\frac{1}{2} \d\delta\rho_{lm} \ , \quad 1=\int[\mathcal{D}\delta\rho]\e^{-\Lambda_\text{uv}\int_{S^2}\d^2z\sqrt{\tilde g}\,\delta\rho^2} \ ,
\end{equation}
where $\Lambda_\text{uv}$ is an ultraviolet cutoff with units of inverse length squared, the one-loop bosonic path integral evaluates to
\begin{equation}\label{eq:bosonGaussianN2}
    \mathcal{Z}'^{\,\text{bos}}_{\text{1-loop}}=\pm \ii \left(2\pi r^2 \Lambda_\text{uv}\right)^\frac{1}{2}\left(\frac{r^2 \Lambda_\text{uv}}{\pi}\right)^2\prod_{l=2}^\infty\left(\frac{4\pi r^2 \Lambda_\text{uv}}{l(l+1)-2}\right)^{l+\frac{1}{2}}\prod_{l=1}^\infty\left(\frac{4\pi r^2 \Lambda_\text{uv}}{l(l+1)}\right)^{l+\frac{1}{2}}~.
\end{equation}
Here, the $\pm\ii$ factor comes from the contour rotation of the negative $\delta\rho_{00}$ mode, while we have omitted the zero modes in both determinants which must be treated separately.\footnote{Nonetheless, we have kept their contribution from the path integral measure in $\left(\frac{r^2 \Lambda_\text{uv}}{\pi}\right)^2$.} Performing a heat-kernel analysis of the bosonic one-loop determinants, see \eg refs.~\cite{Anninos:2021ene, Anninos:2023exn, Anninos:2020hfj}, we formally obtain
\begin{equation}\label{eq:bosonic_heat}
    \begin{aligned}
        &-\frac{1}{2}\sum_{l\geq  2}(2l+1)\log \frac{l(l+1)-2}{4\pi r^2 \Lambda_{\text{uv}}} =  \int_{0}^\infty\frac{\dd t}{2t}\left[\frac{1+\e^{-t}}{1-\e^{-t}}\left(\frac{2\e^{-2t}}{1-\e^{-t}}+ 3\e^{-3t} - 5 \e^{-2t} + 5 \e^{-t}\right)\right]~,\\
&\qquad\quad \ \ -\frac{1}{2}\sum_{l\geq 1}(2l+1)\log \frac{l(l+1)}{4\pi r^2 \Lambda_{\text{uv}}} = \int_{0}^\infty\frac{\dd t}{2t}\left[\frac{1+\e^{-t}}{1-\e^{-t}}\left(\frac{2\e^{-t}}{1-\e^{-t}}+\e^{-t}\right)  \right]~,
    \end{aligned}
\end{equation}
in which we recognize the Harish-Chandra characters of discrete series unitary irreducible representations of $\mathrm{SO}(1,2)$, namely
\begin{equation}
    \chi_{\Delta=2}(t)=\frac{\e^{-2t}}{1-\e^{-t}} \ , \quad \chi_{\Delta=1}(t)=\frac{\e^{-t}}{1-\e^{-t}} \ .
\end{equation}
Evaluating the heat-kernel integrals~\eqref{eq:bosonic_heat} subject to the regularization scheme $\int_0^\infty \frac{\d t}{2t}\e^{-\frac{\varepsilon^2}{4t}-2at}=-\frac{1}{2}\log{\frac{a}{4\pi r^2 \Lambda_\text{uv}}}$, we obtain the following form for the leading divergences
\begin{equation}\label{eq: final boson one loop}
    \log \mathcal{Z}'^{\,\text{bos}}_{\mathrm{1-loop}} =  \frac{4}{\varepsilon^2}  -\frac{8}{3}\log \varepsilon + \text{const.}~,
\end{equation}
where $\varepsilon=\frac{\e^{-\gamma_\text{E}}}{\sqrt{2\pi r^2\Lambda_\text{uv}}}$ for $\gamma_\text{E}$ the Euler-Mascheroni constant.

The fermionic part of the quadratic action~\eqref{eq:S2} appears to depend on $s$, nevertheless we will see that this dependence does not affect the one-loop contribution.
Following \cite{Camporesi:1995fb,Anninos:2023exn} we expand $\delta\psi$ and $\delta\tpsi$ in an eigenbasis of the Dirac operator on the two-sphere $\slashed{D}\psi^{(\pm)}_{\pm lm}=\pm\frac{\ii}{r}(l+1)\psi^{(\pm)}_{\pm lm}$, namely (the $\alpha,\beta$ coefficients are Grassmann-odd variables)
\begin{subequations}\label{eq:Dirac spinors}
    \begin{align}
        \delta\psi = \frac{1}{\sqrt{r}}\sum_{l\geq 0} \sum_{m=0}^l \sum_{\pm} \big(\alpha_{\pm l m}\psi_{\pm l m}^{(+)} + \beta_{\pm l m}\psi^{(-)}_{\pm l m}\big)~,\\
        \delta\tpsi = -\frac{\ii}{\sqrt{r}}\sum_{l\geq 0} \sum_{m=0}^l \sum_{\pm}\big( \beta^*_{\mp l m}\psi_{\pm l m}^{(+)} + \alpha^*_{\mp l m}\psi^{(-)}_{\pm l m}\big)~.
    \end{align}
\end{subequations}
 It is then convenient to define $\boldsymbol{x}_{lm}$ as
\begin{equation}
\boldsymbol{x}_{lm}^T \equiv  \begin{pmatrix} \alpha_{+\,lm}& \alpha_{+\,lm}^* & \alpha_{-\,lm}  & \alpha_{-\,lm}^* & \beta_{+\,lm}  & \beta_{+\,lm}^* & \beta_{-\,lm} & \beta_{-\,lm}^*  \end{pmatrix}~,
\end{equation}
such that the fermionic piece of the quadratic action~\eqref{eq:S2} can be written as
\begin{equation}\label{N2fG}
S_{\text{fer}}^{(2)} = \frac{1}{4\pi} \sum_{l = 0}^\infty\sum_{m=0}^l \boldsymbol{x}_{lm}^TB_{lm}\boldsymbol{x}_{lm}~,
\end{equation}
where $B_{lm}$ is given by
\begin{equation}\label{eq:Bmatrix}
 \begin{pmatrix}  
0 & 0 & 0 & -\frac{1}{2}(l+1) & -\ii(-1)^s \frac{\mu}{2|\mu|} & 0 & 0 & 0 \\ 
0 & 0 & -\frac{1}{2}(l+1) & 0 & 0 & -\ii(-1)^s \frac{\mu^*}{2|\mu|}  & 0 & 0 \\ 
0 & \frac{1}{2}(l+1) & 0 & 0 & 0 & 0 & -\ii(-1)^s \frac{\mu}{2|\mu|}  & 0 \\
\frac{1}{2}(l+1) & 0 & 0 & 0 & 0 & 0 & 0 & -\ii (-1)^s\frac{\mu^*}{2|\mu|}\\
\ii(-1)^s \frac{\mu}{2|\mu|}   & 0 & 0 & 0 &0& 0 & 0 & \frac{1}{2}(l+1) \\ 
0  & \ii(-1)^s \frac{\mu^*}{2|\mu|} & 0 & 0 &0& 0 & \frac{1}{2}(l+1) & 0 \\ 
0  & 0 & \ii(-1)^s \frac{\mu}{2|\mu|}  & 0 &0& -\frac{1}{2}(l+1)  & 0 & 0 \\ 
0  & 0 & 0 & \ii (-1)^s\frac{\mu^*}{2|\mu|} & -\frac{1}{2}(l+1) & 0  & 0 & 0 \\ 
\end{pmatrix}~.
\end{equation}
Note that the above system exhibits zero modes for $l=m=0$. In particular, solving for $B_{00}\boldsymbol{x}_{00}=0$ (\ref{eq:Bmatrix}) we see that the zero mode space is four-dimensional,
\begin{subequations}
\begin{align}
    \alpha_{-00}^* &= -\ii (-1)^s \frac{\mu}{|\mu|}\beta_{+00}~,\quad \beta_{-00}^* = - \ii  (-1)^s \frac{\mu}{|\mu|}\alpha_{+00}~,\\
    \alpha_{+00}^*  &= \ii (-1)^s \frac{\mu}{|\mu|} \beta_{-00}~,\quad \beta_{+00}^* = \ii (-1)^s \frac{\mu}{|\mu|} \alpha_{-00}~.
\end{align}
\end{subequations}
Put differently, by performing the change of variables
\begin{equation}\label{eq: change of variables to c d}
   \begin{aligned}
        \alpha_{+00} &= c_1+d_1~,\quad \beta_{-00}^*= -\ii (-1)^s \frac{\mu}{|\mu|} (c_1-d_1)~,\quad \beta_{+00}= c_2+d_2~,\quad \alpha_{-00}^* = -\ii  (-1)^s \frac{\mu}{|\mu|}(c_2-d_2)~,\cr
    \alpha_{-00} &= c_3+d_3~,\quad \beta_{+00}^* = \ii  (-1)^s \frac{\mu}{|\mu|} (c_3-d_3)~,\quad \beta_{-00} = c_4+d_4~,\quad \alpha_{+00}^* = \ii  (-1)^s \frac{\mu}{|\mu|}(c_4-d_4)~.
   \end{aligned}
\end{equation}
we observe that the quadratic fermionic action becomes,
\begin{equation}\label{eq:x00B00x00}
    \boldsymbol x_{00}^T B_{00} \boldsymbol x_{00} = -4\ii (-1)^s \frac{\mu}{|\mu|} (d_1d_2+d_3d_4)~,
\end{equation}
while the measure of the fermionic zero mode integral transforms as
\begin{equation}\label{eq:measure c d}
   \hspace{-.3cm} \dd\alpha_{\pm 00} \dd\beta_{\pm 00}\dd\alpha_{\pm 00}^* \dd\beta_{\pm 00}^*  = \frac{1}{16}\frac{|\mu|^4}{\mu^4} \dd c_1 \dd c_2 \dd c_3 \dd c_4  \dd d_1 \dd d_2 \dd d_3 \dd d_4 \equiv \frac{1}{16}\frac{|\mu|^4}{\mu^4}\d^4 \boldsymbol c \d^4 \boldsymbol{d} ~.
\end{equation}
From (\ref{eq:x00B00x00}) we conclude that the $c_1,c_2,c_3,c_4$ Grassmann variables span the zero mode sector. 
The all mode fermionic path integration measures are given by
\cite{Anninos:2023exn},
 \begin{equation}
    [\mathcal{D}\delta\psi]=\prod_{l=0}^\infty \prod_{m=0}^l\left(\frac{4\pi}{r^2 \Lambda_\text{uv}}\right)\prod_{\pm} \d \alpha_{\pm lm}\d \beta_{\pm lm} \ , \ [\mathcal{D}\delta\tpsi]=\prod_{l=0}^\infty \prod_{m=0}^l\left(\frac{4\pi}{r^2 \Lambda_\text{uv}}\right)\prod_{\pm} \d \alpha^*_{\pm lm}\d \beta^*_{\pm lm}  \ ,
\end{equation}
where for the $l=0$ modes we take (\ref{eq:measure c d}).
For the Pfaffian, after omitting the $l=0$ modes we obtain
\begin{equation}\label{fermionGaussianN2}
\mathcal{Z}'^{\,\text{fer}}_\text{1-loop} =\left(\frac{4\pi}{r^2\Lambda_{\text{uv}}}\right)^2 \prod_{l= 1}^\infty\left(\frac{(l+1)^2 - (-1)^{2s}}{4\pi r^2 \Lambda_{\text{uv}}}\right)^{2(l+1)} = \left(\frac{4\pi}{r^2\Lambda_{\text{uv}}}\right)^2 \prod_{l= 1}^\infty\left(\frac{l(l+2)}{4\pi r^2 \Lambda_{\text{uv}}}\right)^{2(l+1)}~,
\end{equation}
which is indeed independent of $s$, as promised. A heat-kernel analysis leads to 
\begin{equation}\label{ZgausferN2}
\sum_{l\geq  1}^\infty 2(l+1) \log \frac{l(l+2)}{4 \pi\Lambda_{\text{uv}}r^2}   =-\int_{0}^\infty\frac{\dd t}{2t}\left[\frac{2\e^{-\frac{t}{2}}}{1-\e^{-t}}\left(\frac{4\e^{-\frac{3}{2}t}}{1-\e^{-t}}+ 4\e^{-\frac{t}{2}}- 2\e^{-\frac{3t}{2}}+2 \e^{-\frac{5 t}{2}}\right)\right]~,
\end{equation}
whose leading singular behaviour, after following the same regularization scheme as for the bosonic piece, reads
\begin{equation}\label{eq: final fermion one loop}
    \log \mathcal{Z}'^{\,\text{fer}}_{\mathrm{1-loop}} = -\frac{4}{\varepsilon^2} +\frac{5}{3}\log \varepsilon + \text{const.}~.
\end{equation}
Note that the leading divergences cancel between the bosons and the fermions, expected from general arguments regarding the non-renormalization of the cosmological constant term in supersymmetric theories, while the logarithmic divergences combine with the saddle to the conformal anomaly.

Thus far, for the correlator~\eqref{eq: example AMP PI correlator} we have obtained
\begin{equation}\label{eq:long_AMP_correlator}
    \begin{aligned}
        &\big\langle V^0_{\fb,\fb}(0)V^0_{\fb,\fb}(1)V^0_{\fb,\fb}(\infty)\big\rangle \simeq   \sum_{s\geq 0}\Big(\e^{\sqrt 2 \fb \rho_*^{(s)}}\Big)^3\e^{-S^{\n=2}_\text{L}[\rho_*^{(s)}]} \cr
   \quad &\times\int [\mathcal{D}\delta \rho][\mathcal{D}\delta Y] [\mathcal{D}\delta\psi][\mathcal{D}\delta\tpsi]\e^{-S_{\text{L}}^{(2)}}\e^{\sqrt{2}\fb \delta\rho} (0)\e^{\sqrt{2}\fb \delta\rho}(1)\e^{\sqrt{2}\fb \delta\rho}(\infty) \cr
    &  \times\e^{\frac{1}{4\pi\fb^2r^2}\int \d^2z\sqrt{\tilde g}(\e^{\sqrt 2 \fb \delta\rho}-1-\sqrt 2 \fb \delta\rho-\fb^2\delta\rho^2)}\cr
    &\times\e^{-\frac{1}{4\pi} \int \d^2 z\sqrt{\tilde{g}}\frac{(-1)^s\mu}{2r|\mu|} (\e^{\frac{\fb}{\sqrt{2}}(\delta\rho + \ii \delta Y)} -1)\delta \bpsi \delta \psi}\e^{-\frac{1}{4\pi} \int \d^2 z\sqrt{\tilde{g}}\frac{(-1)^s\mu^*}{2r|\mu|} (\e^{\frac{\fb}{\sqrt{2}}(\delta\rho - \ii \delta Y)}-1)\delta \btpsi \delta \tpsi} \ ,
    \end{aligned}
\end{equation}
where in the last two lines we have included the higher order terms in the fluctuation action around the complex sphere saddles.\footnote{In particular, the exponential interaction terms are subtracted by appropriate terms to account for the contributions that already appear in the quadratic action~\eqref{eq:S2}.} The above correlator can be further expressed as,
\begin{equation}\label{eq: AMP PI final}
    \begin{aligned}
        &\big\langle V^0_{\fb,\fb}(0)V^0_{\fb,\fb}(1)V^0_{\fb,\fb}(\infty)\big\rangle \simeq   \sum_{s\geq 0}\Big(\e^{\sqrt 2 \fb \rho_*^{(s)}}\Big)^3\e^{-S^{\n=2}_\text{L}[\rho_*^{(s)}]} \mathcal{Z}'^{\,\text{bos}}_{\mathrm{1-loop}}\mathcal{Z}'^{\,\text{fer}}_{\mathrm{1-loop}}\cr
    &\times \int\!\! \frac{1}{16}\frac{|\mu|^4}{\mu^4}\d^4 \boldsymbol c\, \d^4 \boldsymbol{d}\int \!\!\prod_{m=0,\pm 1}\d\delta\rho_{1m}\d \delta Y_{00}\, \e^{-\frac{1}{4\pi}\boldsymbol{x}^T_{00}B_{00}\boldsymbol x_{00}}\\ 
    &\times \Big\langle\!\!\Big\langle\;
\prod_{z_i=0,1,\infty}\e^{\sqrt{2}\fb\delta\rho}(z_i)\;\e^{\frac{1}{4\pi\fb^2r^2}\int \d^2z\sqrt{\tilde g}(\e^{\sqrt 2 \fb \delta\rho}-1-\sqrt 2 \fb \delta\rho-\fb^2\delta\rho^2)}\\
& \qquad\times \e^{-\frac{1}{4\pi}\int\d^2z\sqrt{\tilde g}\frac{(-1)^s\mu}{2r|\mu|}
\big(\e^{\frac{\fb}{\sqrt{2}}(\delta\rho+\ii\delta Y)}-1\big)\delta\bpsi\,\delta\psi}
\e^{-\frac{1}{4\pi}\int\d^2z\sqrt{\tilde g}\frac{(-1)^s\mu^*}{2r|\mu|}
\big(\e^{\frac{\fb}{\sqrt{2}}(\delta\rho-\ii\delta Y)}-1\big)\delta\btpsi\delta\tpsi}
\;\Big\rangle\!\!\Big\rangle \ ,
    \end{aligned}
\end{equation}
where $\langle\!\!\langle\,\rangle\!\!\rangle$ schematically represents the normalized Gaussian expectation value over the remaining modes. Using the explicit expression (\ref{eq:x00B00x00}) we obtain
\begin{equation}\label{eq: d integral}
    \int \d^4\boldsymbol{d} \,\e^{-\frac{1}{4\pi}\boldsymbol{x}_{00}^T B_{00} x_{00}} = -\frac{1}{\pi^2} \frac{\mu^2}{|\mu|^2}~.
\end{equation}
We leave a careful evaluation of the higher-order corrections in $\langle\!\!\langle\,\rangle\!\!\rangle$ to future work. One important subtlety that we want to mention is that the above expectation value is only non-vanishing after bringing down fermions from the exponential in the higher-order action. Since in the exponential (\ref{eq:x00B00x00}) the zero modes $c_1,\ldots, c_4$ are absent, the fermionic zero mode integral would otherwise vanish.\footnote{The bosonic zero mode contribution also has to be handled with care, via a Faddeev-Popov treatment.} In particular, we recall the explicit expressions for the Dirac eigenfunctions (\ref{eq:Dirac spinors}) where \cite{Camporesi:1995fb, Anninos:2023exn}
\begin{equation}
\label{eq:ConcreteSpinorHarmonics}
	\psi^{(+)}_{\pm \, lm}(\Omega) = \frac{c_{lm}}{\sqrt{2}} \e^{ \ii \left( m + \frac{1}{2} \right) \phi } \begin{pmatrix} \Phi_{lm}(\theta) \\ \pm \ii \Psi_{lm}(\theta) \end{pmatrix} ~, \qquad \psi^{(-)}_{\pm \, lm} (\Omega)= \frac{c_{lm}}{\sqrt{2}} \e^{ - \ii \left( m + \frac{1}{2} \right) \phi } \begin{pmatrix} \pm \ii \Psi_{lm}(\theta) \\ \Phi_{lm}(\theta) \end{pmatrix}~,
\end{equation}
for
\begin{equation}
\begin{split}
	\Phi_{lm}(\theta) &= \left(\cos \frac{\theta}{2}\right)^{m+1} \left(\sin\frac{\theta}{2}\right)^m \, P_{l-m}^{(m,m+1)}(\cos\theta) ~, \\ 
	\Psi_{lm}(\theta) &= \left(\cos \frac{\theta}{2}\right)^{m} \left(\sin\frac{\theta}{2}\right)^{m+1} \, P_{l-m}^{(m+1,m)}(\cos\theta)  = (-1)^{l-m}\Phi_{lm}(\pi - \theta) ~.
\end{split}
\end{equation}
Here $P_n^{(a,b)}(x)$ are the real-valued Jacobi polynomials which satisfy $P_0^{(0,1)}(x)= P_0^{(1,0)}(x)=1$. The coefficients above read
\begin{equation}
c_{lm}^2 = \frac{(l+m+1)! (l-m)!}{2\pi (l!)^2}~.
\end{equation}
Combining the above explicit expressions with the change of variables (\ref{eq: change of variables to c d}) and restricting to the fermionic zero modes, we find for an arbitrary function $f(z)$
\begin{subequations}\label{eq: f for fermions}
\begin{align}\label{eq: f tbpsi tpsi}
\int\d^2 z\sqrt{\tilde{g}} f(z)\,
\big[\delta\btpsi\,\delta\tpsi\big](z)
&= \frac{r}{2\pi}\,\frac{\mu^{2}}{|\mu|^{2}}\,
\Big[\,-\ii\,\int\d\Omega f\big(c_{1}c_{2}+c_{3}c_{4}\big)
\;+\;\ii\,\int\d\Omega f\,\cos\theta\big(c_{1}c_{4}-c_{2}c_{3}\big) \notag\\
&\hspace{2.2cm}
+\;\int\d\Omega f\,\sin\theta\,\e^{\ii\varphi}\,c_{1}c_{3}
\;-\int\d\Omega f\,\sin\theta\,\e^{-\ii\varphi}\,c_{2}c_{4}\,\Big]\, , \\
\int\d^2 z\sqrt{\tilde{g}} f(z)\,
\big[\delta\bpsi\,\delta\psi\big](z)
&= \frac{r}{2\pi}\,
\Big[\,-\ii\,\int\d\Omega f\big(c_{1}c_{2}+c_{3}c_{4}\big)
\;-\;\ii\,\int\d\Omega f\,\cos\theta\big(c_{1}c_{4}-c_{2}c_{3}\big) \notag\\
&\hspace{1.4cm}
-\;\int\d\Omega f\,\sin\theta\,\e^{\ii\varphi}\,c_{1}c_{3}
\;+\;\int\d\Omega f\,\sin\theta\,\e^{-\ii\varphi}\,c_{2}c_{4}\,\Big]\,,
\end{align}
\end{subequations}
which heavily restrict the class of spherical harmonics appearing in the $\delta \rho,\,\delta Y$ fluctuations~\eqref{eq:spherical_harm_deco} such that (\ref{eq: AMP PI final}) is non-vanishing. From (\ref{eq: d integral}) and (\ref{eq:  f tbpsi tpsi}) we additionally infer that the $\boldsymbol{c}$ and $\boldsymbol{d}$ integrals remove the $\mu,\mu^*$ dependence in the measure in (\ref{eq: AMP PI final}), while the Yukawa vertices pair up to $\mu\mu^*/|\mu|^2=1$. This keeps intact the KPZ scaling in (\ref{eq:corr_tree}). 
\\ \\
Combining the saddle~\eqref{eq:corr_tree} and one-loop~\eqref{eq:bosonGaussianN2}, \eqref{fermionGaussianN2} contributions, while suppressing the residual flat-direction integrals as well as the higher loop contributions obtained by expanding the exponentials in (\ref{eq: AMP PI final}), which we will present in future work \cite{ThemisBeatrixpart2},
leads to\footnote{Here we have discussed the one-loop corrections to the sphere partition function, however these are the same as the one-loop corrections for the correlator~\eqref{eq:corr_tree}, whereas deviations start appearing at two-loops.} 
\begin{equation}\label{eq:non-BPS_NS_1-loop_correlator}
\big\langle V^0_{\fb,\fb}(0)V^0_{\fb,\fb}(1)V^0_{\fb,\fb}(\infty)\big\rangle \simeq  \pm \frac{|\mu|^{2\left(\frac{1}{\fb^2}-3\right)}\e^{\frac{1}{\fb^2} + \left(\frac{1}{\fb^2}-3\right)\log \fb^2}}{2\sin\big(\frac{\pi}{\fb^{2}}\big)} r^{\frac{c}{3}-6}\Lambda_{\mathrm{uv}}^{\frac{1}{2}}~,
\end{equation}
which, as anticipated, is manifestly real. Additionally, the $|\mu|$ dependence is consistent with KPZ scaling~\eqref{eq:wnp_KPZ} and the $r$-scaling follows from the Weyl anomaly, agreeing with the $\n=2$ Liouville central charge reported above (at least up to this loop order).

\subsection{Semiclassical limit of the AMP structure constant}
Having obtained a particular non-BPS correlation function in the NS-sector to one-loop order~\eqref{eq:non-BPS_NS_1-loop_correlator}, our aim now is to compare it with the semiclassical $\fb\to 0$ expansion of the angular momentum preserving $\n=2$ Liouville structure constant~\eqref{eq:N=2-Liouville_structure_constant}. 

Let us start with the $D(\fb,\fb)$ piece in~\eqref{eq:N=2-Liouville_structure_constant}. We have,
\begin{equation}\label{eq:AMP Dbb}
   \begin{aligned}
       D(\fb,\fb) &= {A}(\fb) \left(\frac{|\mu|^2}{4}\fb^{2-2\fb^2}\gamma(\fb^2)\right)^{\frac{1}{\fb^2}-3} \frac{\Upsilon_\fb(\fb)\Upsilon_\fb^3(2\fb)}{\Upsilon_\fb(3\fb-\fb^{-1})\Upsilon_\fb^3(\fb)}\cr
   &={A}(\fb) \left(\frac{|\mu|^2}{4}\fb^{2-2\fb^2}\gamma(\fb^2)\right)^{\frac{1}{\fb^2}-3} \frac{\pi\fb^{2(3-\fb^{-2})}\gamma^2(\fb^2)}{\Gamma^2(\fb^{-2}-2)\gamma(2\fb^2)}\frac{1}{\sin{\big(\frac{\pi}{\fb^2}\big)}} \ ,
   \end{aligned}
\end{equation}
where we used eqs.~\eqref{eq:Ypsilon_shift_identities} and $\Gamma(w)\Gamma(1-w)=\pi\sin^{-1}  {(\pi w)}$, to arrive at the second to last and the last equality, respectively. The semiclassical limit of this expression reads, 
\begin{equation}
    \lim_{\fb\to0}D(\fb,\fb)\simeq {A}(\fb) \frac{\e^{-2\gamma_\text{E}}}{\fb^2} \left(\frac{\e}{4}\right)^{\frac{1}{\fb^2}}\frac{|\mu|^{2\left(\frac{1}{\fb^2}-3\right)}\e^{\frac{1}{\fb^2} + \left(\frac{1}{\fb^2}-3\right)\log \fb^2}}{2\sin\big(\frac{\pi}{\fb^{2}}\big)}\Big(1+\mathcal{O}(\fb^2)\Big)
\end{equation}
Next, we have to evaluate the $\cal W$ integral in~\eqref{eq:N=2-Liouville_structure_constant}, which for our correlator becomes
\begin{equation}
    w_{\fb}\equiv \int_{\mathbb{C}^2} \d^2 y \d^2 z\,|y|^{2(\fb^2-1)}|z|^{2(\fb^2-1)}|1-y|^{-2\fb^2}|1-z|^{-2\fb^2}|y-z|^{-2\fb^2} \ .
\end{equation}
This integral can be explicitly computed via leveraging the formula for the complexified version of Selberg's integral, see \eg refs.~\cite{Dotsenko:1984ad},
\begin{equation}
    \begin{aligned}
        \int_{\mathbb{C}^n}\prod_{j=1}^n\frac{\d z_j\d\bar z_j}{2\ii}&\prod_{j=1}^n|z_j|^{2(\sigma-1)}|1-z_j|^{2(\tau-1)}\prod_{1\leq j < k\leq n}|z_j-z_k|^{4\theta}\\
        &\qquad\qquad=n!\pi^n\prod_{j=1}^n\frac{\gamma\big(\sigma+(j-1)\theta \big)\gamma\big(\tau+(j-1)\theta\big)\gamma(j\theta)}{\gamma\big(\sigma+\tau+(n+j-2)\theta\big)\gamma(\theta)} \ .
    \end{aligned}
\end{equation}
For $n=2,\sigma=\fb^2,\tau=1-\fb^2,\theta=-\fb^2/2$, after some massaging, we get
\begin{equation}
    w_\fb=\frac{1}{32\pi^2}\frac{\gamma^3\big(\frac{\fb^2}{2}\big)}{\gamma\big(\frac{3\fb^2}{2}\big)} \ ,
\end{equation}
whose semiclassical limit yields,
\begin{equation}
    \lim_{\fb\to 0} w_\fb\simeq \frac{3}{8\pi^2}\frac{1}{\fb^4} \Big(1+\mathcal{O}(\fb^2)\Big) \ .
\end{equation}
Combining the two contributions we hence obtain that, for $\fb \rightarrow 0$,
\begin{equation}\label{eq:final_AMP_PI_correlator}
    C(\fb,\fb;0) \simeq \frac{{A}(\fb)\e^{-\gamma_\text{E}}}{\fb^6}  \left(\frac{\e}{4}\right)^{\frac{1}{\fb^2}}\frac{|\mu|^{2\left(\frac{1}{\fb^2}-3\right)}\e^{\frac{1}{\fb^2} + \left(\frac{1}{\fb^2}-3\right)\log \fb^2}}{2\sin\big(\frac{\pi}{\fb^{2}}\big)}\Big(1+\mathcal{O}(\fb^2)\Big)~.
\end{equation}

We can now compare this with the semiclassical path integral expression (\ref{eq:non-BPS_NS_1-loop_correlator}). The $|\mu|^2$ dependence of the two expressions agrees and is consistent with the KPZ scaling~\eqref{eq:wnp_KPZ}, whereas $\Lambda_{\mathrm{uv}}$ is absent from the CFT expression. The cutoff $\Lambda_{\mathrm{uv}}$ is scheme dependent and for a comparison between the path integral and the CFT data we need to allow the rescaling $\Lambda_{\mathrm{uv}}\rightarrow \mathsf s \Lambda_{\mathrm{uv}}$. The parameter $\mathsf s$ can be fixed by comparing higher-loop contributions \cite{Muhlmann:2022duj} between the two sides. Next, we have the exponential term $\e^{\frac{1}{\fb^2}}$ stemming from the on-shell action. The additional term $\left(\frac{\e}{4}\right)^{\frac{1}{\fb^2}}$ in~\eqref{eq:final_AMP_PI_correlator} has its origin in mapping the path integral from the sphere to the disk (this is the frame with respect to which most comparisons in the literature are made). The round two-sphere metric
\begin{equation}
    ds^2 = \frac{4\e^{\sqrt{2}\fb \rho}}{(1+r^2)^2}(\d r^2 + r^2 \d\theta^2) \ ,
\end{equation}
is related to the disk by $\rho \rightarrow \rho - \frac{\sqrt{2}}{\fb}\log\frac{2}{1+r^2}$, where $\theta \sim \theta + 2\pi$ and $r\in (0,\kappa)$. The $\mathcal{N}=2$ Liouville action on the sphere is accordingly related to the action on the disk by \cite{Harlow:2011ny, Anninos:2021ene}
\begin{align}\label{eq:map disk sphere}
    S_{\mathrm{L},S^2}^{\mathcal{N}=2}\rightarrow S_{\mathrm{L},\mathsf{D}}^{\mathcal{N}=2} + \frac{1}{\sqrt{2}\pi \fb}\int_{\partial \mathsf{D}} \d\theta\, \rho + \frac{2}{\fb^2}\log\kappa + \frac{1}{\fb^2} - \frac{1}{\fb^2}\log 4~.
\end{align}
To compare the structure constant to the path integral correlator we therefore need to consider $C(\fb,\fb;0)\left(\frac{\e}{4}\right)^{-\frac{1}{\fb^2}}$ and then the exponentials in~\eqref{eq:non-BPS_NS_1-loop_correlator},~\eqref{eq:final_AMP_PI_correlator} match. We further observe that the semiclassical limit of the exact answer contains the $\sin^{-1}(\pi\fb^{-2})$ term, whose origin in the path integral came from summing the particular complex saddle points in eq.~\eqref{eq:corr_tree}. From that perspective, the defining contour $\mathcal C$ discussed earlier might serve as a starting point for determining the integration cycle of the $\n=2$ Liouville path integral. Lastly, the prefactor $\mathsf{A}(\fb)$ remains an undetermined function of $\fb$; assuming it takes the form of a pure power of $\fb$, fixing the exponent from the path integral would require a careful Faddeev--Popov analysis~\cite{Anninos:2021ene}, which we leave for future work.

\subsection{Semiclassical AMV path integral}
In this subsection we discuss the path integral computation of an angular momentum violating correlator. Specifically, we consider the correlator
\begin{equation}\label{eq:lambda+1 PI}
    \big\langle \bar\psi \psi \e^{2\fb \varphi + \fb\tilde \varphi}(0) \e^{\fb(\varphi + \tilde\varphi)}(1) \e^{\fb(\varphi + \tilde\varphi)} (\infty)\big\rangle =  \big\langle \bar \psi \psi \e^{\frac{3\fb}{\sqrt{2}}\rho + \ii \frac{\fb}{\sqrt{2}}Y}(0) \e^{\sqrt{2}\fb \rho} (1) \e^{\sqrt{2}\fb \rho} (\infty)\big\rangle \ ,
\end{equation}
where we used the change of variables (\ref{eq:rho,Y_def}). The operator with the fermion bilinear is inserted at one point as $:\bar\psi\psi\,\e^{2\fb \varphi + \fb\tilde \varphi}:(z_1)$. This operator corresponds to 
\begin{equation}\label{eq:AMV_insertions_1}
    \alpha_1= \bar\alpha_1= 2\fb~,\quad \tilde\alpha_1 = \bar{\tilde{\alpha}}_1= \fb~,\quad \eta_1 =\bar \eta_1= -1 \quad \Rightarrow  \quad \lambda_1 =1~.
\end{equation}
The other two insertions have
\begin{equation}\label{eq:AMV_insertions_2}
    \alpha_2 = \tilde\alpha_2=\bar \alpha_2= \bar{\tilde{\alpha}}_2=\fb~,\quad \alpha_3= \tilde\alpha_3=\bar \alpha_3= \bar{\tilde{\alpha}}_3=\fb~,\quad \eta_2=\eta_3=\bar\eta_2 = \bar\eta_3=0~,
\end{equation}
and hence the three-point function (\ref{eq:lambda+1 PI}) violates angular momentum conservation by one unit. We will now evaluate it from the path integral and compare our result to the semiclassical structure constant~\eqref{eq:Liouville_AMV_structure_constant} in the next subsection. 

Recall that the $\rho_*$ saddle of the equations of motion of the action in (\ref{eq:lambda+1 PI}) is (as in the AMP case) given by $\rho_*^{(s)}$ (\ref{eq:complex_saddles}), whereas $Y_*=\psi_*= \tilde\psi_*=0$. 
Similarly to the AMP correlator~\eqref{eq:long_AMP_correlator}, we obtain
\begin{equation}
   \begin{aligned}
       &\big\langle \bar \psi \psi \e^{\frac{3\fb}{\sqrt{2}}\rho + \ii \frac{\fb}{\sqrt{2}}Y}(0) \e^{\sqrt{2}\fb \rho} (1) \e^{\sqrt{2}\fb \rho} (\infty)\big\rangle \simeq   \sum_{s\geq 0}\e^{\frac{7}{\sqrt{2}} \fb \rho_*^{(s)}}\e^{-S^{\n=2}_\text{L}[\rho_*^{(s)}]} \cr
   \quad
   &\times\int [\mathcal{D}\delta \rho][\mathcal{D}\delta Y] [\mathcal{D}\delta\psi][\mathcal{D}\delta\tpsi]\e^{-S_{\mathrm{L}}^{(2)} }\delta \bpsi \delta\psi\e^{\frac{3 \fb}{\sqrt{2}} \delta\rho + \ii\frac{\fb}{\sqrt{2}}\delta Y} \!(0)\e^{\sqrt{2}\fb \delta\rho}(1)\e^{\sqrt{2}\fb \delta\rho}(\infty) \cr
&  \times\e^{\frac{1}{4\pi\fb^2r^2}\int \d^2z\sqrt{\tilde g}(\e^{\sqrt 2 \fb \delta\rho}-1-\sqrt 2 \fb \delta\rho-\fb^2\delta\rho^2)}\cr
    &\times\e^{-\frac{1}{4\pi} \int \d^2 z\sqrt{\tilde{g}}\frac{(-1)^s\mu}{2r|\mu|} (\e^{\frac{\fb}{\sqrt{2}}(\delta\rho + \ii \delta Y)} -1)\delta \bpsi \delta \psi}\e^{-\frac{1}{4\pi} \int \d^2 z\sqrt{\tilde{g}}\frac{(-1)^s\mu^*}{2r|\mu|} (\e^{\frac{\fb}{\sqrt{2}}(\delta\rho - \ii \delta Y)}-1)\delta \btpsi \delta \tpsi} \ ,
   \end{aligned}
\end{equation}
where,\footnote{The sum over $s$ is performed in this way because the $(-1)^s$ dependence in the operator insertion will cancel below.}
\begin{subequations}\label{eq: on shell and saddles}
\begin{align}
    &\sum_{s\geq 0}\e^{-S^{\n=2}_\text{L}[\rho_*^{(s)}]} = (|\mu|^2 \fb^2 r^2\e)^{\frac{1}{\fb^2}} \e^{-\frac{\ii \pi}{\fb^2}} \sum_{s\geq 0} \e^{-\frac{2\ii \pi}{\fb^2}s} = (|\mu|^2 \fb^2 r^2\e)^{\frac{1}{\fb^2}} \e^{-\frac{\ii \pi}{\fb^2}} \frac{-\ii}{2\sin(\frac{\pi}{\fb^2})}~,\\
    &\e^{\frac{7\fb}{\sqrt{2}}\rho_*^{(s)}}= -(-1)^s\ii (\fb^2 r^2 |\mu|^2)^{-\frac{7}{2}}~.
 \end{align}
\end{subequations}
Note that compared to the AMP case the saddle comes with an additional imaginary unit, which cancels the $\ii$ coming from the sum over complex saddles. This implies that after performing a Wick rotation for the unsuppressed $\delta\rho_{00}$ mode ($\delta \rho_{00}\rightarrow \pm \ii\delta\rho_{00}$) the path integral still carries an overall $\pm \ii$ phase. This has to do with the chosen path integral operator normalization in~\eqref{eq:lambda+1 PI}, and when the correlator is expressed in terms of the Liouville primaries~\eqref{eq:primariy_holo_Liouville} this imaginary factor will disappear as we elaborate in the next subsection. 

In addition to the above, the $l\geq 1$ for $\delta Y$, the $l \neq 1$ modes for $\delta \rho$ and the non-zero modes for the fermions combine to their respective one-loop determinants (\ref{eq:bosonGaussianN2}) and (\ref{fermionGaussianN2}). As before, keeping track of the zero modes we obtain
\begin{equation}\label{eq: AMV fluctuations}
   \begin{aligned}
       &\big\langle \bar \psi \psi \e^{\frac{3\fb}{\sqrt{2}}\rho + \ii \frac{\fb}{\sqrt{2}}Y}(0) \e^{\sqrt{2}\fb \rho} (1) \e^{\sqrt{2}\fb \rho} (\infty)\big\rangle \simeq   \sum_{s\geq 0}\e^{\frac{7}{\sqrt{2}} \fb \rho_*^{(s)}}\e^{-S^{\n=2}_\text{L}[\rho_*^{(s)}]}\mathcal{Z}'^{\,\text{bos}}_{\mathrm{1-loop}}\mathcal{Z}'^{\,\text{fer}}_{\mathrm{1-loop}}\cr
   &\times \int\!\! \frac{-1}{(4\pi)^2}\frac{|\mu|^2}{\mu^2}\d^4 \boldsymbol c \int \!\!\prod_{m= 0,\pm 1}\d\delta\rho_{1m}\d \delta Y_{00}\,\e^{\ii\frac{\fb}{\sqrt{2}}\delta Y_{00}} \delta \bpsi \delta\psi\, \e^{-\frac{\ii \fb}{\sqrt{2}}\delta Y_{00}}\int \d^2z\sqrt{\tilde{g}} (-1)^s \frac{-1}{4\pi}\frac{\mu^*}{2|\mu|r}\delta \btpsi \delta\tpsi \, \langle\!\!\langle\,...\rangle\!\!\rangle~,
   \end{aligned}
\end{equation}
where we performed the integral over the $d_1,\ldots, d_4$ Grassmann variables.
Compared to (\ref{eq: AMP PI final}) we also brought down the Yukawa coupling in $\langle\!\!\langle\,\rangle\!\!\rangle$ and ignored the $-1$ as it does not survive the $\delta Y_{00}$ integral. The terms in $\langle\!\!\langle\,\rangle\!\!\rangle$ are subleading. Furthermore, we note that the $(-1)^s$ from the Yukawa vertex combines with the saddle (\ref{eq: on shell and saddles}) to $(-1)^s(-1)^s =1$. Again, bringing down the Yukawa vertex is necessary to create a non-vanishing $\delta Y_{00}$ integral. 
Our task for (\ref{eq: AMV fluctuations}) is now to evaluate the Grassmann integral. Compared to the AMP case one of the insertions already carries non-trivial fermion number, hence by bringing down the Yukawa vertex we now have a two-point function. To get a non-trivial $\boldsymbol{c}$ integral the fermions sitting downstairs in (\ref{eq: AMV fluctuations}) need to be in their $l=0$ sector. Using (\ref{eq: f for fermions}) with $f(z)$=1 we obtain
\begin{equation}\label{eq:fermion zero mode integrals AMV}
    \int \dd^2 z\sqrt{\tilde{g}} \delta\bar{\tilde{\psi}}\delta\tilde\psi = -2\ii r \frac{\mu^2}{|\mu|^2} (c_1 c_2+ c_3 c_4)~,\quad \delta\bar \psi \delta\psi = \frac{-\ii}{2\pi r} (c_1 c_2 + c_3 c_4 + \ldots)~.
\end{equation}
where $\ldots$ indicates terms that will not survive the Grassmann integral. This leads to
\begin{equation}
    -\frac{1}{(4\pi)^2}\int\frac{|\mu|^2}{\mu^2}\d^4 \boldsymbol c\, \delta \bpsi \delta\psi \int \dd^2 z\sqrt{\tilde{g}} \delta\bar{\tilde{\psi}}\delta\tilde\psi = \frac{8}{(4\pi)^3}~.
\end{equation}
Combining (\ref{eq:fermion zero mode integrals AMV}) and (\ref{eq: AMV fluctuations}), and suppressing the residual flat-direction integrals we arrive at
\begin{equation}\label{eq:PI AMV +1}
\big\langle \bar \psi \psi \e^{\frac{3\fb}{\sqrt{2}}\rho + \ii \frac{\fb}{\sqrt{2}}Y}(0) \e^{\sqrt{2}\fb \rho} (1) \e^{\sqrt{2}\fb \rho} (\infty)\big\rangle \simeq  \pm\ii\frac{\sqrt{2\pi}\e^{\gamma_E}}{128\pi^4} \frac{\mu^{\frac{1}{\fb^2}-4}(\mu^*)^{\frac{1}{\fb^2}-3}\e^{\frac{1}{\fb^2} + \left(\frac{1}{\fb^2}-\frac{7}{2}\right)\log \fb^2}}{\sin\big(\frac{\pi}{\fb^{2}}\big)} r^{\frac{c}{3}-8}\Lambda_{\mathrm{uv}}^{\frac{1}{2}}~.
\end{equation}
As a consistency check, we indeed obtain the correct KPZ scaling (\ref{eq:Liouville_mu,mu*-dependence}). 
We again leave a higher-loop extension to future work~\cite{ThemisBeatrixpart2}.
\\ \\
Furthermore we note that to obtain the correlator
\begin{equation}\label{eq:lambda-1 PI}
    \langle \bar{\tilde{\psi}} \tilde\psi \e^{\frac{3\fb}{\sqrt{2}}\rho - \ii \frac{\fb}{\sqrt{2}}Y} (0)\e^{\sqrt{2}\fb \rho}(1) \e^{\sqrt{2}\fb \rho}(\infty)\rangle \ ,
\end{equation}
which corresponds to a correlator that violates angular momentum by $-1$ unit, the calculation is basically analogous, but instead of the Yukawa coupling $\delta \bar{\tilde{\psi}}\delta\tilde{\psi} \e^{\frac{\fb}{\sqrt{2}}\delta\rho - \ii\frac{\fb}{\sqrt{2}}\delta Y}$ we have to bring down $\delta \bar \psi\delta{\psi} \e^{\frac{\fb}{\sqrt{2}}\delta\rho + \ii\frac{\fb}{\sqrt{2}}\delta Y}$. This exchanges the powers of $\mu$ and $\mu^*$ compared to (\ref{eq:PI AMV +1}) and leads to
\begin{equation}\label{eq:PI AMV -1}
   \langle \bar{\tilde{\psi}} \tilde\psi \e^{\frac{3\fb}{\sqrt{2}}\rho - \ii \frac{\fb}{\sqrt{2}}Y}(0) \e^{\sqrt{2}\fb \rho} (1)\e^{\sqrt{2}\fb \rho}(\infty)\rangle \simeq  \pm\ii\frac{\sqrt{2\pi}\e^{\gamma_E}}{128\pi^4} \frac{\mu^{\frac{1}{\fb^2}-3}(\mu^*)^{\frac{1}{\fb^2}-4}\e^{\frac{1}{\fb^2} + \left(\frac{1}{\fb^2}-\frac{7}{2}\right)\log \fb^2}}{\sin\big(\frac{\pi}{\fb^{2}}\big)} r^{\frac{c}{3}-8}\Lambda_{\mathrm{uv}}^{\frac{1}{2}}~.
\end{equation}

\subsection{Semiclassical limit of the AMV structure constant}
We now compare the previous discussion to a semiclassical analysis of the AMV structure constant (\ref{eq:Liouville_AMV_structure_constant}). 
We consider three point functions~\eqref{eq:lambda+1 PI},~\eqref{eq:lambda-1 PI} with the corresponding Liouville primaries'~\eqref{eq:primariy_holo_Liouville} parameters given by eqs.~\eqref{eq:AMV_insertions_1},~\eqref{eq:AMV_insertions_2}.

For the structure constant with $\sum_i\lambda_i=+1$, we obtain 
(\ref{eq:Liouville_AMV_structure_constant})
\begin{equation}\label{eq:Cp1}
    \begin{aligned}
        \langle V_{2\fb,\fb}^{-1}(0)V_{\fb,\fb}^0 (1) V_{\fb,\fb}^0(\infty)\rangle  &= \mathsf{A}(\fb) \left(\frac{\mu}{2}\right)^{\frac{1}{\fb^2}-4}\left(\frac{\mu^*}{2}\right)^{\frac{1}{\fb^2}-3} \big(\fb^{2-2\fb^2}\gamma(\fb^2)\big)^{\frac{1}{\fb^2}-\frac{7}{2}}\fb^{5-\frac{2}{\fb^2}}\gamma\left(3-\frac{1}{\fb^2}\right) \frac{\gamma(\fb^2-1)\gamma^2(\fb^2)}{\gamma(3\fb^2-1)}~\cr
    &\simeq 128\e^{-2\gamma_E} \mathsf{A}(\fb)\frac{\mu^{\frac{1}{\fb^2}-4}(\mu^*)^{\frac{1}{\fb^2}-3}\e^{\frac{1}{\fb^2} + \left(\frac{1}{\fb^2}-\frac{11}{2}\right)\log \fb^2}}{\sin\big(\frac{\pi}{\fb^{2}}\big)}\left(\frac{\e}{4}\right)^{\frac{1}{\fb^2}}\big(1+\mathcal{O}(\fb^2)\big) \ ,
    \end{aligned}
\end{equation}
where in the second line we expanded around $\fb\to 0$. Similarly, for the AMV correlator with $\sum_i\lambda_i = -1$, we have
\begin{equation}\label{eq:Cp2}
    \begin{aligned}
        \langle V_{\fb,2\fb}^{1}(0)V_{\fb,\fb}^0 (1) V_{\fb,\fb}^0(\infty)\rangle &=\mathsf{A}(\fb) \left(\frac{\mu}{2}\right)^{\frac{1}{\fb^2}-3}\left(\frac{\mu^*}{2}\right)^{\frac{1}{\fb^2}-4} \big(\fb^{2-2\fb^2}\gamma(\fb^2)\big)^{\frac{1}{\fb^2}-\frac{7}{2}}\fb^{5-\frac{2}{\fb^2}}\gamma\left(3-\frac{1}{\fb^2}\right) \frac{\gamma(\fb^2-1)\gamma^2(\fb^2)}{\gamma(3\fb^2-1)}~\cr
    &\simeq 128\e^{-2\gamma_E} \mathsf{A}(\fb)\frac{\mu^{\frac{1}{\fb^2}-3}(\mu^*)^{\frac{1}{\fb^2}-4}\e^{\frac{1}{\fb^2} + \left(\frac{1}{\fb^2}-\frac{11}{2}\right)\log \fb^2}}{2\sin\big(\frac{\pi}{\fb^{2}}\big)}\left(\frac{\e}{4}\right)^{\frac{1}{\fb^2}}\big(1+\mathcal{O}(\fb^2)\big)
    \end{aligned}
\end{equation}
Comparing to the semiclassical path integral expressions~\eqref{eq:PI AMV +1},~\eqref{eq:PI AMV -1}, we observe the following: the KPZ scaling matches, and after using the sphere-to-disk mapping (\ref{eq:map disk sphere}) so do the exponentials. The sum over saddles in the PI is also correctly captured by the structure constants. The power of $\fb$ is not uniquely fixed by our approach and so we again absorb its ambiguity into $\mathsf{A}(\fb)$. It would be interesting to fix this by performing a Faddeev-Popov analysis. Lastly, we remark that when comparing the correlators $C^{\pm 1}$ with the semiclassical path integral (\ref{eq:PI AMV +1}) and (\ref{eq:PI AMV -1}) we notice an overall imaginary unit discrepancy. This is accounted for by our operator normalization. More specifically, from our conventions (\ref{eq:conventions SUSY}) we infer that $\bar\psi \psi = -2\ii \psi_1 \psi_2 \sim -2\ii \e^{-\ii (K_L+K_R)}$. So the operator $:\bpsi \psi \e^{2\fb \varphi + \fb \tilde\varphi}:$ equals $-2\ii$ times the canonically normalized vertex operators $V_{2\fb,\fb}^{-1}$ and $V^{+1}_{\fb,2\fb}$.

\section{$\frac 12$-BPS correlators and chiral rings}\label{sec:BPS_chiral_rings}

Two-dimensional unitary theories with $\n=2$ supersymmetry typically possess BPS operators saturating the BPS bound $h=|q_R|/2$ (see appendix~\ref{ap:N=2_SCFTs_Basics} for a review). These operators exhibit rich structure, for instance they organize into \textit{chiral rings}~\cite{Lerche:1989uy,Eguchi:1990vz,Witten:1991zz,Witten:1988xj}. In the spectrum of $\n=2$ Liouville theory there exist such BPS states and we will study some of their properties in this section.

Recall from the supercoset discussion in subsection~\ref{sub:supercoset_construction} that the discrete representations of $\mathrm{sl}(2,\mathbb{R})$ correspond to degenerate representations of the $\n=2$ superconformal algebra (see appendix~\ref{ap:degen_reps}). Amongst them, the unflowed ones with $j=|m|$ and $j=|\bar m|$ yield $\frac 12$-BPS states. In particular, in terms of Liouville parameters~\eqref{eq:Liouville-supercoset_dictionary}, a holomorphic chiral and anti-chiral primary obeys, respectively, $\talpha(\fQ-\alpha)+\frac{\eta(\eta-1)}{2}=0$ and $\alpha(\fQ-\talpha)+\frac{\eta(\eta+1)}{2}=0$,\footnote{These conditions for shorter multiplets can also be deduced directly on the Liouville side from the OPE of the primaries with the $G^\pm$ supercurrents~\eqref{eq:Liouville stress tensor and currents}.} and similarly for the anti-holomorphic components. In the full theory we have four types of $\frac 12$-BPS primary operators, see \eg the discussion in appendix~\ref{ap:N=2_SCFTs_Basics}, which are summarized in Table~\ref{tab:BPS_primaries} for the NS-sector where $\eta=\bar\eta=0$.\footnote{As already mentioned, the NS (BPS) primaries with $\eta,\bar\eta\neq0$ can be obtained via integer spectral flow.}    
\begin{table}[ht]
\centering
\small
\setlength{\arrayrulewidth}{.7pt}
\renewcommand{\arraystretch}{1.75}

\begin{tabular}{|c!{\vrule width 1.3pt}c!{\vrule width 1.3pt}c|}
\hline

\rule{0pt}{3.5ex}
BPS primaries
& Conditions
& Vertex operators \\

\hhline{|=|=|=|}

\rule{0pt}{3.5ex}
$(c,c)$
& $\tilde{\alpha}=\bar{\tilde{\alpha}}=0~,\ \ \alpha=\bar{\alpha}$
& $\mathsf V_{\alpha}^{(c,c)}=\e^{\alpha\varphi}$ \\

\hhline{|=|=|=|}

\rule{0pt}{3.5ex}
$(c,a)$
& $\tilde{\alpha}=\bar{\alpha}=0~,\ \ \alpha=\bar{\tilde{\alpha}}$
& $\mathsf V_{\alpha}^{(c,a)}=\e^{\alpha(\varphi_L+\tvarphi_R)}$ \\

\hhline{|=|=|=|}

\rule{0pt}{3.5ex}
$(a,c)$
& $\alpha=\bar{\tilde{\alpha}}=0~,\  \ \tilde{\alpha}=\bar{\alpha}$
& $\mathsf V_{\tilde{\alpha}}^{(a,c)}=\e^{\tilde{\alpha}(\tvarphi_L+\varphi_R)}$ \\

\hhline{|=|=|=|}

\rule{0pt}{3.5ex}
$(a,a)$
& $\alpha=\bar{\alpha}=0~,\ \  \tilde{\alpha}=\bar{\tilde{\alpha}}$
& $\mathsf V_{\tilde{\alpha}}^{(a,a)}=\e^{\tilde{\alpha}\tvarphi}$ \\

\hline
\end{tabular}

\caption{$\frac 12$-BPS primaries of $\mathcal{N}=2$ Liouville theory in the NS-sector ($\eta=\bar{\eta}=0$). The type of the BPS primary is identified from the parenthesis, \eg $(c,a)$ implies that the primary is chiral in the holomorphic sector and anti-chiral in the anti-holomorphic sector.}
\label{tab:BPS_primaries}

\end{table}
In writing down the explicit expressions for the vertex operators we leveraged the constraint $\alpha+\talpha=\bar\alpha+\bar\talpha$, which further forces these BPS operators to be spinless. Note that for the discrete representations of (the universal cover of) $\mathrm{SL}(2,\mathbb R)$ $j$ has to be real and in the interval $\frac 12<j<\frac{\fk+1}{2}$, which translates into $1<p+\tilde p<1+\fb^2$ for the Liouville parameters $\alpha=p\fb^{-1},\talpha=\tilde p\fb^{-1}$. This implies that, in the semiclassical $\fb\to 0$ limit of the theory, the spectrum does not contain any BPS states.\footnote{Relatedly, recall that in the semiclassical analysis of the cigar CFT (discussed in subsection~\ref{subsec:cigar_target}) these states were indeed absent. This is to be expected since discrete representations correspond to bound states of the theory.} Nonetheless, as usual such semiclassical correlators can be obtained formally via analytic continuation in $\alpha,\talpha$ away from the spectrum.

As is well known~\cite{Lerche:1989uy}, $\n=2$ BPS primaries of the same type form \textit{chiral rings} under fusion at coincident points. We will focus on the \textit{chiral--chiral} $(c,c)$ ring, being mirror dual to the $(c,a)$ supercoset ring, albeit similar considerations apply to the remaining chiral rings of the theory. The multiplication of the ring is given by the non-singular OPE of $(c,c)$ primaries,
\begin{equation}\label{eq:def_chiral_ring}
   \mathsf V^{(c,c)}_{\alpha_i}\mathsf V^{(c,c)}_{\alpha_j}=\sum _{\alpha_k}\mathsf C_{ij}^k    \mathsf V^{(c,c)}_{\alpha_k} \ ,
\end{equation}
where $\mathsf C_{ij}^k$ are the chiral ring coefficients. Matching dimensions and R-charges imposes $\alpha_i+\alpha_j=\alpha_k$. One can evaluate the chiral ring coefficients via correlators with $(a,a)$ primaries, as
\begin{equation}\label{eq:C_ijk-relation}
    \big\langle \mathsf V^{(c,c)}_{\alpha_i}(0)\mathsf V^{(c,c)}_{\alpha_j}(1)\mathsf V^{(a,a)}_{\talpha_{\tilde k}}(\infty)\big\rangle=\sum _{\alpha_k}\mathsf C_{ij}^k\big\langle \mathsf V^{(c,c)}_{\alpha_k}(0)\mathsf V^{(a,a)}_{\talpha_{\tilde k}}(\infty)\big\rangle \ .
\end{equation}
The two-point function on the rhs is (up to a normalization factor) the Zamolodchikov metric~\cite{Cecotti:1991me}, which in our case is diagonal in the $\alpha_k,\talpha_{\tilde k}$ indices due to R-charge conservation. Notice that since we are not inserting the BPS primaries at the same point in~\eqref{eq:C_ijk-relation} their OPE on the rhs will generically contain super-descendants, whose contribution however gets killed once we evaluate their two-point function with $\mathsf V^{(a,a)}(\infty)$. More specifically, given any super-descendant $\mathcal{O}$ due to the state-operator map we have $\big\langle \mathsf V^{(a,a)}_{\talpha_{\tilde k}}(\infty) \mathcal{O}(0)\big\rangle=\big\langle \mathsf V^{(a,a)}_{\talpha_{\tilde k}}\big| \mathcal{O}\big\rangle=0$, since $L_0,\bar L_0,J_0,\bar J_0$ are Hermitian and thus states with different scaling dimension and R-charge are orthogonal.

Let us now explicitly evaluate the coefficients $\mathsf C_{ij}^k$ through~\eqref{eq:C_ijk-relation}, by computing the corresponding two and three-point functions. In fact, as already anticipated from the discussion after eq.~\eqref{eq:supercoset_2pt_func}, both correlators will exhibit LSZ poles and we will evaluate their residue. To begin, recall that \eg R-charge conservation demands $\alpha_i+\alpha_j=\talpha_{\tilde k}$ for the above three-point function, which for the BPS primaries in question coincides with the angular momentum preserving constraint appearing in the structure constant~\eqref{eq:N=2-Liouville_structure_constant}. Accordingly, the $D(\alpha_i,\talpha_i)$ piece in~\eqref{eq:N=2-Liouville_structure_constant} reads,
\begin{equation}
D(\alpha_i,0,\alpha_j,0,0,\tilde\alpha_{\tilde k};0)={A}(\fb)\left(\frac{|\mu|^2}{4}\fb^{2-2\fb^2}\gamma(\fb^2)\right)^{\frac{1}{\fb^2} -\frac{\tilde\alpha_{\tilde k}}{\fb}}\frac{\Upsilon_\fb(\fb)\Upsilon_\fb(\talpha_{\tilde k})}{\Upsilon_\fb(\talpha_{\tilde k}-\fb^{-1})\Upsilon_\fb\big(\frac 12 (\alpha_i+\alpha_j-\talpha_{\tilde k})\big)} \ ,
\end{equation}
which exhibits a pole due to the vanishing of $\Upsilon_\fb\big(\frac 12 (\alpha_i+\alpha_j-\talpha_{\tilde k})\big)$ in the denominator, whose residue is
\begin{equation}
    \begin{aligned}
    &\Res_{\fb(\alpha_i+\alpha_j)=\fb\talpha_{\tilde k}}D(\alpha_i,0,\alpha_j,0,0,\tilde\alpha_{\tilde k};0)=2\fb A(\fb)\left(\frac{|\mu|^2}{4}\fb^{2-2\fb^2}\gamma(\fb^2)\right)^{\frac{1}{\fb^2} -\frac{\tilde\alpha_{\tilde k}}{\fb}}\frac{\Upsilon_\fb(\talpha_{\tilde k})}{\Upsilon_\fb(\talpha_{\tilde k}-\fb^{-1})}\\
        &\qquad\qquad\qquad\qquad\qquad\quad=2A(\fb)\left(\frac{|\mu|^2}{4}\fb^{2-2\fb^2}\gamma(\fb^2)\right)^{\frac{1}{\fb^2} -\frac{\tilde\alpha_{\tilde k}}{\fb}} \fb^{\frac{2\tilde\alpha_{\tilde k}}{\fb} -\frac{2}{\fb^2}}\gamma\left(\fb^{-1}(\talpha_{\tilde k}-\fQ)\right) \ ,
    \end{aligned}
\end{equation}
where we used that $\mathrm{Res}_{x=0}\Upsilon_\fb^{-1}(x) =\Upsilon_\fb^{-1}(\fb)$, which follows from the third identity in eq.~\eqref{eq:Ypsilon_shift_identities}. The $\mathcal W$ integral of eq.~\eqref{eq:N=2-Liouville_structure_constant} in this case becomes,
\begin{equation}
    \begin{aligned}
        \mathcal{W}_\fb&\equiv \int_{\mathbb{C}^2}\d^2y\d^2z\,|y|^{2(\fb\alpha_j-1)}|z|^{-2}|1-z|^{-2\fb\alpha_i}|y-z|^{-2\fb\alpha_j}\\
        &=\int_\mathbb{C}\d^2u\,|u|^{2(\fb\alpha_j-1)}|1-u|^{-2\fb\alpha_j}\int_\mathbb{C}\d^2z\,|z|^{-2}|1-z|^{-2\fb\alpha_i}\\
        &=\pi^2\Gamma(\tfrac\sigma 2)\Gamma(\tfrac\tau 2) \ ,
    \end{aligned}
\end{equation}
where in the second equality we used the change of variables $y\to uz$, while in the third the formula for the complex Beta integral~\eqref{eq:complex_Beta_int} and regularized the poles of the integrals with $\sigma=-\fb\varepsilon,\tau=\fb\varepsilon$, where $\varepsilon=\alpha_i+\alpha_j-\talpha_{\tilde k}$. This quantity is also divergent, with residue $\Res_{\sigma=0,\tau=0} \mathcal{W}_\fb=4\pi^2$. The above imply the following residue for the three-point function,
\begin{equation}\label{eq:3pt_res}
    \Res_{\fb(\alpha_i+\alpha_j)=\fb\talpha_{\tilde k}}\big\langle \mathsf V^{(c,c)}_{\alpha_i}\mathsf V^{(c,c)}_{\alpha_j}\mathsf V^{(a,a)}_{\talpha_{\tilde k}}\big\rangle=8\pi^2{A}(\fb)\left(\frac{|\mu|^2}{4}\fb^{2-2\fb^2}\gamma(\fb^2)\right)^{\frac{1}{\fb^2} -\frac{\tilde\alpha_{\tilde k}}{\fb}} \fb^{\frac{2\tilde\alpha_{\tilde k}}{\fb} -\frac{2}{\fb^2}}\gamma\left(\fb^{-1}(\talpha_{\tilde k}-\fQ)\right) \ .
\end{equation}
Similarly we can compute the residue of the two-point correlator. In particular, from the Liouville two-point function~\eqref{eq:Liouville_2-pt-function} including the $m$-basis dependence from the $\mathcal{W}$-integral (recall also~\eqref{eq:supercoset_2pt_func}), we find 
\begin{equation}
    \big\langle \mathsf V^{(c,c)}_{\alpha_k}\mathsf V^{(a,a)}_{\talpha_{\tilde k}}\big\rangle=\frac{4\pi^2}{\fb} A(\fb)\left(\frac{|\mu|^2}{4\fb^2}\fb^{2-2\fb^2}\gamma(\fb^2)\right)^{\frac{1}{\fb^2} -\frac{\tilde\alpha_{\tilde k}}{\fb}}\gamma\left(\fb^{-1}(\talpha_{\tilde k}-\fQ)\right)\Gamma\Big(\frac \delta 2\Big) \ ,
\end{equation}
where we introduced the regulator $\delta=\fb(\alpha_k-\talpha_{\tilde k})$ for the LSZ pole. The corresponding residue is,
\begin{equation}\label{eq:2pt_res}
\begin{aligned}
    \Res_{\delta=0}\big\langle \mathsf V^{(c,c)}_{\alpha_k}\mathsf V^{(a,a)}_{\talpha_{\tilde k}}\big\rangle=\frac{8\pi^2}{\fb} A(\fb)\left(\frac{|\mu|^2}{4}\fb^{2-2\fb^2}\gamma(\fb^2)\right)^{\frac{1}{\fb^2} -\frac{\tilde\alpha_{\tilde k}}{\fb}}\fb^{\frac{2\talpha_{\tilde{k}}}{\fb}-\frac{2}{\fb^2}}\gamma\left(\fb^{-1}(\talpha_{\tilde k}-\fQ)\right) \ ,
\end{aligned}    
\end{equation}
where we used $\Res_{z=0}\Gamma(z)=1$. Putting everything together, we obtain the following chiral ring coefficient~\eqref{eq:C_ijk-relation},
\begin{equation}\label{eq:C_ijk-answer}
    \mathsf C_{ij}^k=\fb \,\delta_{\alpha_i+\alpha_j,\talpha_{\tilde k}} \ .
\end{equation}
Amusingly, since most of the factors in the correlators~\eqref{eq:3pt_res},~\eqref{eq:2pt_res} cancel, we observe that the chiral ring coefficient depends linearly on $\fb$ and, in particular, does not receive quantum corrections.\footnote{\label{ft:C_ijk-numerical_value}Nonetheless, apart from the particular operator normalization we are using for $\mathsf V^{(c,c)}_{\alpha}$, the precise numerical value of $\mathsf C_{ij}^k$ depends also on the convention for taking the residue of divergent correlation functions.} This rhymes nicely with the fact that chiral rings are protected topological subsectors of the full $\n=2$ SCFT~\cite{Eguchi:1990vz,Witten:1991zz}.

One could deduce this chiral ring coefficient also from the semiclassical path integral of the theory, as follows. We choose $\frac 12$-BPS vertex operators from Table~\ref{tab:BPS_primaries} with $\alpha_i,\talpha_{\tilde j}\sim\mathcal{O}(1)\,\,\text{or} \,\,\mathcal{O}(\fb)$, so that we do not disturb the complex saddles~\eqref{eq:complex_saddles}, and then evaluate eq.~\eqref{eq:C_ijk-relation} at tree-level, resulting in
\begin{equation}\label{eq:C_ijk-semiclassical}
    \mathsf C_{ij}^k=\frac{\big\langle\e^{\alpha_i\varphi}\e^{\alpha_j\varphi}\e^{\talpha_{\tilde k}\tvarphi}\big\rangle}{\big\langle\e^{\alpha_k\varphi}\e^{\talpha_{\tilde k}\tvarphi}\big\rangle}
    \simeq 
    \frac{\sum_{s\geq 0}\e^{\frac{\alpha_i}{\sqrt 2}\rho_\star^{(s)}}\e^{\frac{\alpha_j}{\sqrt 2}\rho_\star^{(s)}}\e^{\frac{\talpha_{\tilde k}}{\sqrt 2}\rho_\star^{(s)}}\e^{-S^{\n=2}_\text{L}[\rho_*^{(s)}]}}{\sum_{s\geq0}\e^{\frac{\alpha_k}{\sqrt 2}\rho_\star^{(s)}}\e^{\frac{\talpha_{\tilde k}}{\sqrt 2}\rho_\star^{(s)}}\e^{-S^{\n=2}_\text{L}[\rho_*^{(s)}]}}\simeq 1\ ,
\end{equation}
after imposing $\alpha_i+\alpha_j-\talpha_{\tilde k}=\alpha_k-\talpha_{\tilde k}=0$. The path integral result agrees at tree-level with the exact answer~\eqref{eq:C_ijk-answer}, recall the comment in footnote~\ref{ft:C_ijk-numerical_value}, which further predicts that the loop expansion in $\fb$ cancels exactly (up to the linear in $\fb$ term) between the numerator and denominator correlators in~\eqref{eq:C_ijk-semiclassical}.

\section{Discussion}

The main aim of this work has been to initiate a systematic study of the structure constants of spacelike $\n=2$ Liouville theory on $S^2$. This has been achieved by leveraging the mirror duality of the theory with the $\mathrm{SL}(2)_\fk/\mathrm{U}(1)$ supercoset model. In the angular momentum preserving and violating sectors, our proposal is that the structure constant is given by eqs.~\eqref{eq:N=2-Liouville_structure_constant} and~\eqref{eq:Liouville_AMV_structure_constant}, respectively. In section~\ref{sec:Liouville_semiclassical}, we tested this proposal against the semiclassical expansion of the $\n=2$ Liouville path integral for non-BPS three-point functions of light operators, finding detailed agreement up to one-loop order. In section~\ref{sec:BPS_chiral_rings}, we further studied the chiral-chiral ring of $\frac{1}{2}$-BPS operators and found that the resulting chiral ring coefficient reduces to a protected quantity consistent with semiclassical considerations. Our calculations provide further evidence in favor of the (already well-established) duality between $\n=2$ Liouville and the $\mathrm{SL}(2)_\fk/\mathrm{U}(1)$ supercoset~\cite{Hori:2001ax,Creutzig:2010bt}. We now offer some concluding thoughts on future directions.

\paragraph{$\n=2$ Liouville path integral}
In section~\ref{sec:Liouville_semiclassical} we discussed the semiclassical path integral of $\mathcal{N}=2$ Liouville theory in the angular momentum preserving and violating cases. Keeping track of the fermionic zero modes and performing a complete Faddeev-Popov (super)determinant analysis, it would be very interesting to examine the behavior of higher-loop corrections and, in particular, to determine which cancellations occur. We plan to present these issues in a separate work~\cite{ThemisBeatrixpart2}. Moreover, it would be interesting to understand how the situation changes if we no longer consider Liouville SCFT but Liouville quantum supergravity, where \eg  one must divide the volume of the supergroup $\mathrm{OSp}(2|2,\mathbb C)$~\cite{Anninos:2023exn}. Finally, from the supercoset model we inferred that the three-point function can violate winding number by at most one unit.~It is desirable to understand how this is captured on the Liouville side.

\paragraph{$\n=2$ analytic superconformal bootstrap}
As already mentioned, a successful strategy for obtaining the structure constants of Liouville theory with $\n=0,1$ supersymmetry has been Teschner's analytic (super)conformal bootstrap~\cite{Teschner:1995yf}. In the bosonic case, this approach offers a derivation of the DOZZ formula~\cite{Dorn:1994xn,Zamolodchikov:1995aa}, while in the $\n=1$ theory it similarly leads to the corresponding super-Liouville structure constant~\cite{Poghossian:1996agj,Belavin:2007gz,Muhlmann:2025ngz,Rangamani:2025wfa}. $\n=2$ Liouville theory is different, since it does not enjoy the celebrated $\fb$ to $\fb^{-1}$ self-dual symmetry, and therefore Teschner's argument cannot be applied in a straightforward way. Nevertheless, it would still be interesting to study the analytic properties of the resulting single shift equation and whether there exist further bootstrap-type constraints that can pin down the structure constant uniquely. In this regard, recall that in this work we made extensive use of the $H^+_3$ model structure constants that were derived using the conformal bootstrap~\cite{Teschner:1997ft}. This theory does not exhibit any type of self-dual symmetry; nonetheless, due to the Kač-Moody symmetry of the theory, the additional ingredient is provided by the KZ equation which, combined with the existence of Virasoro degenerate representations, is sufficient to ensure a unique solution of the crossing shift equations. One could imagine a version of this super-conformal bootstrap applied explicitly to the supercoset via combining the super-KZ equation with the $\n=2$ degenerate representations reviewed in appendix~\ref{ap:degen_reps}. It would also be intriguing to study whether such an approach can be implemented directly on the Liouville side. Since these theories contain operators with spin, the standard analytic bootstrap will have to be generalized, perhaps along the lines of~\cite{Migliaccio:2017dch}.

Ultimately, one would also like to prove crossing symmetry of the $\n=2$ Liouville structure constants~\eqref{eq:N=2-Liouville_structure_constant},~\eqref{eq:Liouville_AMV_structure_constant}. Since the four-point functions of the $H^+_3$ model are known to be crossing symmetric~\cite{Teschner:2001gi}, a simple argument in that direction might be possible, at least in the BPS subsector of the theory in which $\n=2$ superconformal blocks are simply related to Virasoro conformal blocks~\cite{Lin:2016gcl}.

\paragraph{Toward $\n=2$ timelike Liouville theory} 
Thus far, we have considered $\n=2$ Liouville theory with central charge $c>3$. Using the standard arguments implemented for bosonic Liouville, as presented \eg in ref.~\cite{Ribault:2015sxa}, we expect that the structure constants~\eqref{eq:N=2-Liouville_structure_constant},~\eqref{eq:Liouville_AMV_structure_constant} will remain valid under analytic continuation for arbitrary complex values of the central charge $c\in\mathbb C$ -- except in the timelike regime $c\leq 3$ (recall figure~\ref{fig:figure intro}). This particular regime is not accessible via analytic continuation (similarly to the $\n=0,1$ cases), because \eg infinitely many poles associated with degenerate fields cross the integration contour in the conformal block decomposition of the four-point function.\footnote{Despite this, there appear to be notable exceptions \cite{Anninos:2021ene}. In the explicit AMP and AMV examples studied here, the structure constants simplify to ordinary Gamma functions, see \eg eqs.~(\ref{eq:AMP Dbb}), (\ref{eq:Cp1}), and (\ref{eq:Cp2}), closely resembling the exceptional cases identified in \cite{Anninos:2021ene}.} Nevertheless, in bosonic Liouville theory it was observed that the DOZZ formula can be analytically continued to rational values of the central charge in the bosonic timelike regime $c\leq 1$~\cite{Runkel:2001ng,McElgin:2007ak}. For instance, continuing the (spacelike) DOZZ formula to $c=1$ one arrives at the Runkel–Watts theory~\cite{Runkel:2001ng}. The $\n=2$ analogue of this would be to continue the central charge to $c=3$, \ie to take $\fb,\fk\to\infty$. This corresponds to the strongly coupled regime of Liouville theory, but also to the strict classical limit of strings propagating on the cigar geometry. Therefore, in the dual description the theory reduces in this limit to the classical phase space of strings on the cigar; investigating the dual Liouville behavior at that point offers a promising (and preliminary) window into $\n=2$ timelike Liouville.

To our knowledge, further properties of $\n=2$ timelike Liouville theory have not been studied in the literature.\footnote{Apart from the loop expansion of the sphere path integral discussed in~\cite{Anninos:2023exn}.} Since our approach here relied heavily on the mirror duality between $\n=2$ Liouville and the supercoset~\cite{Hori:2001ax}, one would have to examine carefully whether an extension of the original Hori-Kapustin arguments to the timelike regime exists, possibly yielding a dual description of $\n=2$ timelike Liouville. Overall, it would be particularly interesting to study the timelike theory, both from a SCFT, as well as a solvable model of dS$_2$ quantum (super)gravity point of view.

\paragraph{More supersymmetry?} Another natural generalization of the present work is to consider Liouville theory with more supersymmetry. For the specific value $\fb=\sqrt 2 $, the central charge of $\n=2$ Liouville becomes $c=6$ and the theory describes an $A_1$ singularity of an ALE space~\cite{Eguchi:1988vra,Eguchi:2000cj,Eguchi:2004ik}. In turn, this implies that the symmetry of the theory is enhanced from $\n=2$ to the small $\n=4$ superconformal algebra (at level $1$). Away from this point little is known about $\n=4$ Liouville theory, see however ref.~\cite{Eguchi:2016cyh}, offering an interesting arena to explore. (See also ref.~\cite{Johnson:2024tgg} for an $\n=4$ JT supergravity perspective.)

\paragraph{The super-Schwarzian limit} In ref.~\cite{Mertens:2017mtv}, it was shown
that the Schwarzian theory arises as the large-central charge limit of Liouville theory between two ZZ-branes, with Schwarzian correlators of local operators obtained by combining the $\fb\to 0$ limit of the ZZ-brane wavefunctions with the Liouville structure constants. This was shown explicitly in~\cite{Mertens:2017mtv} for the cases with $\n=0,1$ supersymmetry. In addition, the $\n=2$ density of states was determined in~\cite{Mertens:2017mtv,Turiaci:2023jfa}, while various $\n=2$ Schwarzian correlators were computed by alternative methods in~\cite{Lin:2022zxd}. It would be interesting to combine the $\n=2$ ZZ-brane boundary states of~\cite{Eguchi:2003ik,Ahn:2004qb} with the bulk structure constants discussed here, and check that both the spectrum of~\cite{Turiaci:2023jfa} and the
correlators of~\cite{Lin:2022zxd} are recovered in the small-$\fb$ limit of $\n=2$ Liouville theory.\footnote{We thank J. Turiaci for a captivating discussion on this point.}

\section*{Acknowledgments}
It is a great pleasure to acknowledge useful interactions with Pietro Benetti Genolini, Andreas Blommaert, Lorenz Eberhardt, Lisa Fleischer, Gregory Kafanelis, Grigoria Kiosse, Manki Kim, Thomas Mertens, Juan Maldacena, Victor Rodriguez, Bruno Sivilotti, Joaquin Turiaci, Edward Witten and Xi Yin. We are especially grateful to Al.~Zamolodchikov for correspondence, and to Dionysios Anninos and Davide Gaiotto for stimulating discussions.
We used OpenAI's ChatGPT $5.5$ and ChatGPT Pro, and Anthropic's Claude Opus $4.8$ and Fable 5, for help with various computations and explanations.

B.M. acknowledges fruitful discussions during the workshop ``Observers, wormholes and complex saddles in cosmology", organized at the Bernoulli Center for Fundamental Studies (EPFL, Lausanne) from 18--22 May 2026. B.M. gratefully acknowledges funding provided by the Leinweber foundation at the Institute for Advanced Study and the National Science Foundation with grant number PHY-2514611. Research at Perimeter Institute is supported in part by the Government of Canada through the Department of Innovation, Science and Economic Development Canada and by the Province of Ontario
through the Ministry of Colleges and Universities.

\appendix

\section{Special functions}\label{ap:special_functions}

The Barnes double gamma function $\Gamma_b(w)$ has the following integral representation,
\begin{equation}
    \log\Gamma_b(w)= \int_{\mathbb{R_+}}\frac{\d s}{s} \left(\frac{\e^{-ws}-\e^{-\frac{1}{2}(b+b^{-1})s}}{(1-\e^{-bs})(1-\e^{-b^{-1}s})}-\frac{1}{2}\Big(\frac{1}{2}(b+b^{-1})-w\Big)^2\e^{-s}-\frac{\frac{1}{2}(b+b^{-1})-w}{s}\right) \ .
\end{equation}
As a function of $w$, $\Gamma_b(w)$ is meromorphic, has no zeroes and poles at $w=-rb-sb^{-1},\, r,s\in\mathbb{N}$. The Ypsilon function of Liouville theory is then defined to be,
\begin{equation}\label{eq:ypsilon_from_double_gamma}
    \Upsilon_b(w)\equiv\frac{1}{\Gamma_b(w)\Gamma_b(b+b^{-1}-w)} \ ,
\end{equation}
with the corresponding integral representation
\begin{equation}
    \log \Upsilon_b(w)=\int_{\mathbb{R}_+}\frac{\d s}{s}\Bigg(\Big(\frac{1}{2}(b+b^{-1})-w\Big)^2\e^{-s}-\frac{\sinh^2{\big[\big(\frac{1}{2}(b+b^{-1})-w\big)\frac{s}{2}\big]}}{\sinh{\frac{bs}{2}}\sinh{\frac{b^{-1}s}{2}}}\Bigg) \ .
\end{equation}
Accordingly, the Ypsilon function $\Upsilon_b(w)$ is an entire function of $w$ (for $\text{Re}\,b>0$) and 
\begin{equation}\label{eq:Y-zeros}
    \mathrm{zeros~of}~\Upsilon_b(w): \quad w=-rb-sb^{-1},\,w=(r+1)b+(s+1)b^{-1}~,\quad r,s\in \mathbb{N}~.
\end{equation}
Moreover, the Ypsilon function obeys the following shift identities,
\begin{equation}\label{eq:Ypsilon_shift_identities}
    \begin{aligned}
        &\Upsilon_b(w+b)=\gamma(bw)\,b^{1-2bw}\,\Upsilon_b(w) \ , \\ &\Upsilon_b(w+b^{-1})=\gamma(b^{-1}w)\,b^{-1+2b^{-1}w}\,\Upsilon_b(w) \ , \\
        &\Upsilon_b(w-b)=\gamma^{-1}(bw-b^2)\,b^{2bw-2b^2-1}\,\Upsilon_b(w) \ , \\
        &\Upsilon_b(w-b^{-1})=\gamma^{-1}(b^{-1}w-b^{-2})\,b^{-2b^{-1}w+2b^{-2}+1}\,\Upsilon_b(w) \ ,
    \end{aligned} 
\end{equation}
where as usual $\gamma(w)=\frac{\Gamma(w)}{\Gamma(1-w)}$, as well as exhibits self-dual symmetry under $b\leftrightarrow b^{-1}$, \ie
\begin{equation}
    \Upsilon_{b^{-1}}(w)=\Upsilon_b(w) \ .
\end{equation}
Another useful identity, which follows \eg from eq.~\eqref{eq:ypsilon_from_double_gamma}, is
\begin{equation}
    \Upsilon_b(w)=\Upsilon_b(b+b^{-1}-w) \ ,
\end{equation}
which in particular implies $\Upsilon_b(b^{-1})=\Upsilon_b(b)$.

\section{Integral bestiary}\label{app:bestiary}
In this appendix we evaluate various integrals. A useful integral throughout this discussion is the complex Beta integral,
\begin{equation}\label{eq:complex_Beta_int appendix}
    \int_{\mathbb{C}} \d^2u\,
u^{\alpha-1}\bar u^{\bar\alpha-1}
(1-u)^{\beta-1}(1-\bar u)^{\bar\beta-1}
=\pi
\frac{\Gamma(\alpha)\Gamma(\beta)\Gamma(1-\bar\alpha-\bar\beta)}
{\Gamma(1-\bar\alpha)\Gamma(1-\bar\beta)\Gamma(\alpha+\beta)} \ .
\end{equation}

\paragraph{Integral \ref{eq:J_integral}}
We evaluate the integral
\begin{equation}
\mathrm{WNP}_{\mathcal{J}}:\quad \mathcal J_{\mathrm{VNP}}(j_i,m_i,\bar m_i)
\equiv
\int \prod_{i=1}^3 \d^2x_i\,
x_i^{j_i+m_i-1}\bar x_i^{j_i+\bar m_i-1}\,
|x_{12}|^{-2j_{123}}|x_{23}|^{-2j_{231}}|x_{13}|^{-2j_{312}} \ .
\end{equation}
 To do so, we implement the change of variables,
\begin{equation}\label{eq:J_integral_variables_change}
\begin{cases}
    x_1=x \ ,\\
x_2=xy \ ,\\
x_3=xz \  \ ,
\end{cases}\Rightarrow \ 
\begin{cases}
    \d^2x_1 \d^2x_2 \d^2x_3=|x|^4 \d^2x \d^2y \d^2z \ ,\\
    x_{12}=x(1-y) \ , \\
x_{13}=x(1-z) \ , \\
x_{23}=x(y-z) \ ,
\end{cases}
\end{equation}
and we get
\begin{equation}
\begin{aligned}
\mathcal J_{\mathrm{VNP}}(j_i,m_i,\bar m_i)
&=
\int_{\mathbb{C}^3} \d^2x \d^2y \d^2z\,
|x|^4
\frac{x^{a_1}(xy)^{a_2}(xz)^{a_3} 
\bar x^{\bar a_1}(\bar x\bar y)^{\bar a_2}(\bar x\bar z)^{\bar a_3}}{
|x|^{2(j_{123}+j_{231}+j_{312})} 
|1-y|^{2j_{123}}|1-z|^{2j_{312}}|y-z|^{2j_{231}}} \ ,
\end{aligned}
\end{equation}
but since $j_{123}+j_{231}+j_{312}=j_1+j_2+j_3$, one can define $M_{123}\equiv m_1+m_2+m_3,\,
\bar M_{123}\equiv\bar m_1+\bar m_2+\bar m_3$ and re-write the integral as
\begin{equation}\label{eq:some_interm_int}
\mathcal J_{\mathrm{VNP}}(j_i,m_i,\bar m_i)
=
\left(
\int_\mathbb{C} \d^2x\,x^{M_{123}-1}\bar x^{\bar M_{123}-1}
\right)\,
\mathcal{W}(j_i,m_i,\bar m_i) \ ,
\end{equation}
for
\begin{equation}
\mathcal W(j_i,m_i,\bar m_i)
=
\int_{\mathbb{C}^2} \d^2y \d^2z\,
y^{a_2}\bar y^{\bar a_2}
z^{a_3}\bar z^{\bar a_3}
|1-y|^{-2j_{123}}|1-z|^{-2j_{312}}|y-z|^{-2j_{231}} \ .
\end{equation}
The first integral in~\eqref{eq:some_interm_int} was evaluated previously in eq.~\eqref{eq:delta_Mellin}, yielding
\begin{equation}
\mathcal J_{\mathrm{VNP}}(j_i,m_i,\bar m_i)
=
\tilde{\delta\,}\Big(\sum_{i=1}^3m_i\Big)\,
\mathcal W(j_i,m_i,\bar m_i) \ .
\end{equation}

\paragraph{Integral \ref{eq:main-text_WNV_int}} The integral
\begin{equation}
    \mathrm{WNV}_{\mathcal{J}}^{-1}:~ \mathcal{J}^{-1}_{\mathrm{WNV}}(j_i,m_i,\bar m_i) = \int_{\mathbb C^3}\prod_{i=1,3,4}\d^2x_i\,x_i^{j_i+m_i-1}\bar x_i^{j_i+\bar m_i-1}(x_{43}+zx_{31})^r(\bar x_{43}+\bar z\bar x_{31})^r  \ ,
\end{equation}
can be evaluated as follows; first set,
\begin{equation}\label{eq:variable_change_integral}
    \begin{cases}
        x_1=\rho u \ ,\\
        x_3=\rho v \ ,\\
        x_4=\rho \ .
    \end{cases}
    \Rightarrow 
    \begin{cases}
        \d^2x_1\d^2x_3\d^2x_4=|\rho|^4\d^2\rho \d^2u\d^2v \ ,\\
        x_{43}+zx_{31}=\rho\big(1+(z-1)v-zu\big) \ ,\\
    \end{cases}
\end{equation}
from which the integral becomes,
\begin{equation}
\int_\mathbb{C}\d^2\rho\,|\rho|^4\rho^{a_1+a_3+a_4+r}\bar\rho^{\bar a_1+\bar a_3+\bar a_4+r}\int_{\mathbb C^2}\d^2u\d^2v\, u^{a_1}\bar u^{\bar a_1}v^{a_3}\bar v^{\bar a_3} \big(1+(z-1)v-zu\big)^r\big(1+(\bar z-1)\bar v-\bar z\bar u\big)^r \ ,
\end{equation}
where again $a_i=j_i+m_i-1$. The exponents in the $\rho$ integral can be massaged to give,
\begin{equation}
   a_1+a_3+a_4+r=m_1+m_3+m_4+\frac{\fk+2}{2} -3 \ , 
\end{equation}
hence, from eq.~\eqref{eq:delta_Mellin}, this integral evaluates to
\begin{equation}\label{eq:BC_deltas}
    \int_\mathbb{C}\d^2\rho\,\rho^{m_1+m_3+m_4+\frac{\fk+2}{2}-1}\bar\rho^{\bar m_1+\bar m_3+\bar m_4+\frac{\fk+2}{2}-1}=\tilde\delta\Big(m_1+m_3+m_4+\frac{\fk+2}{2} \Big) \ .
\end{equation}
The second $u,v$ integral above can be split into a product of two complex Beta integrals;
define 
\begin{equation}
    L(v)\equiv 1+(z-1)v\rightarrow1+(z-1)v-zu=L(v)-zu \ ,
\end{equation}
implying that the second integral can be written as,
\begin{equation}
    \int_{\mathbb C^2}\d^2u\d^2v\, u^{a_1}\bar u^{\bar a_1}v^{a_3}\bar v^{\bar a_3}(L-zu)^r(\bar L-\bar z \bar u)^r \ .
\end{equation}
Setting, now,
\begin{equation}
    t=\frac{zu}{L} \ \rightarrow \ L-zu=L(1-t) \ , \ \d^2u=\Big|\frac L z \Big|^2\d^2t \ ,
\end{equation}
the $u$ integral becomes,
\begin{equation}
    \int_{\mathbb C^2}\d^2u\, u^{a_1}\bar u^{\bar a_1} (L-zu)^r(\bar L-\bar z\bar u)^r=z^{-a_1-1}\bar z^{-\bar a_1-1}L^{a_1+r+1}\bar L^{\bar a_1+r+1}\underbrace{\int_\mathbb{C}\d^2t\,t^{a_1}\bar t^{\bar a_1}(1-t)^r(1-\bar t)^r}_{B^1_\mathbb{C}} \ ,
\end{equation}
which is a complex Beta integral that will be explicitly evaluated shortly. The remaining $v$ integral is then
\begin{equation}
    z^{-a_1-1}\bar z^{-\bar a_1-1}B^1_\mathbb{C}\int_\mathbb{C}\d^2v\, v^{a_3}\bar v^{\bar a_3}\big(1+(z-1)v\big)^{a_1+r+1}\big(1+(\bar z-1)\bar v\big)^{\bar a_1+r+1} \ .
\end{equation}
This last integral can be computed via setting,
\begin{equation}
    s= -(z-1)v \ \rightarrow \ 1+(z-1)v=1-s \ , 
\end{equation}
yielding
\begin{equation}
    (z-1)^{-a_3-1}(\bar z-1)^{-\bar a_3-1}\underbrace{\int_\mathbb{C}\d^2s\,s^{a_3}\bar s^{\bar a_3}(1-s)^{a_1+r+1}(1-\bar s)^{\bar a_1+r+1}}_{B^2_\mathbb{C}} \ ,
\end{equation}
namely another complex Beta integral. Both complex Beta integrals can be evaluated using~\eqref{eq:complex_Beta_int appendix}, with respectively $\alpha=a_1+1,\,\beta=r+1$ and $\alpha=a_3+1,\,\beta=a_1+r+2$, resulting in
\begin{equation}\label{B1B2}
    \begin{aligned}
B^1_\mathbb{C}B^2_\mathbb{C}&=\left(\pi\frac{\Gamma(j_1+m_1)\Gamma(r+1)\Gamma(-j_1-\bar m_1-r)}{\Gamma(1-j_1-\bar m_1)\Gamma(-r)\Gamma(j_1+m_1+r+1)}\right)\left(\pi\frac{\Gamma(a_3+1)\Gamma(a_1+r+2)\Gamma(-\bar a_1-\bar a_3-r-2)}{\Gamma(-\bar a_3)\Gamma(-a_1-r-1)\Gamma(a_1+a_3+r+3)}\right)\\
        &=\pi^2 \frac{1}{\gamma\big(j_1+j_3+j_4-\frac{\fk+2}{2} \big)}\prod_{i=1,3,4}\frac{\Gamma(j_i+m_i)}{\Gamma(1-j_i-\bar m_i)}   \ ,
    \end{aligned}
\end{equation}
where in the second equality we went on-shell of the delta constraints~\eqref{eq:BC_deltas}.

\paragraph{Integral \ref{eq:def_wnv_+1}} The goal is to compute the following integral transform of eq.~\eqref{eq:def_wnv_+1}. This procedure is very similar to the evaluation of eqs.~\eqref{eq:def_wnv_-1},~\eqref{eq:main-text_WNV_int}, hence our presentation in this paragraph will be brief. After taking the $x_2,\bar x_2\to 0$ limit and through a similar change of variables as in~\eqref{eq:variable_change_integral}, eq.~\eqref{eq:def_wnv_+1} contains a $\rho$ integral of the form,
\begin{equation}
    \int_\mathbb{C}\d^2\rho\,\rho^{m_1+m_3+m_4-\frac{\fk+2}{2}-1}\bar\rho^{\bar m_1+\bar m_3+\bar m_4-\frac{\fk+2}{2} -1}=\tilde\delta\Big(m_1+m_3+m_4-\frac{\fk+2}{2}\Big) \ ,
\end{equation}
which again is simply the statement that the degenerate field carries winding number $w=-1$. Setting further
\begin{equation}
    P\equiv m_1+j_3+j_4-j_2 \ , \ Q\equiv m_3+j_1+j_4-j_2 \ , \ R\equiv m_4+j_1+j_3-j_2 \ ,
\end{equation}
and employing similar changes of variables as before, we arrive at the $u$ integral 
\begin{equation}
    \int_{\mathbb C}\d^2v\,v^{Q-1}\bar v^{\bar Q-1}(L+Mv)^r(\bar L+\bar M\bar v)^r=L^{Q+r}\bar L^{\bar Q+r}M^{-Q}\bar M^{-\bar Q}\underbrace{\int_\mathbb{C}d^2s\,s^{Q-1}\bar s^{\bar Q-1}(1-s)^r(1-\bar s)^r}_{\tilde B^1_\mathbb{C}} \ ,
\end{equation}
and, correspondingly, at the $v$ integral
\begin{equation}
    (1-z)^{Q+r}(1-\bar z)^{\bar Q+r}z^{P+r}\bar z^{\bar P+r}\underbrace{\int_\mathbb{C}\d^2t\,t^{P+Q+r-1}\bar t^{\bar P+\bar Q+r-1}(1-t)^{-Q}(1-\bar t)^{-\bar Q}}_{\tilde B^2_\mathbb{C}} \ .
\end{equation}
Evaluating the complex Beta integrals through~\eqref{eq:complex_Beta_int appendix}, we get
\begin{equation}
    \tilde B^1_\mathbb{C}\tilde B^2_\mathbb{C}=\frac{\pi^2}{\gamma\big(j_1+j_3+j_4-\frac{\fk+2}{2}\big)}\prod_{i=1,3,4}\frac{\Gamma(j_i-m_i)}{\Gamma(1-j_i+\bar m_i)} \ ,
\end{equation}
which is precisely what we had before, after sending $m\mapsto-m$. In arriving at this relation we made use of the Gamma function property $\Gamma(z)\Gamma(1-z)=\frac{\pi}{\sin{(\pi z)}}$ and also that, since $Q-\bar Q=m_3-\bar m_3\in\mathbb Z$, it is true that $\sin{(\pi \bar Q)}=(-1)^{Q-\bar Q}\sin{(\pi Q)}$. The remaining $z$-dependence combines into
\begin{equation}
     f_{-m}(z_i)= z_{21}^{m_1}z_{23}^{m_3}z_{42}^{m_4}z_{43}^{\hat h_1+\hat h_2-\hat h_3-\hat h_4+m_1}z_{41}^{\hat h_2+\hat h_3-\hat h_1-\hat h_4+m_3}z_{31}^{\hat h_2+\hat h_4-\hat h_1-\hat h_3+m_4} \ .
\end{equation}
Combining the above, one arrives at~\eqref{eq:ful_m_basis_wnv+1_sl2}.

\section{Review of the $H^+_3$ model}\label{ap:H^+_3_model}

Here we review aspects of the bosonic $H^+_3$ WZNW model at level $\fk+2$, following~\cite{Teschner:1997ft,Teschner:1999ug,Teschner:2001gi,Maldacena:2001km}.

\subsection{Conventions and unflowed correlation functions}

First, we remark that Teschner's original discussion of the model~\cite{Teschner:1997ft,Teschner:1999ug,Teschner:2001gi} concerned solely unflowed $\mathrm{sl}(2)_{\fk+2}$ representations. Having said that, note that $H^+_3\simeq\mathrm{SL}(2,\mathbb{C})/\mathrm{SU}(2)$, namely the target space is Euclidean AdS$_3$. As a diagonal $2$-dimensional CFT, $H^+_3$ enjoys a $\mathrm{sl}(2)_{\fk+2} \times\overline{\mathrm{sl}(2)}_{\fk+2}$ current symmetry algebra. It has an action of the form,
\begin{equation}
    S_{H^+_3}=(\fk+2)\int \d^2 z\,\Big(\partial\phi\bar{\partial}\phi+\e^{2\phi}\partial\beta\bar{\partial}\bar{\beta}\Big) \ ,
\end{equation}
where $\beta$ parametrizes the complex planes $\mathbb{C}$ that foliate $H^+_3$ along the radial direction $\phi$. 

Affine primary operators $\Phi_j$ in the $H^+_3$ model are labeled by the Casimir eigenvalue $j$ having conformal dimensions $\hat h=\bar{\hat{h}}=-\frac{j(j-1)}{\fk}$,\footnote{Note that the Casimir eigenvalue $j$ used here differs from the conventions in~\cite{Teschner:1997ft,Teschner:1999ug} as $j_\text{here}=-j_\text{there}$.} and, in the isospin $x$ coordinate representation, have the following OPEs with the currents
\begin{equation}
    j^a(z) \Phi_j(x;w)\sim-\frac{D_x(t^a)\Phi_j(x;w)}{z-w} \ ,
\end{equation}
namely the currents act on the primaries as differential operators, with\footnote{Note that this $\pm,0$ basis is different from the $\pm,3$ basis we used in constructing the supercoset in section~\ref{sec:supercoset}.}
\begin{equation}\label{eq:sl(2)_diff-op}
    D^-=\partial_x \ , \ D^+=x^2\partial_x +2jx \ , \ D^0 =x\partial_x+j \ .
\end{equation}
In this basis, the $\mathrm{sl}(2)_{\fk+2}$ current commutation relations read
\begin{equation}\label{eq:current_commutators}
    \begin{aligned}
        [j^0_n,j^\pm_m]&=\pm j^\pm_{n+m}\ , \\
        [j^+_n,j^-_m]&=-2j^0_{n+m}+(\fk+2)n\delta_{n+m,0} \ ,\\
        [j^0_n,j^0_m]&=-\frac{\fk+2}{2} n \delta_{n+m,0} \ ,
    \end{aligned}
\end{equation}
while from the Sugawara construction, the modes of the stress-tensor are built out of the current modes as,
\begin{equation}\label{eq:Squgawara_sl2}
    L_n=\frac{\kappa_{ab}}{2\fk}\sum_{m\in\mathbb Z}:j^a_{n-m}j^b_m:=\frac{1}{2\fk
    }\sum_{m\in\mathbb Z}:\big(-2j^0_{n-m}j^0_m+j^+_{n-m}j^-_m+j^-_{n-m}j^+_m\big):\ .
\end{equation}
Moreover, since we will use it later, note that the corresponding current Ward identity reads,
\begin{equation}\label{eq:current_Ward_identity}
    \big\langle(j^a_{-1}\Phi_{j})(x;z))\prod_{j\neq i} \Phi_{j_j}(x_j;z_j)\big\rangle=\sum_{i}\frac{D^a_{j_i}}{z_i-z}\big\langle \Phi_{j}(x;z)\prod_i \Phi_{j_i}(x_i;z_i)\big\rangle \ ,
\end{equation}
where from now on we will abbreviate $D_{x_j}(t^a)$ as $D^a_{j}$.

Semiclassically, the affine primaries $\Phi_j(x;z)$ can be thought of as wavefunctions  on EAdS$_3$,\footnote{For simplicity, we are suppressing the dependence of the primaries $\Phi_j$ on the anti-holomorphic coordinates.} 
\begin{equation}\label{eq:Phi-j_wavefunction}
    \begin{aligned}
        &\Phi_j(x;z)\propto\big(|\beta-x|^2\e^{\phi}+\e^{-\phi}\big)^{-2j} \ ,
    \end{aligned}
\end{equation}
where $z$ is a worldsheet coordinate (\eg $\phi=\phi(z,\bar z)$ etc.) and $x$ is an auxiliary complex variable that can be thought of as parameterizing the boundary of EAdS$_3$. The structure constants of the theory are known. In particular, the three-point functions read~\cite{Teschner:1997ft,Ribault:2014hia}
\begin{equation}\label{eq:Phij_3-point_func}
\big\langle
\Phi_{j_1}(x_1;z_1)\Phi_{j_2}(x_2,;z_2)\Phi_{j_3}(x_3;z_3)
\big\rangle=\frac{D(j_1,j_2,j_3)}{
|z_{12}|^{2\hat h_{123}}
|z_{23}|^{2\hat h_{231}}
|z_{13}|^{2\hat h_{312}}
|x_{12}|^{2j_{123}}
|x_{23}|^{2j_{231}}
|x_{13}|^{2j_{312}}} \ ,
\end{equation}
where the structure constants are\footnote{In refs.~\cite{Teschner:1997ft,Teschner:1999ug} the structure constants were given in terms of a function $G(j)$, that is related to the Ypsilon function as $G(j)=b^{-b^2j^2+j(b^2+1)}\,\Upsilon^{-1}_b(j)$, where $b^2=\fk^{-1}$.}~
\begin{equation}\label{eq:H3+_D(j1,j2,j3)}
    D(j_1,j_2,j_3)={A}(\fk^{\frac{1}{2}}) {f}(\fk^{\frac{1}{2}})^{\frac{1}{\fk}(1-\tilde j)}\frac{\Upsilon_{\sqrt{\fk}}(\fk^{-\frac{1}{2}})\prod_{i=1}^3\Upsilon_{\sqrt{\fk}}(2\fk^{-\frac{1}{2}}j_i)}{\Upsilon_{\sqrt{\fk}}(\fk^{-\frac{1}{2}}(\tilde{j \,}-1))\prod_{\text{cyc}}\Upsilon_{\sqrt{\fk}}(\fk^{-\frac{1}{2}}j_{ijk})} \ ,
\end{equation}
where $f(\fk^\frac{1}{2})\equiv\fk^{1-\fk}\gamma(\fk)$, $\tilde{j \,}=j_1+j_2+j_3$, $j_{ijk}=j_i+j_j-j_k$ and the cyclic product is taken over cyclic permutations of $i,j,k\in\lbrace{1,2,3\rbrace}$.
Taking the (distributional) limits $j_3\rightarrow 0$ and $j_3\rightarrow 1$, respectively, we infer
\begin{subequations}
    \begin{align}
        \lim_{j_3\rightarrow 0} D(j_1,j_2,j_3)&=2\pi  {A}(\fk^{\frac{1}{2}})\big(\fk^{-1}{f}(\fk^{\frac{1}{2}})\big)^{\frac{1-2j_1}{\fk}}\gamma\left(\frac{2j_1-1}{\fk}\right)\delta(j_1-j_2)~, \\
        \lim_{j_3\rightarrow 1} D(j_1,j_2,j_3)&=\pi {A}(\fk^{\frac{1}{2}}) (\fk^{-1}{f(\fk^{\frac{1}{2}}}))^{-\frac{1}{\fk}}\gamma(\fk^{-1})\delta(j_1+j_2-1)~.
    \end{align}
\end{subequations}
Therefore, the corresponding two-point function takes the form,
\begin{equation}\label{eq:H3+_2pt-func}
    \big\langle\Phi_{j_1}(x_1;z_1)\Phi_{j_2}(x_2;z_2)\big\rangle=\frac{1}{|z_{12}|^{4\hat h_{1}}}\left[ T(\fk)\delta^{(2)}(x_{12})\delta(j_1+j_2-1)+\frac{B(j_1)}{|x_{12}|^{4j_1}}\delta(j_1-j_2)\right] \ ,
\end{equation}
for
\begin{equation}\label{eq:B(j),T(k)_relations}
    B(j)=2\pi  \mathrm{A}(\fk^{\frac{1}{2}})\big(\fk^{-1}{f}(\fk^{\frac{1}{2}})\big)^{\frac{1-2j}{\fk}}\gamma\left(\frac{2j-1}{\fk}\right) \ , \quad {T}(\fk)=\pi {A}(\fk^{\frac{1}{2}}) (\fk^{-1}{f}(\fk^{\frac{1}{2}}))^{-\frac{1}{\fk}}\gamma(\fk^{-1}) \ ,
\end{equation}
and, hence, the reflection coefficient reads
\begin{equation}\label{eq:R(j)_relations}
    R(j)=\frac{B(j)}{ T(\fk)}=2\big(\fk^{-1} f(\fk^{\frac{1}{2}})\big)^\frac{2-2j}{\fk}\frac{\gamma\left(\frac{2j-1}{\fk}\right)}{\gamma(\fk^{-1})} \ .
\end{equation}
A more careful definition of the reflection coefficient actually requires the shadow transform
\begin{equation}
    \Phi_{j}(x;z)=(2j-1)R(j)\int\d^2x'\,|x-x'|^{-4j}\Phi_{1-j}(x';z) \ ,
\end{equation}
which yields the reflection identity $R(j)R(1-j)=1$ up to $j$-independent factors.

Note that the structure constants~\eqref{eq:H3+_D(j1,j2,j3)} were obtained for the continuous representations of $\mathrm{sl}(2,\mathbb{R})$, namely when $j=\frac{1}{2}+i\mathbb{R}_+$. These are the only representations in the spectrum of the $H^+_3$ model. The correlation functions for the discrete representations can be obtained from the same formula via analytic continuation in $j$, as was explained \eg in~\cite{Maldacena:2001km}. For the spectrally flowed representations of $\mathrm{sl}(2)_{\fk+2}$ the situation is more complicated, since these are generally not highest-weight representations (recall discussion around eq.~\eqref{eq:bosonic_Sugawara_flowed_generatros}). This is an issue because Teschner derived the structure constants through analytic conformal bootstrap methods, by for instance leveraging the Knizhnik–Zamolodchikov (KZ) equations for affine primaries~\cite{Teschner:1997ft}. Nonetheless, it was shown in ref.~\cite{Ribault:2005ms} that the KZ equations for spectrally flowed representations are equivalent to the ordinary KZ equations when spectral flow is preserved or violated by one unit. It is in this sector where a simple relation between correlators of flowed and unflowed primaries exists~\cite{Ribault:2005ms,Maldacena:2001km}.

We emphasize, lastly, that the operator normalization that we use here is different from the one that is typically used in discussions of string theory on AdS$_3$, with the two being related schematically by the reflection coefficient \ie $\Phi_j^\text{there}(x;z)\sim R(j)\Phi_j^\text{here}(x;z)$. We choose the different normalization here to make contact with the semiclassical expansion of the $\n=2$ Liouville theory path integral.

\subsection{Constraints on winding number violation}
Here we review the argument given in ref.~\cite{Maldacena:2001km} regarding the $\mathrm{sl}(2)_{\fk+2}$ representation theory constraints on winding number violation of correlation functions. (An argument using the Coulomb gas formalism was discussed in~\cite{Giribet:2001ft}.)

Consider the general bosonic $\mathrm{sl}(2)_{\fk+2}$ state
\begin{equation}
    \ket \Psi\equiv \prod_{i=1}^{N_c}\Phi_{\mathcal C}^{w_i}\prod_{j=1}^{N_+}\Phi_{\mathcal D^+}^{w_j}\prod_{k=1}^{N_-}\Phi_{\mathcal D^-}^{w_k} \ket 0 \ ,
\end{equation}
where $\Phi^{w_i}_{\mathcal C,\mathcal D^\pm}\in\big\lbrace\mathcal C^{\alpha,w}_j,\mathcal D^{\pm.w}_j\big\rbrace$, namely are continuous or discrete primary spectrally flowed representations, acting on the non-normalizable vacuum $\ket 0$ of the theory. Then, take the following operators
\begin{equation}
    \begin{aligned}
        &j_P\equiv\oint\frac{\d z}{2\pi\ii}\prod_{i=1}^{N_c}(z-z_i)^{-w_i+1}\prod_{j=1}^{N_+}(z-z_j)^{-w_j+1}\prod_{k=1}^{N_-}(z-z_k)^{-w_k}j^+(z)=j_a^++c_1j^+_{a-1}+c_2j^+_{a-2}+\dots \ ,\\&j_N\equiv\oint\frac{\d z}{2\pi\ii}\prod_{i=1}^{N_c}(z-z_i)^{w_i+1}\prod_{j=1}^{N_+}(z-z_j)^{w_j}\prod_{k=1}^{N_-}(z-z_k)^{w_k+1}j^-(z)=j^-_b+\tilde c_1j^-_{b-1}+\tilde c_2j^-_{b-2}+\dots \ ,
    \end{aligned}
\end{equation}
for $a=-\sum_{i=1}^{N-1}w_i+N_c+N_+,\,b=\sum_{i=1}^{N-1}w_i+N_c+N_-,\,(N-1)=N_c+N_++N_-$. Using~\eqref{eq:currents_spectral-flow}, it is straightforward to show that both $j_P,j_N$ annihilate $\Psi$. We can further decompose $\Psi$ into a (direct) sum of affine representations with definite spectral flow $\mathrm w$, schematically as
\begin{equation}\label{eq:Phi_w_deco}
    \ket \Psi=\sum_{R,\mathrm w} \ket{\Psi^{\mathrm{w}}_R} \ ,
\end{equation}
with the representation $R\in\big\lbrace\mathcal{\hat{ C}^{\alpha,\mathrm{w}}_j},\mathcal{\hat{ D}^{\pm.\mathrm{w}}_j}\big\rbrace$. Therefore, it must be that $j_P\Psi^{\mathrm{w}}_R=j_N\Psi^{\mathrm{w}}_R=0$, for all $\mathrm w$ appearing in the decomposition~\eqref{eq:Phi_w_deco}. Suppose, now, that there exists a state in this decomposition with $\mathrm w\leq -a,-a-1$ (there are two possibilities, respectively, depending on whether the state $\ket{\Psi^\mathrm{w}_R}$ $\in$ $\mathcal{\hat{C}}^\mathrm{\alpha,w}_j,\mathcal{\hat{D}}^{+,\mathrm{w}}_j$ or $\mathcal{\hat{D}}^{-,\mathrm{w}}_j$). Then, take the state with the lowest conformal dimension at fixed $j^3_0$, denoted by $\ket h$. From earlier, we have 
\begin{equation}
    j_P\ket{\Psi^\mathrm{w}_R}=j^+_a\ket h+c_1j^+_{a-1}\ket h+\dots=0 \ ,
\end{equation}
and since the various terms in the sum have different dimensions the only possibility is $j^+_a\ket h=0$, as this contribution cannot be canceled against these other terms. Since $j^+_a$ is a creation operator for $\ket h$ and there are no null states in the physical spectrum of the $\mathrm{SL}(2)_{\fk+2}$ WZW model, we have arrived at a contradiction. Hence, we conclude that $\mathrm{w}\geq -a+1,-a$. 

Similarly, one can obtain constraints from the condition $j_N\Psi=0$, which result in the upper bounds $w\leq b-1,b$ depending on whether $\Psi^\mathrm{w}_R\in\mathcal C^\mathrm{\alpha,w}_j,\mathcal D^{-,\mathrm{w}}_j$ or $\mathcal{D}_j^{+,\mathrm{w}}$, respectively. Overall, we have the following constraints on the decomposition~\eqref{eq:Phi_w_deco}
\begin{equation}\label{eq:WNV_constraints_Psi}
    \begin{aligned}
        &\mathcal C_j:\qquad-N_+-N_c+1\leq \mathrm w -\sum_{i=1}^{N-1}w_i\leq N_-+N_c-1 \ ,\\&\mathcal D^+_j:\qquad-N_+-N_c+1\leq \mathrm w -\sum_{i=1}^{N-1}w_i\leq N_-+N_c \ ,\\&\mathcal D^-_j:\qquad-N_+-N_c\leq \mathrm w -\sum_{i=1}^{N-1}w_i\leq N_-+N_c-1 \ .
    \end{aligned}
\end{equation}
To deduce a constraint on $N$-point correlation functions, consider the setup
\begin{equation}\label{eq:correlator_w_violation_rules}
    \bra 0 \Phi^{w_N}_{R_N} \ket \Psi \ ,
\end{equation}
where $\Phi^{w_N}_{R_N}$ should be thought of as a conjugate primary, with conjugation acting on the representations as $(\mathcal C^{\alpha,w}_j)^\dagger=\mathcal C^{-\alpha,-w}_j,(\mathcal D^{\pm,w}_j)^\dagger=\mathcal D^{\mp,-w}_j$. A necessary condition for the correlator~\eqref{eq:correlator_w_violation_rules} to be non-vanishing is $-w_N=\mathrm w$, namely there should exist a representation $R_N^\dagger$ with spectral flow $-w_N$ in the decomposition~\eqref{eq:Phi_w_deco}. Combining this with the constraints~\eqref{eq:WNV_constraints_Psi}, one finds that $N$-point functions can fail to preserve winding number only up to $N-2$ units (see~\cite{Maldacena:2001km} for a more careful treatment).

\subsection{Null state equation for the degenerate field $\Phi_{\frac{\fk+2}{2}}(x;z)$}

The primary $\Phi_{\frac{\fk+2}{2}}(x;z)$ is degenerate, because
\begin{equation}
   \langle j|j^-_1j^+_{-1}|j\rangle =\langle j|j^+_{-1}j^-_{1}+2j^0_0+\fk+2|j\rangle=-2j+\fk+2 \ .
\end{equation}
Note that, since $D^0=x\partial_x+j$, the relation used above $j^0_0|j\rangle=-j|j\rangle$ is valid only at $x=0$. To get the null state condition for general $x$ we can utilize the corresponding translation operator, which since $D^-=\partial_x$ is simply $\e^{-xj^-_0}$. This implies
\begin{equation}\label{eq:BCH}
    \e^{-xj^-_0}j^+_{-1}\e^{xj^-_0}=j^+_{-1}-x[j^-_0,j^+_{-1}]+\frac 12 x^2\big[j^-_0,[j^-_0,j^+_{-1}]\big]+\dots=j^+_{-1}-2xj^0_{-1}+x^2j^-_{-1} \ ,
\end{equation}
where we used the BCH lemma, the commutators~\eqref{eq:current_commutators} and that the nested commutators in the dots vanish because $[j^-_0,j^-_{-1}]=0$. Hence, the null state condition becomes
\begin{equation}\label{eq:null_state_before_BCH}
    (j^+_{-1}-2xj^0_{-1}+x^2j^-_{-1})\Phi_{\frac{\fk+2}{2}}(x;z) = 0 \ .
\end{equation}
The next step is to turn~\eqref{eq:null_state_before_BCH} into a differential equation for the four-point function with three arbitrary affine primary insertions and a degenerate $\Phi_{\frac{\fk+2}{2}}$ insertion. Specifically, combining eqs.~\eqref{eq:null_state_before_BCH},~\eqref{eq:BCH} with the Ward identity~\eqref{eq:current_Ward_identity}, we have
\begin{equation}
    \begin{aligned}
        \big\langle(j^+_{-1}-2xj^0_{-1}&+x^2j^-_{-1})\Phi_{\frac{\fk+2}{2}}(x;z)\prod_i\Phi_{j_i}(x_i;z_i)\big\rangle\\
        &=\sum_i\frac{D^+_i-2xD^0_i+x^2D^-_i}{z_i-z}\big\langle\Phi_{\frac{\fk+2}{2}}(x;z)\prod_i\Phi_{j_i}(x_i;z_i)\big\rangle=0 \ .
    \end{aligned}
\end{equation}
For the differential operators~\eqref{eq:sl(2)_diff-op} this last equation yields,
\begin{equation}
    \label{eq:null_state_diff_eq_for_correlator}
    \sum_i\frac{(x_i-x)^2\partial_{x_i}+2j_i(x_i-x)}{z_i-z}\big\langle\Phi_{\frac{\fk+2}{2}}(x;z)\prod_i\Phi_{j_i}(x_i;z_i)\big\rangle=0 \ .
\end{equation}

We would like to express~\eqref{eq:null_state_diff_eq_for_correlator} as a differential equation for the corresponding conformal block. For that, we choose the following parameterization for a general four-point function,
\begin{equation}\label{eq:G-4-pt-func}
    G\equiv\big\langle\Phi_{\frac{\fk+2}{2}}(x_2;z_2)\prod_{i=1,3,4}^4\Phi_{j_i}(x_i;z_i)\big\rangle=\big|P_z(z_i)\big|^2\big|P_x(x_i)\big|^2\frac{D(j_1,j_2,j_s)D(j_s,j_3,j_4)}{B(j_s)}\big|\mathcal{F}(x,z)\big|^2 \ ,
\end{equation}
where $j_2=\frac{\fk+2}{2}$, $B(j)$ is the two-point function normalization, the conformal cross ratios read $x=\frac{x_{21}x_{43}}{x_{31}x_{42}},\,z=\frac{z_{21}z_{43}}{z_{31}z_{42}}$, and
\begin{equation}
    \begin{aligned}
        &P_z(z_i)=z_{42}^{-2\hat h_2}z_{41}^{\hat h_2+\hat h_3-\hat h_1-\hat h_4}z_{43}^{\hat h_1+\hat h_2-\hat h_3-\hat h_4}z_{31}^{\hat h_4-\hat h_1-\hat h_2-\hat h_3} \ ,\\
        &P_x(x_i)=x_{42}^{-2j_2}x_{41}^{j_2+j_3-j_1-j_4}x_{43}^{j_1+j_2-j_3-j_4}x_{31}^{j_4-j_1-j_2-j_3} \ ,
    \end{aligned}
\end{equation}
where the $x_i,z_i$-dependence is fixed by the isometries of the target and the conformal invariance of the worldsheet, respectively. In writing~\eqref{eq:G-4-pt-func} we assumed that there is a single fusion channel in the $\Phi_{j_1}\Phi_{\frac{\fk+2}{2}}$ OPE, something that will be demonstrated later. In the limit where $x_1,z_1\to 0,\,x_2,z_2\to x,z,\,x_3,z_3\to 1,\,x_4,z_4\to\infty$, the four-point  function $G$ reduces to the conformal block $\big|\mathcal F(x;z)\big|^2$.

We now insert~\eqref{eq:G-4-pt-func} into~\eqref{eq:null_state_diff_eq_for_correlator}. As usual, the operator insertion at infinity is defined as
\begin{equation}
    \Phi_{j_4}(\infty;\infty)\equiv\lim_{x_4,z_4\to\infty}x_4^{2j_4}z_4^{2\hat h_4}\Phi_{j_4}(x_4;z_4) \ ,
\end{equation}
thus we can define the corresponding four-point function
\begin{equation}
    \hat G\equiv x_4^{2j_4}z_4^{2\hat h_4} G \ ,
\end{equation}
where we are going to suppress the anti-holomorphic dependence in the following since it is equivalent to the holomorphic one. Note also that,
\begin{equation}\label{eq:x_hat_lim}
    x\,\hat=\,x_2+\frac{x_2(x_2-1)}{x_4}+\mathcal{O}(x_4^{-2}) \ ,
\end{equation}
where from now on the hat symbol above equal signs denotes that we are setting $x_1=0,\,x_3=1$, while we further define
\begin{equation}
    a=j_1+j_2-j_3-j_4 \ , \
    b=-2j_2 \ , \
    c=j_2+j_3-j_1-j_4 \ , \
    d=j_4-j_1-j_2-j_3 \ ,
\end{equation}
from which we have 
\begin{equation}
    P_x(x_i)=x_{43}^ax_{42}^bx_{41}^cx^d_{31} \ ,
\end{equation}
and lastly
\begin{equation}
    \tilde D(j_1,j_3,j_4)\equiv \frac{D(j_1,j_2,j_s)D(j_s,j_3,j_4)}{B(j_s)} \ .
\end{equation}
Now, in the $x_4\to\infty $ limit, the numerator of the $i=4$ term in eq.~\eqref{eq:null_state_diff_eq_for_correlator} remains finite. To see this, we recall~\eqref{eq:x_hat_lim} and expand the conformal block for large-$x_4$ as,
\begin{equation}
    \mathcal{F}(x,z)=\mathcal{F}(x_2,z)+\frac{x_2(x_2-1)}{x_4}\partial_{x_2}\mathcal{F}(x,z)\big|_{x=x_2}+\dots \ ,
\end{equation}
from which the numerator of the fourth term in question becomes
\begin{equation}
    \begin{aligned}
        &(x_4-x_2)^2\Big(\partial_{x_4}-\frac{2j_4}{x_4}\Big)+2j_4(x_4-x_2)\hat G=\Bigg(\frac{2j_4}{x_4}x_2(x_4-x_2)+(x_4-x_2)^2\partial_{x_4}\Bigg)\hat G\\
        &\quad\hat = \,z_4^{2\hat h_4}P_z\tilde D\Bigg(\frac{2j_4}{x_4}x_2(x_4-x_2) + (x_4-x_2)^2\partial_{x_4}\Bigg)\Bigg(\mathcal F(x_2,z)+\frac{\lambda_1(x_2)}{x_4}+\mathcal{O}(x_4^{-2})\Bigg) \ .
    \end{aligned}
\end{equation}
Here $\lambda_1(x_2)$ is a function independent of $x_4$. This expression is finite for large-$x_4$ and the subsequent $z_4\to\infty$ limit makes the full term vanish, hence only the $i=1,3$ terms contribute. To evaluate them we have,
\begin{equation}\label{eq:partial_logPx}
    \partial_{x_1}x\,\hat =\,x-1 \ , \ \partial_{x_3}x\,\hat =\, -x \ , \ 
    \partial_{x_1}\log{P_x(x_i)}\,\hat =\, -d \ , \ \partial_{x_3}\log{P_x(x_i)}\,\hat = \,d \ ,
\end{equation}
implying then,
\begin{equation}
    \begin{aligned}
        &\partial_{x_1}\big(P_x(x_i)\mathcal F(x,z)\big)=P_x\big(\partial_{x_1}\log{P_x}\mathcal F+\partial_{x_1}x\partial_x\mathcal F\big)=P_x\big(-d\mathcal F+(x-1)\partial_x \mathcal F\big) \ ,\\
        &\partial_{x_3}\big(P_x(x_i)\mathcal F(x,z)\big)=P_x\big(\partial_{x_3}\log{P_x}\mathcal F+\partial_{x_3}x\partial_x\mathcal F\big)=P_x\big(d\mathcal F-x\partial_x \mathcal F\big) \ .
    \end{aligned}
\end{equation}
Due to these, the full null state equation~\eqref{eq:null_state_diff_eq_for_correlator} for the insertions at $0,1,\infty$, becomes
\begin{equation}
    \begin{aligned}
        &\qquad\qquad\quad \left(\frac{x^2\partial_{x_1}-2j_1x}{z}+\frac{(x-1)^2\partial_{x_3}-2j_3(x-1)}{z-1}\right)\big(P_x\mathcal F\big)=0\\
        &\quad\Rightarrow\quad  x(x-1)(x-z)\partial_x\mathcal F=\Big[d(x^2-2xz+z)+2j_1x(1-z)+2j_3z(1-x)\Big]\mathcal F \ .
    \end{aligned}
\end{equation}
A similar equation holds for the anti-holomorphic conformal block, hence solving for $\mathcal F(x,z)$ yields the full explicit expression for $|\mathcal F(x,z)|^2$.

\subsection{KZ equation for the degenerate field $\Phi_{\frac{\fk+2}{2}}(x;z)$}

In this subsection we will derive the Knizhnik–Zamolodchikov (KZ) equation for an $N$-point correlation function of $\Phi_{k/2}$ and $N-1$ primary operators $\Phi_j$.

To achieve that, we will follow the usual steps of deriving the KZ equation and then substitute the null state condition~\eqref{eq:BCH}. From the Sugawara construction~\eqref{eq:Squgawara_sl2}, we have
\begin{equation}
    \begin{aligned}
        \partial_z\Phi_j\equiv L_{-1}\Phi_j&=\frac{1}{2\fk}\sum_{m\in\mathbb{Z}}:\big(-2j^0_{-1-m}j^0_m+j^+_{-1-m}j^-_m+j^-_{-1-m}j^+_m\big):\Phi_j\\
        &=\frac{1}{\fk}\big(2j^0_{-1}(x\partial_x+j)-j^+_{-1}\partial_x-j^-_{-1}(x^2\partial_x+2jx)\big)\Phi_j \ ,
    \end{aligned}
\end{equation}
whereas taking a derivative of the null state condition~\eqref{eq:BCH},
\begin{equation}
    j^+_{-1}\partial_x\Phi_{\frac{\fk+2}{2}}(x;z)=(2j^0_{-1}+2xj^0_{-1}\partial_x-2xj^-_{-1}-x^2j^-_{-1}\partial_x)\Phi_{\frac{\fk+2}{2}}(x;z) \ ,
\end{equation}
and combining the two, we arrive at
\begin{equation}
    \partial_z\Phi_{\frac{\fk+2}{2}}(x;z)=(j^0_{-1}-xj^-_{-1})\Phi_{\frac{\fk+2}{2}}(x;z) \ .
\end{equation}
Using the Ward identity~\eqref{eq:current_Ward_identity} this can be translated to
\begin{equation}\label{eq:null_KZ_for_correlator}
    \partial_{z}G=\sum_i\frac{(x_i-x)\partial_{x_i}+j_i}{z_i-z}G \ ,
\end{equation}
where from before $G$ denotes four-point function~\eqref{eq:G-4-pt-func}.\footnote{This equation is in fact valid for any $N$-point function.}

We will now turn this into a differential equation for the conformal block.  Recalling the parameterization~\eqref{eq:G-4-pt-func}, we successively evaluate both sides of eq.~\eqref{eq:null_KZ_for_correlator}. The lhs reads,
\begin{equation}
    \partial_{z_2}G=\tilde DP_x\partial_{z_2}(P_z\mathcal F) \ , \ \text{where} \ \partial_{z_2}(P_z\mathcal F)=P_z(\partial_{z_2}\log{P_z}\mathcal F+\partial_{z_2}z\partial_z\mathcal F) \ .
\end{equation}
Moreover, it is simple to check that $\lim_{z_4\to\infty}\partial_{z_2}z\,\hat =\, 1$ and,
\begin{equation}
    P_z\,\hat =\, (z_4-1)^{\hat h_1+\hat h_2-\hat h_3-\hat h_4}(z_4-z_2)^{-2\hat h_2}z_4^{\hat h_2+\hat h_3-\hat h_1-\hat h_4}\rightarrow\lim_{z_4\to\infty}\partial_{z_2}\log{P_z}\,\hat=\,\lim_{z_4\to\infty}\frac{2\hat h_2}{z_4-z_2}=0 \ .
\end{equation}
Hence, we conclude that in the large-$z_4$ limit,
\begin{equation}
    \partial_{z_2}G=P_zP_x\tilde D\partial_z\mathcal F \ .
\end{equation}
The rhs of~\eqref{eq:null_KZ_for_correlator}, on the other hand, becomes
\begin{equation}
    P_zP_x\tilde D\sum_{i=1,3,4}\frac{(x_i-x_2)P^{-1}_x\partial_{x_i}(P_x\mathcal F)+j_i\mathcal F}{z_i-z_2} \ .
\end{equation}
As in the case of the null state differential equation~\eqref{eq:null_state_diff_eq_for_correlator}, the $i=4$ term in the sum vanishes in the large-$z_4$ limit. The derivation of this fact is very similar to the one we gave earlier. The other two terms become,
\begin{equation}
    P_x^{-1}\partial_{x_i}(P_x\mathcal F)=\partial_{x_i}\log{P_x}\mathcal F+\partial_{x_i}x\partial_x\mathcal F \ ,
\end{equation}
while from before~\eqref{eq:partial_logPx},
\begin{equation}
    \partial_{x_1}x\,\hat=\,x-1 \ , \ \partial_{x_3}x\,\hat=\,-x \ , \ \partial_{x_1}\log{P_x}\,\hat=\,-d \ , \ \partial_{x_3}\log{P_x}\,\hat=\,d \ ,
\end{equation}
which imply that,
\begin{equation}
    P^{-1}_x\partial_{x_1}(P_x\mathcal F)\,\hat =\,-d\mathcal F+(x-1)\partial_x \mathcal F \ , \ P^{-1}_x\partial_{x_3}(P_x\mathcal F)\,\hat\, =d\mathcal F-x\partial_x \mathcal F \ .
\end{equation}
Putting everything together, eq.~\eqref{eq:null_KZ_for_correlator} takes the form,
\begin{equation}\label{eq:null_KZ_for_F}
   \begin{aligned}
       & \partial_z \mathcal F=\left(\frac{x(x-1)}{z}-\frac{x(x-1)}{z-1}\right)\partial_x \mathcal F-\left(\frac{j_1+dx}{z}+\frac{d(1-x)+j_3}{z-1}\right)\mathcal F \ , \\
       &\ \ \Rightarrow\quad \Big[z(z-1)\partial_z+x(x-1)\partial_x\Big]\mathcal F=\Big[dx+j_1-z\Big(j_4-\frac{\fk+2}{2}\Big)\Big]\mathcal F \ .
   \end{aligned}
\end{equation}

\subsection{Determining the degenerate conformal block}

We therefore have the following system of equations for the conformal block $\mathcal{F}(x,z)$,
\begin{equation}\label{eq:F_system}
    \begin{cases}
        x(x-1)(x-z)\partial_x\mathcal F=\Big[d(x^2-2xz+z)+2j_1x(1-z)+2j_3z(1-x)\Big]\mathcal F \ , \\
        \Big[z(z-1)\partial_z+x(x-1)\partial_x\Big]\mathcal F=\Big[dx+j_1-z\Big(j_4-\frac{\fk+2}{2}\Big)\Big]\mathcal F \ .
    \end{cases}
\end{equation}
The beautiful fact is that these equations admit a unique solution (at least locally, \ie up to monodromies), completely determining the four-point function $G$~\eqref{eq:G-4-pt-func}.

The null state differential equation can be written as,
\begin{equation}\label{eq:partialx_F/F}
    \frac{\partial_x\mathcal F}{\mathcal F}=\frac{2j_3+d}{x}+\frac{2j_1+d}{x-1}+\frac{r}{x-z} \ ,
\end{equation}
for $r=\frac{\fk+2}{2}-j_1-j_3-j_4$, which integrates to
\begin{equation}\label{eq:x_F}
    \mathcal F(x,z)=g(z)x^{2j_3+d}(x-1)^{2j_1+d}(x-z)^r \ ,
\end{equation}
where $g(z)$ is an arbitrary function of $z$. Its precise form can be determined via inserting~\eqref{eq:x_F} into the KZ equation. More specifically, given eq.~\eqref{eq:partialx_F/F} and the fact that,
\begin{equation}
    \frac{\partial_z \mathcal F}{\mathcal F}=\frac{g'(z)}{g(z)}-\frac{r}{x-z} \ ,
\end{equation}
we can substitute back in the KZ eq.~\eqref{eq:null_KZ_for_F}, to get
\begin{equation}
    \frac{g'(z)}{g(z)}=\frac{j_1}{z}+\frac{j_3}{z-1} \ ,
\end{equation}
which forces $g(z)$ to be of the form
\begin{equation}
    g(z)=z^{j_1}(z-1)^{j_3} \ .
\end{equation}
In conclusion, the homogeneous system~\eqref{eq:F_system} locally admits a unique solution for the conformal block, up to an overall multiplicative constant, which reads
\begin{equation}\label{eq:degen_conformal_block_F}
    \mathcal F(x,z)=z^{j_1}(z-1)^{j_3}x^{2j_3+d}(x-1)^{2j_1+d}(x-z)^{r} \ ,
\end{equation}
where, recall
\begin{equation}
    d=j_4-j_1-\frac{\fk+2}{2}-j_3 \ , \ r=\frac{\fk+2}{2} -j_1-j_3-j_4 \ .
\end{equation}
Notice that the conformal block~\eqref{eq:degen_conformal_block_F} has the expected branch cut behavior when $x,z\to0,1$. The branch cut when $x\to z$ is somewhat surprising from the CFT point of view, but can be interpreted as some sort of worldsheet instanton contribution (from the AdS$_3$ string theory point of view), where the worldsheet extends out to the asymptotic boundary of AdS~\cite{Maldacena:2001km}.

\section{Background on $\n=2$ SCFTs}\label{ap:N=2_SCFTs}

The purpose of this appendix is to review basic aspects of two-dimensional $\n=2$ superconformal field theories, as well as the degenerate representations of the $\n=2$ superconformal algebra.

\subsection{Preliminaries of $\n=2$ SCFTs}\label{ap:N=2_SCFTs_Basics}

A $(2,2)$ superconformal field theory possesses a stress-tensor $T$, two spin-$3/2$ fermionic supercurrents $G^1,G^2$ and a $\mathrm U(1)_R$ R-symmetry current $J$ that rotates them -- and similarly for the anti-holomorphic sector. It is useful to combine the supercurrents into the following complex supercurrents $G^\pm \equiv \tfrac{1}{\sqrt{2}}(G^1 \pm \ii G^2)$. The corresponding Laurent expansions read,
\begin{equation}
    T=\sum_{n\in\mathbb{Z}}\frac{L_n}{z^{n+2}} \ , \quad J=\sum_{a\in\mathbb{Z}}\frac{J_a}{z^{a+1}}  \ , \quad G^\pm=\sum_{r\in\mathbb{Z}+\nu}\frac{G^\pm_r}{z^{r+\frac{3}{2}}} \ ,
\end{equation}
where $\nu=1/2$ in the NS-sector and $\nu=0$ in the Ramond sector. These Laurent modes obey the standard $\mathcal{N}=2$ superconformal algebra (SCA),
\begin{equation}\label{eq:N=2-SCA}
    \n=2 \, \text{SCA} \begin{cases}
        [L_n,L_m]=(n-m)L_{n+m}+\frac{c}{12}(n^3-n)\delta_{n+m,0} \ ,\\
        [L_n,G^\pm_r]=(\tfrac{1}{2}n-r)G^\pm_{n+r} \ ,\\
        [L_n,J_a]=-aJ_{n+a} \ , \\
        [J_a,J_b]=\frac{c}{3}a\delta_{a+b,0} \ ,\\
        [J_a,G^\pm_r]=\pm G^\pm_{a+r} \ ,\\
        \lbrace G^+_r,G^-_s\rbrace=2L_{r+s}+(r-s)J_{r+s}+\frac{c}{3}(r^2-\tfrac{1}{4})\delta_{r+s,0} \ ,\\
        \lbrace G^\pm_r,G^\pm_s\rbrace=0 \ ,
    \end{cases}
\end{equation}
with trivial (anti-)commutation relations between the holomorphic and anti-holomorphic modes. In particular, the first commutation relation implies that the theory is conformally invariant, the second and the third that $G^\pm,J$ are primaries with dimensions $\frac{3}{2},1$ respectively, the fourth that $J$ generates a $\mathrm U(1)_R$ symmetry, the fifth that $G^\pm$ is charged under $\mathrm U(1)_R$ and the sixth that successive supersymmetry transformations result in a conformal transformation and a $\mathrm U(1)_R$ rotation. The $\n=2$ SCA~\eqref{eq:N=2-SCA} can as usual be derived from the $\n=(2,2)$ superspace. Furthermore, note that the global $\n=2$ superconformal algebra consists of the following $8$ generators
\begin{equation}
    L_0,J_0,L_{\pm1},G^\pm_{\pm\frac{1}{2}} \ ,
\end{equation}
and is termed $\mathrm{osp}(2|2)$, with the corresponding global superconformal group being $\mathrm{OSP}(2|2,\mathbb C)$.

\paragraph{$\n=2$ super-Verma module}
The Cartan subalgebra of the $\n=2$ SCA is spanned by $L_0,J_0$ and therefore representations of the super-Verma module will be labeled by their scaling dimension $h$ and their R-charge $q_R$. The $\n=2$ super-primaries $|h,q_R\rangle$ are then defined in the obvious way, namely
\begin{equation}
    \n=2\,\, \text{super-primaries}\,\ \begin{cases}
        L_0|h,q_R\rangle=h|h,q_R\rangle \ ,\\
        J_0|h,q_R\rangle=q|h,q_R\rangle \ ,\\
        L_n|h,q_R\rangle=J_a|h,q_R\rangle=G^\pm_r|h,q_R\rangle=0 \ , \forall n,a,r>0 \ .
    \end{cases}
\end{equation}
The corresponding super-Verma module can be obtained by acting on $|h,q_R\rangle$ with the generators $L_{n},J_{-a},G^\pm_{-r}$, for $n,a,r>0$.

\paragraph{BPS operators} The full $(2,2)$ theory enjoys four Poincaré supercharges $\mathcal{Q}^\pm,\bar{\mathcal{Q}}^\pm$, obtained via integrating the global supercurrent modes $G^\pm_{-\frac{1}{2}},\bar G^\pm_{-\frac{1}{2}}$. As usual, BPS operators are in one-to-one correspondence with the cohomology classes of the supercharges. One can, therefore, define (\eg holomorphic) \textit{chiral} and \textit{anti-chiral primaries} as~\cite{Lerche:1989uy}
\begin{equation}
    G^+_{-\frac{1}{2}}|h,q_R\rangle=0 \quad \text{(chiral)}~, 
    \qquad 
    G^-_{-\frac{1}{2}}|h,q_R\rangle=0 \quad \text{(anti-chiral)}~.
\end{equation}
Then, the $\mathcal{N}=2$ algebra~\eqref{eq:N=2-SCA} implies
\begin{equation}
    \{G^+_{-\frac{1}{2}},G^-_{\frac{1}{2}}\} = 2L_0 - J_0 \ , 
    \qquad 
    \{G^+_{\frac{1}{2}},G^-_{-\frac{1}{2}}\} = 2L_0 + J_0 \ ,
\end{equation}
and requiring $(G^\pm_{-1/2})^\dagger = G^\mp_{1/2}$, we get
\begin{equation}
    \big\|G^+_{-\frac{1}{2}}|h,q_R\rangle\big\|^2 = (2h - q_R)\| |h,q_R\rangle\|^2 \ , 
    \qquad 
    \big\|G^-_{-\frac{1}{2}}|h,q_R\rangle\big\|^2 = (2h + q_R)\| |h,q_R\rangle\|^2 \ .
\end{equation}
Assuming that the theory is unitary, \ie requiring the norms being positive semi-definite, yields the BPS bound
\begin{equation}\label{eq:BPS_bound}
    h \geq \frac{|q_R|}{2} \ ,
\end{equation}
which is saturated iff the primary is (anti-)chiral. Similar considerations apply for the anti-holomorphic sector as well. Consequently, we overall have three types of primary operators in the theory; non-BPS operators, $\frac{1}{4}$-BPS operators that are either holomorphic (anti-)chiral or anti-holomorphic (anti-)chiral, and $\frac{1}{2}$-BPS operators that are holomorphic (anti-)chiral and anti-holomorphic (anti-)chiral. The latter are further partitioned into four categories,
\begin{equation}
    (c,c) \ , \quad (c,a) \ , \quad (a,c) \ , \quad (a,a) \ ,
\end{equation}
with the entries in the parenthesis denoting chiral ($c$) or anti-chiral ($a$) representations, in the holomorphic and anti-holomorphic sectors, respectively.

It is straightforward to show~\cite{Lerche:1989uy} that the OPE of $\frac{1}{2}$-BPS operators of the same type has no singular terms, while forming a ring at coincident insertion points -- commonly referred to as the \textit{chiral ring} (see \eg eq.~\eqref{eq:def_chiral_ring}). For selection rules on what can generically appear in the OPE of $\frac{1}{2}$-BPS primaries and further discussion, see \eg ref.~\cite{Lin:2016gcl}.

\paragraph{Spectral flow} The $\mathcal{N}=2$ superconformal algebra~\eqref{eq:N=2-SCA} has an outer automorphism, which goes by the name of \textit{spectral flow}~\cite{Schwimmer:1986mf}. Spectral flow is labeled by a real parameter $\beta$ and acts on the modes of the algebra as,
\begin{equation}\label{eq:spectral flow1}
    L_n \mapsto L_n + \beta J_n + \frac{c}{6}\beta^2 \delta_{n,0}~,\quad J_a \mapsto J_a + \frac{c}{3}\beta \delta_{a,0}~,\quad G_r^\pm \mapsto G_{r\pm \beta}^\pm~,
\end{equation}
\ie if the modes $\{L_n,J_a,G_r^\pm\}$ satisfy~(\ref{eq:N=2-SCA}) then so do the flowed generators. On a primary state $|h,q_R\rangle$ spectral flow acts as 
\begin{equation}\label{eq:spectral flow}
    h \mapsto h + \beta q_R +\frac{c}{6}\beta^2~,\quad q_R \mapsto q_R + \frac{c}{3}\beta~.
\end{equation}
For instance, half-integer spectral flow maps NS-sector states to R-sector states, and vice versa. This means that when \eg $\beta=-\frac{1}{2}$ chiral primaries get mapped to Ramond ground states.

\paragraph{Mirror symmetry} Another outer automorphism of the combined left and right-moving $\n=2$ algebras is the following,
\begin{equation}
    \lbrace L_n,J_a,G^\pm_r\rbrace\mapsto \lbrace L_n,J_a,G^\pm_r \rbrace\ , \qquad \lbrace \bar L_n,\bar J_a,\bar G^\pm_r\rbrace\mapsto \lbrace \bar L_n,-\bar J_a,\bar G^\mp_r \rbrace \ ,
\end{equation}
which, exchanges the sign of the anti-holomorphic R-charge, \ie $\bar q_R\to -\bar q_R$, while leaving the holomorphic R-charge and conformal dimensions intact. Therefore, under this automorphism the $(c,c)$ chiral ring becomes the $(c,a)$ chiral ring, and similarly for their Hermitian conjugate rings $(a,a)$ and $(a,c)$. This automorphism is in fact the worldsheet statement of \textit{mirror symmetry}~\cite{Hori:2000kt,Hori:2003ic}.

\subsection{$\n=2$ degenerate representations}\label{ap:degen_reps}

As in the case of $\n=0,1$ (S)CFTs in  two-dimensions, the degenerate representations of the (super) Virasoro algebra can be read off from the zeroes of the Kač determinant. The $\n=2$ Kač determinant is known~\cite{Nam:1985qe, Boucher:1986bh}. In the NS-sector, up to an overall unimportant constant, it reads
\begin{equation}
    \det{M_N^\text{NS}}\propto 
    \prod_{\substack{
        r,s\in\mathbb{Z}_{\geq 1} \\
        rs\leq 2N \\
        s\in 2\mathbb{Z}_{\geq 1}
    }}
    \big(h-h^\text{NS}_{\rsr}\big)^{
        p_{\tiny{\text{NS}}}\big(N-\frac{rs}{2},m\big)
    }
    \prod_{p\in\mathbb{Z}+\frac{1}{2}}
    \big(h-\tilde{h}^\text{NS}_{p}\big)^{
        \tilde{p}_{\tiny{\text{NS}}}
        \big(N-|p|,m-\text{sgn}(p);p\big)
    } \ ,
\end{equation}
where,
\begin{equation}
    \begin{aligned}
        &h^\text{NS}_{\rsr}=\frac{q^2_R\fb^2}{4}+\frac{1}{4\fb^2}-\frac{\fb^2}{16}\Big(\frac{2r}{\fb^2}+s\Big)^2\ ,\\
&\tilde{h}^\text{NS}_{p}=pq_R-\frac{1}{\fb^2}\Big(p^2-\frac{1}{4}\Big) \ ,
\end{aligned}
\end{equation}
for the $\n=2$ Liouville central charge parameterization $c=3(1+2\fb^{-2})$, whilst $p_\text{NS},\,\tilde p_\text{NS}$ count the number of NS sector descendants in each case, see~\cite{Nam:1985qe,Boucher:1986bh} for explicit expressions, and $m$ is the relative $\mathrm{U}(1)_R$ charge. From the Kač determinant we learn that, for any choice of $r\in\mathbb{Z}_{\geq 1},s\in2\mathbb{Z}_{\geq 1},p\in\mathbb{Z}+\frac{1}{2}$, a NS primary $|h^\text{NS}_{\rsr},q_R\rangle$ has a null state at level $\frac{rs}{2}$ with the same R-charge $q_R$, while the primary $|h^\text{NS}_{p},q_R\rangle$ has a null state at level $|p|$ with R-charge $q_R+\text{sgn}(p)$. In particular, the $\n=2$ BPS primaries correspond to degenerate representations with $p=\pm\frac{1}{2}$. The degenerate representations with dimensions $h^\text{NS}_{\rsr}$ are similar in spirit to those of $\n=0,1$ (super) Virasoro, while the difference in the $\n=2$ case comes from the degenerate representations with dimension $h^\text{NS}_p$ where the corresponding null states have different R-charge.

Similarly, the R-sector $\n=2$ Kač determinant is given by,
\begin{equation}
    \det{M_{N}^\text{R}}\propto 
    \prod_{\substack{
        r,s\in\mathbb{Z}_{\geq 1} \\
        rs\leq 2N \\
        s\in 2\mathbb{Z}_{\geq 1}
    }}
    \big(h-h^{\text{R},\pm}_{\rsr}\big)^{
        p_{\tiny{\text{R}}}\big(N-\frac{rs}{2},m\big)
    }
    \prod_{p\in\mathbb{Z}}
    \big(h-\tilde{h}^{\text{R},\pm}_{p}\big)^{
        \tilde{p}_{\tiny{\text{R}}}
        \big(N-|p|,m-\text{sgn}(p);p\big)
    } \ ,
\end{equation}
with,
\begin{equation}
    \begin{aligned}
        &h^{\text{R},\pm}_{\rsr}=\frac{q_R^2\fb^2}{4}+\frac{1}{4\fb^2}-\frac{\fb^2}{16}\Big(\frac{2r}{\fb^2}+s\Big)^2+\frac{1}{8}\equiv h^\text{NS}_{\rsr}+\frac 18\ ,\\
&\tilde{h}^{\text{R},\pm}_{p}=pq_R-\frac{1}{\fb^2}\Big(p^2-\frac{1}{4}\Big)+\frac 18\equiv h^\text{NS}_p+\frac 18 \ ,
\end{aligned}
\end{equation}
where, as before $p_\text{R},\,\tilde p_\text{R}$ enumerate the R-sector descendants and $m$ is the relative R-charge, with the $\pm$ superscripts indicating the double degeneracy of R-sector highest weight representations. From the determinant formula it follows that primaries $|h^{\text{R},\pm}_{\rsr},q_R\rangle$ have a null state at level $\frac{rs}{2}$ with R-charge $q_R\mp\frac 12$, while primaries with dimension $h^{\text{R},\pm}_p$ have a null state at level $|p|$ with $q_R\mp\frac 12 +\text{sgn}(p)$ R-charge. For instance, the degenerate representations with $h^{\text{R},\pm}_{p=0}=\frac{c}{24}$ correspond to the Ramond ground states, which is expected since spectral flow maps these states to NS-sector (anti-)chiral primaries that we already argued correspond to degenerate representations of the $\n=2$ algebra.

\bibliographystyle{JHEP}
\bibliography{bib}
\end{document}